\documentclass[aps,prd,twocolumn]{revtex4-2}
\usepackage{blindtext}
\usepackage{mathtools,mathalfa,amssymb,amsmath,amsfonts,amsthm,bm}
\usepackage{appendix}
\usepackage{calrsfs}
\usepackage{color}
\usepackage[mathscr]{euscript}

\renewcommand{\d}{\mathrm{d}}
\newcommand{\re}{\mathrm{Re}}
\newcommand{\im}{\mathrm{Im}}

\newcounter{definition}
\newenvironment{definition}
  {\refstepcounter{definition}
   \vspace{1 em}
   \noindent{\bf Definition~\thedefinition:}
   \begin{em}}
  {\end{em}}

\newcommand{\pullback}[1]{\hbox{\lower0.5ex\hbox{${}_{\leftarrow}$}}\kern-1.9ex{#1}}
\newcommand{\pullbacklong}[1]{\hbox{\lower0.85ex\hbox{${}_{\longleftarrow}$}}\kern-3.0ex{#1}}
\newcommand{\pullbackllong}[1]{\hbox{\lower0.85ex\hbox{${}_{\longleftarrow\!\!-\!\!-\!\!-\!\!-}$}}\kern-6.4ex{#1}}

\begin{document}
\title{Tidal deformations of slowly spinning isolated horizons}
\author{Ariadna Ribes Metidieri}
\email{ariadna.ribesmetidieri@ru.nl}
\affiliation{Institute for Mathematics, Astrophysics and Particle Physics, Radboud University, Heyendaalseweg 135, 6525 AJ Nijmegen, The Netherlands}

\author{B\'eatrice Bonga}
\email{bbonga@science.ru.nl}
\affiliation{Institute for Mathematics, Astrophysics and Particle Physics, Radboud University, Heyendaalseweg 135, 6525 AJ Nijmegen, The Netherlands}

\author{Badri Krishnan}
\email{badri.krishnan@ru.nl}
\affiliation{Institute for Mathematics, Astrophysics and Particle Physics, Radboud University, Heyendaalseweg 135, 6525 AJ Nijmegen, The Netherlands}
\affiliation{Albert-Einstein-Institut, Max-Planck-Institut f{\"u}r Gravitationsphysik, Callinstra{\ss}e 38, 30167 Hannover, Germany}
\affiliation{Leibniz Universit{\"a}t Hannover, 30167 Hannover, Germany}

\begin{abstract}

  It is generally believed that tidal deformations of a black hole in
  an external field, as measured using its gravitational field
  multipoles, vanish.  However, this does not mean that the black hole
  horizon is not deformed.  Here we shall discuss the deformations of
  a black hole horizon in the presence of an external field using a
  characteristic initial value formulation. Unlike existing methods,
  the starting point here is the black hole horizon itself. The effect
  of, say, a binary companion responsible for the tidal deformation is
  encoded in the geometry of the spacetime in the vicinity of the
  horizon.  The near horizon spacetime geometry, i.e. the metric, spin
  coefficients, and curvature components, are all obtained by
  integrating the Einstein field equations outwards starting from the
  horizon. This method yields a reformulation of black hole
  perturbation theory in a neighborhood of the horizon.  By
  specializing the horizon geometry to be a perturbation of Kerr, this
  method can be used to calculate the metric for a tidally deformed
  Kerr black hole with arbitrary spin.  As a first application, we
  apply this formulation here to a slowly spinning black hole and
  explicitly construct the spacetime metric in a neighborhood of the
  horizon.  We propose natural definitions of the electric and
  magnetic surficial Love numbers based on the Weyl tensor component
  $\Psi_2$.  From our solution, we calculate the tidal perturbations
  of the black hole, and we extract both the field Love numbers and
  the surficial Love numbers which quantify the deformations of the
  horizon.

\end{abstract}

\maketitle

\section{Introduction}
\label{sec:intro}

The response of a system to an external perturbation depends on its
constitution. Therefore, understanding this response allows us to
infer the constitutive properties of a system. This applies equally to
atoms and molecules, as well as to stars.  In a gravitationally bound
binary system, each of the binary components is tidally deformed by
the gravitational field of its companion.  Within the linear
approximation, the quadrupolar deformation is proportional to the
strength of the external quadrupolar field, and the constant of
proportionality determines the so-called (quadrupolar) Love number.
This tidal deformation also leaves its imprint in various observations
of the binary.  In the case of a binary system consisting of two
neutron stars, this tidal deformation leads to modifications of the
emitted gravitational wave signal, which can be used to deduce the
equation of state of the nuclear matter making up the neutron stars
\cite{Flanagan:2007ix}.  This method has been employed in the analysis
of gravitational wave data from binary neutron star merger events to
constrain the equation of state of neutron star matter and to
determine neutron star radii (see
e.g. \cite{LIGOScientific:2018hze,De:2018uhw,Capano:2019eae}).  Black
holes, within standard general relativity, are found to have vanishing
Love numbers
\cite{Poisson:2021yau,Binnington:2009bb,LeTiec:2020bos,LeTiec:2020spy,Charalambous:2021mea,Pani:2015hfa,Pani:2018inf,Damour:2009vw,Damour:2009wj,Gurlebeck:2015xpa}.
Thus, gravitational wave observations by themselves can potentially
allow us to distinguish between black holes and neutron stars.

Tidal perturbations also play an important role in extreme mass ratio
systems, consisting of a supermassive black hole with a stellar mass
companion. The spacetime is, to an excellent approximation (away from
the location of the stellar mass companion), well-modeled by that of a
tidally perturbed black hole.  Such systems are important targets for
the LISA detector \cite{Babak:2017tow}.  The stellar mass effectively
maps the spacetime of the larger black hole, thereby providing a very
sensitive probe of possible deviations from the Kerr spacetime and
general relativity \cite{Ryan:1995wh,Barack:2006pq}.

When talking about tidal deformations within general relativity, one
needs to distinguish between field and surficial deformations,
i.e. deformations of the asymptotic gravitational field of the object
in question, versus deformations of the shape of the object itself.
Within Newtonian gravity, due to its linearity, both of these
different ways of quantifying tidal deformations are equivalent. This
is not the case in general relativity, and one needs to distinguish
between field and surficial Love numbers. In other words, calculating
the multipole moments of the gravitational field in Newtonian gravity
is equivalent to calculating the source multipole moments of the mass
distribution within the star.  This simple correspondence does not
hold in general relativity and the two sets of multipole moments can
be quite different \cite{Ashtekar:2004gp,Damour:2009va,Landry:2014jka}.  The
statement that the Love number of a black hole vanishes refers to the
asymptotic field moments. In fact, the shape of a black hole is
explicitly seen to change in the presence of an external field.  This
is confirmed by known solutions (see
e.g. \cite{Geroch:1982bv,Fairhurst:2000xh}), perturbative calculations
(e.g. \cite{Damour:2009va,Hartle:1973zz,Hartle:1974gy,OSullivan:2014ywd,OSullivan:2015lni,Cabero:2014nza})
and numerical simulations of binary mergers (see
e.g. \cite{Gupta:2018znn,Prasad:2020xgr,Prasad:2021dfr,Prasad:2024vsz}).  On the
other hand, the field Love numbers are believed to appear in the
gravitational wave signal (see
e.g. \cite{Damour:2009wj,Damour:2009wj,Flanagan:2007ix,Hinderer:2009ca}). However,
in the late inspiral phase of a binary merger when the two compact
objects are very close to each other, the surficial Love numbers will
provide a more economical description of the near horizon metric.
Thus, one might conjecture that in this regime of the black hole
merger process, these Love numbers might be measurable in the
gravitational wave signal as well; this will be discussed further
in Sec.~\ref{sec:lovenumbers}.

The starting point for the calculation of the Love numbers is to
determine the response of a compact object of mass $M$ immersed in an
external gravitational field.  If $\mathscr{R}$ is the local radius of
curvature of the external gravitational field at the location of the
compact object, and if we assume the black hole mass $M$ is much
smaller than $\mathscr{R}$, the dimensionless small parameter
$M/\mathscr{R}$ determines the perturbations of the local spacetime
geometry and of the matter field configuration within the compact
object (if any matter fields are present).  One strategy for
calculating the local gravitational field in the vicinity of the
compact object can be summarized as follows
\cite{Thorne:1984mz,DEath:1975jps,Manasse:1963zz,Manasse:1963a,Poisson_2010}.
We start with the spacetime metric $g_{ab}^{(0)}$; this is the
background metric on which the black hole moves.  Consider then a
world-line located at the position of the compact object.  The
spacetime metric $g_{ab}^{(0)}$ in the vicinity of the world-line can
then be expanded in powers of $r/\mathscr{R}$ \cite{Poisson:2011nh}.
In the presence of the compact object, the spacetime metric $g_{ab}$
will be modified away from $g_{ab}^{(0)}$, and can be expanded in
powers of $M$.  On the other hand, the metric can also be written as
that of a perturbed black hole, e.g. as a perturbation of the
Schwarzschild or Kerr metric. Matching these two approximations and
using the Einstein equations then yields the tidally deformed black
hole metric, and also the values of the Love numbers.  The black hole
horizon is generally also perturbed away from its original coordinate
location, and the location and geometrical/physical properties of the
perturbed horizon needs to be calculated explicitly using the tidally
deformed black hole metric obtained from the above calculation.

Tidal perturbations have been extensively studied using the above
formalism for non-spinning, i.e. Schwarzschild black holes and slowly
spinning Kerr black holes \cite{Poisson:2014gka}. More recently it has
also been applied to arbitrary spinning Kerr black holes
\cite{LeTiec:2020bos,LeTiec:2020spy,Charalambous:2021mea}.  These
calculations are sufficiently involved that alternate approaches can
provide additional insight.  An important alternate approach to this
problem is the use of Effective Field Theory techniques (see
e.g. \cite{Ivanov:2022qqt,Ivanov:2022hlo}).  Here we shall present yet
another alternate approach to tidal perturbations which starts from
the horizon structure and allows one to treat a general deformed Kerr
horizon.  It also allows one to incorporate external matter fields and
potentially also alternate theories of gravity (as long as there is a
horizon structure available).

We rely on two key ingredients.  The first is that the geometry of
black hole horizons has been thoroughly studied in a quasi-local
framework which leads to the notions of isolated and dynamical
horizons
\cite{Ashtekar:1998sp,Ashtekar:1999yj,Ashtekar:2000hw,Ashtekar:2000sz,Ashtekar:2001is,Ashtekar_2002,Adamo:2009cd,Lewandowski:2006mx,Lewandowski:2000nh}.
These notions allow one to study horizons without assuming global
stationarity and symmetries. Thus, for isolated horizons where the
black hole is not absorbing energy and is time-independent, the rest
of the universe is allowed to be dynamical.  The second ingredient is
a construction of the spacetime in the vicinity of an isolated
horizon.  Working within a characteristic initial value formulation,
we start with the intrinsic horizon geometry and integrate the
Einstein field equations outwards
\cite{Friedrich:1983vi,Lewandowski:1999zs,Lewandowski:2018khe,Dobkowski-Rylko:2018ahh,Krishnan:2012bt}.
A tidal perturbation of the horizon leads to corresponding
perturbations of the near horizon geometry.  Our goal in this work is
to carry through this calculation in detail and to obtain the near
horizon geometry for a general distorted rotating black hole.  We
present this formalism for black hole perturbation theory and
illustrate it for the well-known case of a tidally perturbed
Schwarzschild black hole, allowing for small spins.  Subsequent work
will apply this method to perturbations of a Kerr black hole with
arbitrary spin.

The main feature of our approach will be the centrality of the horizon
geometry itself.  As mentioned above, requiring the inner boundary to
be an isolated horizon assumes that there is no infalling radiation.
Is this a valid assumption, or at least a useful starting point?
Numerical simulations of binary black hole mergers show that the two
individual horizons are isolated to a good approximation, even very
close to the time when the common horizon forms
\cite{PhysRevLett.123.171102}. One might therefore expect this to be a
good starting point (though it should be noted that the infalling flux
is not vanishing and can be numerically measured
\cite{Prasad:2020xgr,Prasad:2024vsz}).  It has also been found in previous studies
that tidally perturbed black hole horizons are indeed isolated at
leading order, and that the fluxes of infalling radiation can be
calculated at linear order in perturbation theory \cite{Vega_2011}.
Given this evidence, we shall take as a working hypothesis that the
horizon is isolated and we shall investigate the near horizon
geometries compatible with this assumption in greater detail than done
before.  For example, we shall show generally that including a tidal
horizon perturbation on a Kerr black hole implies that the neighboring
spacetime must be radiative with a non-vanishing Weyl tensor component
$\Psi_4$ (transverse to the horizon), thereby connecting the algebraic
properties of the Weyl tensor to tidal perturbations.  In this paper,
we present detailed calculations for slowly spinning horizons, but
this statement is in fact true for a general Kerr black hole.

As we shall see, in the context of black hole perturbation theory, our
assumption of requiring the black hole to be \emph{exactly} isolated
corresponds to algebraically special perturbations.  This should be
viewed as a first approximation which can, and will, be relaxed in
future work.  Useful starting points in this direction are provided by
\cite{Ashtekar:2021kqj,Booth:2003ji,Booth:2012xm}: i) First,
\cite{Ashtekar:2021kqj} sets up the mathematical framework for
discussing perturbed isolated horizons and fluxes across it. ii) Going
to more dynamical situations, slowly evolving horizons (where the
horizon area increase is comparatively small) are discussed in
\cite{Booth:2003ji}.  iii) Finally, \cite{Booth:2012xm} constructs the
near horizon geometry in the vicinity of a fully non-perturbative
dynamical horizon.  Each of these notions will have useful
applications in the context of tidally perturbed black holes, even in
the late inspiral stage of a binary black hole merger.

The plan for the rest of this paper is the following.
Sec.~\ref{sec:prelim} introduces the basic definitions of isolated
horizons and the main results in the formalism.  This includes the
constraint equations on the horizon and the notions of mass, angular
momentum, and higher multipole moments, which will be used later.
Sec.~\ref{sec:nearhorizon} outlines the procedure for constructing a
near horizon geometry within a characteristic initial value
formulation of the Einstein equations as pioneered by Friedrich and
Stewart \cite{Friedrich:1983vi}. This section uses the Newman-Penrose
formalism and also presents two examples of the construction, namely
the usual Schwarzschild metric in ingoing null coordinates, and the
Robinson-Trautman solution as an example of a radiative solution.
Sec.~\ref{sec:ihperturb} then discusses a perturbed horizon.  This
involves a perturbative analysis of the constraint equations on the
horizon. Sections \ref{sec:radial-perturb} and \ref{sec:tidal-bh} 
incorporate the perturbations in the construction of the near horizon
geometry and thereby obtain the metric of a tidally perturbed black
hole.  Finally, Sec.~\ref{sec:lovenumbers} discusses some aspects of
our calculations related to different notions of tidal Love numbers,
as relevant for gravitational wave astronomy.  We conclude in
Sec.~\ref{sec:conclusion}. The Appendices clarify some notation and
provide a short compendium of useful equations and results. There, we also
present some additional details not covered in the main text.

We conclude the introduction by providing a summary of the main
results presented in this paper.  i) We provide a construction of the
near horizon metric and spin coefficients of a tidally perturbed
isolated black hole analogous to the well-known Bondi construction
near null-infinity. This provides an unambiguous choice for the null
tetrad near the horizon, which allows for an unambiguous computation of the Weyl
tensor components in the Newman-Penrose formalism.  In this paper, we
apply this construction to a slowly spinning black hole.  ii) We
identify the Weyl tensor component $\Psi_2$ as the one that encodes
the information of the tidal perturbation.  At the horizon, it tells us
the distortion of the horizon electric and magnetic multipole moments,
while far away from the black hole (or in the limit when the black
hole mass is taken to be infinitesimally small), it also contains all
the information about the external tidal field.  This then allows us
to relate the source and field multipole moments for a tidally
perturbed black hole. iii) This leads to a natural definition of the
surficial Love numbers including also the magnetic surficial Love
numbers which, to our knowledge, has not been discussed previously in
the literature. Finally, iv) we note that there is an inherent
systematic uncertainty in the definitions of the field multipole
moments which follows from the procedure of matched asymptotic
expansions commonly employed in the literature. This uncertainty is
not new: it was already pointed out in the pioneering work by Hartle
and Thorne in 1984.  The surficial Love numbers do not suffer from the
same ambiguity, and thus provide a clearer construction of the
near-horizon geometry.

\section{Preliminaries}
\label{sec:prelim}

There is extensive literature on the properties of isolated horizons
covering mathematical, quantum, and physical aspects. This is part of
the still larger body of work on quasi-local horizons applicable to
time-dependent situations (see
e.g. \cite{Ashtekar:2004cn,Booth:2005qc,hayward2013black}).  The goal
of this section is to collect the main pre-requisites, concepts and
results, necessary for describing the geometry of tidally distorted
black hole horizons and the next section will deal with its near
horizon geometry.

\subsection{Basic Definitions}
\label{subsec:definitions}

The well-known Kerr-Newman black hole solutions within general
relativity have horizons with time-independent geometries.  Thus,
their area, angular momentum, charge, and in fact all higher moments
are time-independent.  This is hardly surprising since these
spacetimes are all globally stationary, and there are no fluxes of
infalling matter or radiation across the horizon.  While black holes
in our universe will not be exactly stationary, there are numerous
situations of black holes in dynamical spacetimes (such as in a binary
system) where time-dependent effects can be treated perturbatively.
However, it is important to not assume the notion of global
stationarity as in the Kerr-Newman black holes.  Thus, we should not
identify the ADM mass of the entire spacetime with the black hole
mass, and similarly for the angular momentum. This is evidently true
for a binary black hole system where the ADM mass and angular momentum
will include contributions from both black holes, and also other
contributions such as kinetic energies, radiation, and the interaction
energy between the black holes.

When the separation between the two black holes is sufficiently large,
one could attempt to identify the asymptotic regions of each black
hole and obtain approximate masses, spins, and higher multipole
moments.  However, this is perhaps not always viable in the late
inspiral stage when the separation between the two black holes would
be small (or at the very least, the systematic errors in the physical
parameter would grow). We shall discuss this further in
Sec.~\ref{sec:lovenumbers}.

In this work, we shall use the framework of quasi-local horizons,
restricted to the case of isolated horizons, to model a tidally
distorted black hole.  In general, this framework is based on the
notion of marginally trapped surfaces (to be discussed below), and it
provides a useful way of studying fully dynamical black holes without
reference to global notions such as event horizons and asymptotic
flatness.  It allows a clear formulation of the laws of black hole
mechanics
\cite{Ashtekar:1998sp,Ashtekar:1999yj,Ashtekar:2001is,Ashtekar:2000hw}
and black hole entropy calculations in quantum gravity (see
e.g. \cite{Ashtekar:1997yu,Ashtekar:2000eq}). It has proven to be
especially useful in numerical relativity when dealing with binary
black hole mergers (see
e.g. \cite{Dreyer:2002mx,Schnetter:2006yt,Gupta:2018znn,Owen:2009sb}).  In
simulations of binary black hole mergers, this allows one to calculate
mass, angular momentum, and higher multipole moments for each black
hole individually without reference to asymptotic infinity and without
reference to event horizons which cannot be located in real time.  See
e.g.  \cite{Ashtekar:2004cn,Booth:2001gx,Gourgoulhon:2005ng} for more
complete reviews.

Quasi-local horizons are capable of dealing
with a general time-varying horizon in a non-perturbative setting.
There are several important examples where black holes involved in
dynamical processes are almost isolated, and it makes sense to
consider a perturbative framework.  This occurs in binary systems not
only when the binary companion is far away (compared to the size of
the black hole), but is also valid surprisingly close to the merger.
See for example Figure 2 in \cite{PhysRevLett.123.171102}: It is seen
that in a head-on collision of two black holes, the area increase of
the two individual black holes is relatively moderate even when the
common horizon is formed.  An even more dramatic example is provided
by the Robinson-Trautman solutions
\cite{Robinson:1962zz,Chrusciel:1992rv} which will be discussed
further in Sec.~\ref{sec:robinson-trautman}.  In these black hole
solutions, we can have radiation arbitrarily close to the horizon. This
radiation is however transverse to the horizon and is not infalling,
and the horizon itself remains time-independent.
 
The black holes in all of the above examples are well modeled within
the isolated horizon framework or as perturbations thereof. The basic
mathematical objects to be understood are null, 3-dimensional
hypersurfaces in a spacetime.  We denote by $\Delta$ such a
hypersurface.  The intrinsic metric $q_{ab}$ on $\Delta$ is degenerate
and has signature $(0,+,+)$. Unlike spacelike or timelike manifolds,
we need to take some care in projecting tensor fields onto $\Delta$,
and care must be taken in the position of indices. The intrinsic
metric $q_{ab}$ is simply the restriction of the spacetime metric
$g_{ab}$: $q_{ab}X^aY^b = g_{ab}X^aY^b$ for any vector fields
$X^a,Y^b$ tangent to $\Delta$.  This is the pull-back of the spacetime
metric to $\Delta$: $q_{ab}=\pullbacklong{g_{ab}}$, where an
under-arrow indicates the pullback of the indices.  A null vector
$\ell^a$ tangent to $\Delta$ is said to be a null-normal to $\Delta$
if $q_{ab}\ell^b = 0$.  Since $\ell^a$ is null and also surface
orthogonal, its integral curves are geodesics so that
\begin{equation}
  \ell^a\nabla_a\ell^b = \kappa\ell^b\,,
\end{equation}
with $\kappa$ being the acceleration of $\ell^a$, i.e. the surface
gravity; $\nabla_a$ the spacetime derivative operator compatible
with the 4-metric $g_{ab}$.  We shall always take $\ell^a$ to be
future-directed.

Being degenerate, the inverse $q^{ab}$ is not unique but all of our
constructions will be insensitive to this ambiguity.  If $q^{ab}$ is
an inverse in the sense that $q_{am}q_{bn} q^{mn} = q_{ab}$, then so
is $q^{ab} + V^{(a}\ell^{b)}$ with $V^a$ being tangent to
$\Delta$. Given a null normal $\ell^a$ to $\Delta$, its expansion
$\Theta_{{(\ell)}}$ is defined as
\begin{equation}
  \Theta_{(\ell)} := q^{ab}\nabla_a\ell_b\,.
\end{equation}
This is insensitive to the non-uniqueness of $q^{ab}$.  Note that
$q_{ab}$, being degenerate, does not uniquely specify a derivative
operator. In fact, without additional assumptions or geometric
structures, there is not a unique torsion-free derivative operator on
$\Delta$ compatible with $q_{ab}$.

We shall be exclusively concerned with the case when $\Delta$ is ruled
by the integral curves of $\ell^a$ and has spherical cross-sections.
Thus it has topology $S^2\times\mathbb{R}$, as is the case for the
Schwarzschild or Kerr event horizons.  On every cross-section,
$q_{ab}$ induces a Riemannian 2-metric which we shall denote
$\widetilde{q}_{ab}$, and a corresponding volume 2-form
$\widetilde{\epsilon}_{ab}$.  Thus, the area of cross-sections is
measured by integrating $\widetilde{\epsilon}$. The above notions are
of course also applicable to the well-known Schwarzschild and Kerr
event horizons, which are stationary in the sense that the area is a
constant. It is easy to verify that for a Kerr black hole, every
cross-section of the horizon (as long as it is a complete sphere) has
the same area and one can therefore sensibly talk about the area as a
geometric invariant of the Kerr event horizon.  Since the area is
constant, the black hole can be considered ``isolated'' also in the
sense that it is in equilibrium and not interacting with its
surrounding spacetime and matter fields; any infalling matter or
radiation would lead to an increase in the area following the area
increase law. The framework of isolated horizons provides a systematic
treatment of this situation.

Isolated horizons are conveniently introduced in a series of three
definitions, starting from the weakest and imposing increasingly
stronger conditions.  We can now state the first definition:

\begin{definition}
  \label{def_1}
  A sub-manifold $\Delta$ of a space-time $(\mathcal{M},g_{ab})$ is
  said to be a \emph{non-expanding horizon} (NEH) if  
  \begin{enumerate}
  \item $\Delta$ is topologically $S^2\times \mathbb{R}$ and null.
    For the projection map
    $\Pi: S^2\times \mathbb{R} \rightarrow S^2$, the fiber
    $\Pi^{-1}(p)$ for any $p\in S^2$ are null curves in $\Delta$.
    \label{cond_1_1}
  \item Any null normal $\ell^a$ of $\Delta$ has vanishing expansion,
    $\Theta_{(\ell)}=0$. This condition is insensitive to the
    rescaling $\ell^a\rightarrow f\ell^a$ with $f$ being a positive
    definite function.
    \label{cond_1_2}
  \item All equations of motion hold at $\Delta$ and the
    stress energy tensor $T_{ab}$ is such that $-T^{a}{}_{b}\ell^b$ is
    future-causal for any future-directed null normal $\ell^a$.
    \label{cond_1_3}
  \end{enumerate}
\end{definition}

The second condition above is the critical one: it requires all
cross-sections of $\Delta$ to be marginally outer trapped
surfaces (MOTS). The last condition will not be relevant for us since we
shall work with vacuum spacetimes, but we keep it for completeness.

The shear of $\ell^a$, $\sigma^{ab}$ is
\begin{equation}
  \sigma_{ab} := \pullbackllong{\nabla_{(a} \ell_{b)}} - \frac{1}{2}
  \Theta_{{(\ell)}} q_{ab} \,.
\end{equation}
Using the Raychaudhuri equation and the energy condition,
condition~\ref{cond_1_2} can be shown to yield $\sigma_{ab}=0$.  Thus,
we conclude that $\pullbackllong{\nabla_{(a} \ell_{b)}} = 0$, which
also means that $\mathcal{L}_\ell q_{ab} = 0$.

\begin{figure}
  \begin{center}
    \includegraphics[angle=-0,width=\columnwidth]{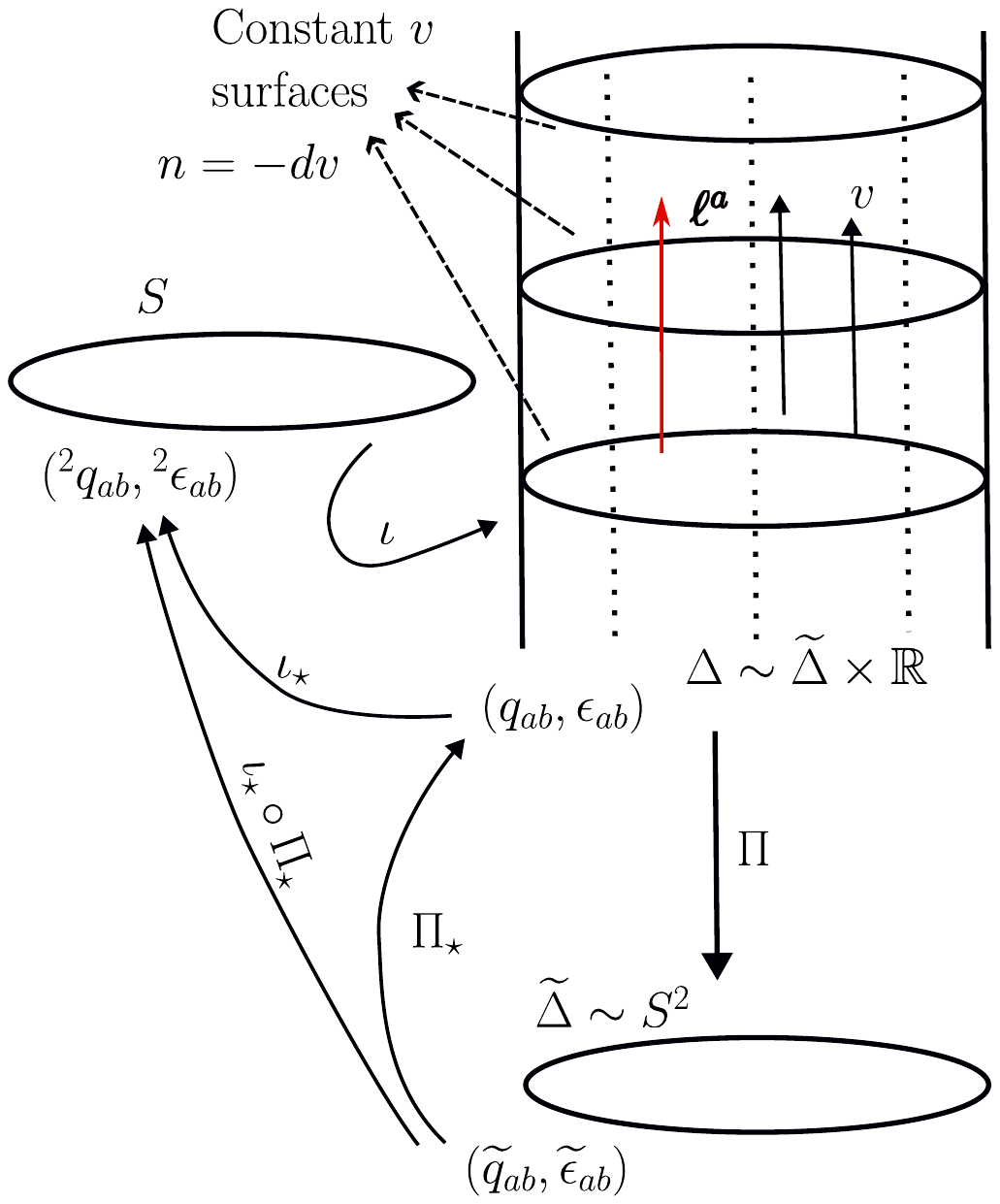}
    \caption{The projection map $\Pi$ and the foliation of the
      horizon. The NEH $\Delta$ is topologically $S^2\times\mathbb{R}$
      and projects to the base-space $\widetilde{\Delta}$ of spherical
      topology. The fields $(\widetilde{q}_{ab},\widetilde{\omega}_a)$
      live on the base space and can be pulled back through
      $\Pi_\star$ to the fields $(q_{ab},\omega_a)$ on $\Delta$. These
      are guaranteed to satisfy $\mathcal{L}_\ell q_{ab} = 0$ and
      $\mathcal{L}_\ell \omega = 0$.  For a choice of affine parameter
      $v$ along the null-normal $\ell^a$ (satisfying
      $\ell^a\nabla_av = 1$), the constant $v$ surfaces yield a
      foliation of $\Delta$ which are shown as cross-sections in the
      figure. Each of these cross-sections can be considered as an
      embedding of a manifold $S$ (again with topology $S^2$) into
      $\Delta$.  We can identify $S$ with $\widetilde{\Delta}$ in a
      natural way using the composition
      $\Pi\circ\iota:S\rightarrow \widetilde{\Delta}$. Thus, with a
      slight abuse of notation, we shall use the notation
      $\widetilde{q}_{ab}$ and $\widetilde{\epsilon}_{ab}$ instead of
      ${}^2q_{ab}$ and ${}^2\epsilon_{ab}$.  }\label{fig:projections}
  \end{center}
\end{figure}
We can introduce a derivative operator $\mathcal{D}$ on a NEH
$\Delta$. As mentioned before, the degeneracy of $q_{ab}$ implies that
there are an infinite number of torsion-free derivative operators
that are compatible with it. However, on an NEH, the property
$\pullbackllong{\nabla_{(a} \ell_{b)}} =0$ can be used to construct a
unique (torsion-free) derivative operator. It can be shown that this
condition signifies that the space-time connection $\nabla$ induces a
unique torsion-free derivative operator $\mathcal{D}$ on $\Delta$
which is compatible with $q_{ab}$~\cite{Ashtekar:2000hw}; thus
$\mathcal{D}_a = \pullbacklong{\nabla_a}$.  We thus need to specify
the pair $(q_{ab},\mathcal{D}_a)$ to fully characterize the geometry
of $\Delta$, and our strategy will be to strengthen the notion of a NEH
by imposing restrictions on various components of $\mathcal{D}_a$.

Some of the various relevant geometric objects and manifolds are
indicated in Fig.~\ref{fig:projections}.  This figure shows the
different kinds of geometric objects in our problem, and it will be
worthwhile to elaborate on these briefly; details may be found in
\cite{Ashtekar_2002}.  Since $\Delta$ is a null surface, it is
non-trivial to raise and lower indices and it is important to keep
track of these.  We can project $\Delta$ to a topological sphere (the
``base-space'' $\widetilde{\Delta}$) by identifying points on $\Delta$
connected by a null generator.  We get in this way a natural
projection
$\Pi: \widetilde{\Delta} \times \mathbb{R} \rightarrow
\widetilde{\Delta}$.  It is straightforward to generalize
$\widetilde{\Delta}$ to be a compact manifold without boundary, but we
shall restrict ourselves to a sphere in this work.  We equip
$\widetilde{\Delta}$ with a Riemannian metric $\widetilde{q}_{ab}$
which gives us the derivative operator, volume element, and scalar
curvature $\widetilde{\mathcal{D}}_a$, $\widetilde{\epsilon}_{ab}$ and
$\widetilde{\mathcal{R}}$, respectively.  We can pull-back these
fields to $\Delta$ using the differential $\Pi_\star$ to obtain a
degenerate metric $q_{ab}$ and a 2-form $\epsilon_{ab}$ on $\Delta$:
\begin{equation}
  q_{ab} = \Pi_\star \widetilde{q}_{ab}\,,\quad   \epsilon_{ab} = \Pi_\star \widetilde{\epsilon}_{ab}\,. 
\end{equation}
These are evidently seen to satisfy
$\mathcal{L}_\ell q_{ab} = 0 = \mathcal{L}_\ell\epsilon_{ab}$, and
$q_{ab}\ell^b = 0 = \epsilon_{ab}\ell^b$.  

The foliation of the horizon requires a function $v$ whose level sets
give the leaves of the foliation.  We shall tie the null-normal to the
foliation by $\ell^a\nabla_a v = 1$ and $n_a=-\mathcal{D}_av$ (so that
$\ell\cdot n = -1$) is the 1-form orthogonal to the foliation.  A
given sphere of the foliation can be considered to be an embedding of
a sphere $S$ into $\Delta$, i.e. $\iota:S \rightarrow \Delta$.  This
map allows us to pull-back various fields to $S$; in the literature
one often uses the notation ${}^2q_{ab} = \iota_\star q_{ab}$ and
${}^2\epsilon_{ab} = \iota_\star\epsilon_{ab}$.  To avoid notational
clutter, we shall however generally not not use this notation, and we
shall use instead $\widetilde{q}_{ab},\widetilde{\epsilon}_{ab}$.
Thus, we shall use $\widetilde{q}_{ab}$ to refer to both the metric on
a cross-section $S$ and also on the base-space $\widetilde{\Delta}$,
and it shall be clear from the context which is meant.  This
discussion also makes clear that just as for the Kerr event horizon,
any complete spherical cross-section of $\Delta$ has the same
area. This area is a geometric invariant of $\Delta$, and we can talk
sensibly about ``the area $A$ of $\Delta$'', and its area radius
$R=\sqrt{A/4\pi}$.

When embedded in a spacetime manifold, we can consider $n_a$ to be the
pullback of a spacetime 1-form corresponding to a future-directed null
vector $n^a$; this is the ingoing null normal to $\Delta$. Finally, we
can complete $(\ell^a,n^a)$ to a null tetrad
$(\ell^a,n^a,m^a,\bar{m}^a)$ by introducing a complex null vector
$m^a$ tangent to the leaves of the foliation, such that
$\ell\cdot m = n\cdot m = 0$, and $m\cdot\bar{m} = 1$.  As we shall
see, this tetrad can be extended to a neighborhood of $\Delta$ and
tensor fields can be decomposed in terms of
$(\ell^a,n^a,m^a,\bar{m}^a)$. This forms the basis of the
Newman-Penrose formalism
\cite{Newman:1961qr,Penrose:1985jw,Chandrasekhar:1985kt,Stewart:1991}.
which we will summarize it in Sec.~\ref{sec:npsummary}.

On a NEH, there is no canonical scaling of the null generators:
$\ell^a$ and $f\ell^a$ (for any positive non-vanishing function $f$)
are both perfectly acceptable.  In the standard Schwarzschild/Kerr
solutions, we have globally defined timelike and rotational Killing
vectors available to us.  For a Schwarzschild black hole, the timelike
Killing vector is also a null generator of the horizon. Thus, for that
solution, we get a preferred null generator by normalizing the
timelike Killing vector to have unit norm at infinity.  A similar
strategy is also available in Kerr.  This strategy is generically not
viable because the spacetime in the vicinity of the isolated horizon
will generally not be stationary; thus we will not have access to
spatial infinity where the Killing vector could be normalized.  As we
shall see, it is nonetheless possible to single out a preferred class
of null normals on an isolated horizon.

Two null normals $\ell^a$ and $\tilde{\ell}^a$ to an NEH $\Delta$ are
said to belong to the same equivalence class $[\ell ]$ if
$\tilde\ell^a = c\ell^a$ for some positive \emph{constant} $c$.
Weakly isolated horizons are characterized by the property that, in
addition to the metric $q_{ab}$, the connection component
$\mathcal{D}_a\ell^b$ is also `time-independent'.  From the properties
of $\ell^a$ discussed above, it is easy to show that there must exist
a connection 1-form $\omega_a^{(\ell)}$ associated with any given
$\ell^a$ such that
\begin{equation}
  \mathcal{D}_a\ell^b = \omega^{(\ell)}_a\ell^b\,.
\end{equation}
The acceleration is given by
$\kappa_{(\ell)} = \ell^a\omega^{(\ell)}_a$.  It can be easily
verified that when $\ell^a\rightarrow f\ell^a$, $\omega_a$ undergoes a
gauge transformation:
\begin{equation}
  \omega^{(f\ell)}_a = \omega^{(\ell)}_a + \mathcal{D}_a\ln f\,.
\end{equation}
However, $\omega$ is invariant under constant rescalings, a fact which
will be useful for our next definition.

\begin{definition}
  \label{def_2}
  The pair $(\Delta, [\ell])$ is said to constitute a \emph{weakly
  isolated horizon} (WIH) provided $\Delta$ is an NEH and each null
  normal $\ell^a$ in $[\ell]$ satisfies
  \begin{equation}
    \label{wihcond}
    \mathcal{L}_\ell\omega_a = 0\,.
  \end{equation}
\end{definition}

On a weakly isolated horizon, since we are allowed only constant
rescalings, $\omega_a$ is invariant and we can drop the reference to
$\ell^a$ on $\omega^{(\ell)}$.  A WIH does not represent a
real physical restriction on a NEH.  We can always choose the
equivalence class $[\ell]$ on a NEH, but there is no unique
choice~\cite{Ashtekar_2002}. In numerous applications, a WIH is
sufficient and there is no need to impose any further
restrictions. The laws of black hole mechanics can be shown to hold
for WIHs \cite{Ashtekar:2000hw,Ashtekar:2001is} and they are also
sufficient for numerous applications in numerical relativity
simulations of black holes for calculating mass, angular momentum and
higher multipole moments (see
e.g. \cite{Dreyer:2002mx,Schnetter:2006yt,Gupta:2018znn}). The zeroth
law will in fact be useful for us.  This is the result that the
surface gravity $\kappa_{(\ell)}= \omega_a\ell^a$ is constant on
$\Delta$.

The condition $\mathcal{L}_\ell \omega_a = 0$ can be written as
\begin{equation}
  \left[\mathcal{L}_{\ell},\mathcal{D}\right]\ell^a = 0 \, .
\end{equation}
This form makes more explicit that this is a restriction on
$\mathcal{D}_a$.  An obvious generalization of this condition would be
to require that all components of $\mathcal{D}_a$ should be
`time-independent'. This leads us to our third definition:

\begin{definition}
  \label{def_3}
  The pair $(\Delta, [\ell])$ is said to constitute an \emph{isolated
    horizon} (IH) provided $\Delta$ is an NEH and each null normal
  $\ell^a$ in $[\ell]$ satisfies
  \begin{equation}
    \label{ihcond}
    \left[\mathcal{L}_{\ell},\mathcal{D}\right] = 0\,.
  \end{equation}
\end{definition}

If an equivalence class $[\ell]$ can be found that satisfies
Eq.~\eqref{ihcond} then the NEH is said to admit an IH structure.  We
shall later summarize the steps required for finding an
admissible $[\ell]$ on a NEH.

\subsection{Mass, angular momentum and higher multipoles}
\label{subsec:multipoles}

To define the physical parameters of a black hole, and for the laws of
back hole mechanics to hold on the horizon, it is sufficient to
consider a WIH.  Unlike other treatments of this topic where the basic
variables of a black hole are mass and angular momentum and the area
is a derived quantity, here it is more natural to begin with the area
and angular momentum.  We have already seen that the area $A$ (and
correspondingly, the radius $R$) is a geometric invariant on a NEH.
Expressions for angular momentum and mass are based on Hamiltonian
calculations within a suitable phase space.  Here the phase space
consists of a spacetime with a WIH as an inner-boundary. It is
possible to carry out the detailed calculation in either metric or
connection variables
\cite{Ashtekar:1999yj,Ashtekar:2000hw,Ashtekar:2001is,Booth:2001gx,Booth:2005ss,Booth:2003jp}.
Angular momentum is the Hamiltonian which generates rotations, while
energy is the generator of time translations.  In the context of a
diffeomorphism invariant theory like general relativity, the relevant
Hamiltonians are all integrals over the boundary 2-surfaces which in
our case, are cross-sections of a WIH. This allows a clear
identification of the energy and angular momentum of an axisymmetric
WIH.  Let us consider a WIH in a vacuum spacetime with an axial
symmetry $\varphi^a$, i.e.
\begin{equation}
  \mathcal{L}_\varphi q_{ab} = 0\,,\quad \mathcal{L}_\varphi\omega_a = 0\,.
\end{equation}
Then, the angular momentum is
\begin{equation}\label{eq:J-ih}
  J = -\frac{1}{8\pi}\oint_S \left(\varphi^a\omega_a\right)\widetilde{\epsilon}\,,
\end{equation}
where $S$ is a cross-section of $\Delta$.  It can be shown that any
cross-section $S$ will yield the same value of $J$ and thus, just like
the area, $J$ is a geometric invariant we can talk sensibly about the
value of $J$ for an axisymmetric WIH.

Turning now to notions of energy, here we will need a suitable time
translation Killing vector on $\Delta$. This is taken to be of the
form $A\ell^a - \Omega\varphi^a$, where $A,\Omega$ are constant on a
given WIH but vary over phase space. In particular $\Omega$ is the
angular velocity.  Hamiltonian considerations lead to an expression
for the mass as
\begin{equation}
  \label{eq:mass-ih}
      M = \frac{1}{2R} \sqrt{R^4 + 4J^2}\,.
\end{equation}
Note that for a non-spinning black hole this reduces to the
Schwarzschild expression $M = R/2$. The Hamiltonian
analysis of \cite{Ashtekar:2000hw,Ashtekar:2001is} also yields
expressions for the surface gravity and angular velocity in terms of
$(A,J)$ (in fact, the important point is that the
analysis of \cite{Ashtekar:2000hw,Ashtekar:2001is} shows that these
quantities can depend \emph{only} on $A$ and $J$). We
shall need the expression for the surface gravity later:
\begin{equation}
  \label{eq:kerrkappa}
  \widetilde{\kappa}(A, J) =
  \frac{R^4-4J^2}{2R^3\sqrt{R^4+4J^2}},
\end{equation}
This is the usual expression for surface gravity for a Kerr metric and
for Schwarzschild this becomes
$\widetilde{\kappa} = (2R)^{-1} = (4M)^{-1}$. 

The expression for the angular momentum can also be expressed in terms
of a Weyl tensor component.  In terms of the null tetrad
$(\ell^a,n^a,m^a,\bar{m}^a)$, the Weyl tensor can be decomposed into 5
complex scalar quantities denoted $\Psi_0,\Psi_1,\Psi_2,\Psi_3$, and
$\Psi_4$.  These will be described more fully in Sec.~\ref{sec:npsummary}, but for now, we only
need the expression for $\Psi_2$ in terms of the Weyl tensor
$C_{abcd}$:
\begin{equation}
  \Psi_2 = C_{abcd}\ell^am^b\bar{m}^cn^d\,. 
\end{equation}
We shall see that $\Psi_2$ is a geometric invariant on a WIH.  Its
real part yields the scalar curvature $\widetilde{\mathcal{R}}$ of
$\widetilde{q}_{ab}$, and its imaginary part is related to the
exterior derivative of $\omega_a$:
\begin{equation}
  \label{eq:psi2-omega}
  \widetilde{\mathcal{R}} =-4\textrm{Re}\left[\Psi_2\right]\,,\quad d\omega = 2\im\left[\Psi_2\right] \epsilon\,.
\end{equation}
The angular momentum can then be rewritten as
\begin{equation}
  \label{eq:J-Psi2}
  J  = -\frac{1}{4\pi}\oint_S \zeta\mathrm{Im}[\Psi_2]\,\widetilde{\epsilon}\,,
\end{equation}
where $\zeta$ is a ``potential'' for the $\varphi^a$ in the following
sense:
\begin{equation}
  \varphi^a\widetilde{\epsilon}_{ab} = \partial_b\zeta\,,\quad \oint_S \zeta \, \widetilde{\epsilon} = 0\,.
\end{equation}
(For a Kerr black hole, in terms of the usual spherical coordinates, it turns out that $\zeta = \cos\theta$).

Beyond the mass and angular momentum, the geometry of a WIH can be
expressed in terms of multipole moments.  The basic idea is to express
$\Psi_2$ as an infinite set of numbers by decomposing it in terms of
spherical harmonics.  However, which spherical coordinates should we
use, and how can we compare two different calculations which might
employ different coordinate systems?  As shown in
\cite{Ashtekar:2004gp}, on an axisymmetric WIH one can define a set of
invariant coordinates and orthonormal spherical harmonics $Y^{m}_\ell$
which can be used to decompose $\Psi_2$.  In this way, we get a set of
mass and spin multipole moments associated with the real and imaginary
parts of $\Psi_2$, respectively:
\begin{equation}
  \label{eq:geometric-multipoles}
  I_\ell + iL_\ell = -\oint \Psi_2\,Y^{0}_\ell(\zeta)\widetilde{\epsilon}\,.
\end{equation}
The zeroth mass moment $I_0$ is a topological invariant:
$I_0 = \sqrt{\pi}$.  Assuming there are no conical singularities, the
mass-dipole moment vanishes $I_1 = 0$. Similarly, if $\omega$ has no
singularities corresponding to a magnetic monopole, then $L_0 =
0$. From \eqref{eq:J-Psi2}, we see also that $L_1$ is proportional to
the angular momentum.

The importance of $\Psi_2$ for us resides in the fact that it also
encodes tidal deformations.  Thus, for a black hole with a binary
companion, when its horizon is deformed due to the tidal field of its
companion, this deformation is a perturbation of $\Psi_2$ and thus
changes these multipole moments from their Kerr values.  In
astrophysical applications where tidally perturbed black holes are
expected to be close to Kerr, we use the horizon area and $L_1$ to
identify the Kerr parameters.  These Kerr parameters identify uniquely
all the higher moments, and any deviations from these Kerr values are
to be interpreted as tidal perturbations of the horizon geometry.

Finally, we note that there are alternative definitions of multipole moments
available in the literature.  After all, different choices of
spherical coordinates are possible, which lead to different spherical
harmonics and thus to different multipole moments.  One should thus be
careful in interpreting the multipole moments and corresponding Love
numbers.  We mention in particular the multipole moments defined in
\cite{Ashtekar:2021kqj} which exploits the conformal geometry of the
horizon cross-sections.

\subsection{Constraint equations on an isolated horizon}
\label{subsec:constraints}

As discussed in the previous section, the geometry of $\Delta$ is
completely specified by the degenerate metric $q_{ab}$, and the
derivative operator $\mathcal{D}_a$.  Since $q_{ab}\ell^b=0$, the
``non-degenerate part'' of $q_{ab}$ is simply $\widetilde{q}_{ab}$
constructed above.  The information within $\mathcal{D}_a$ is
conveniently written in terms of the ingoing null-normal $n_a$ to the
horizon, which satisfies the normalization condition
$\ell\cdot n = -1$. Starting from an initial cross-section $S_0$ and
its normal $n_a$, we can extend this everywhere on $\Delta$ by
requiring $\mathcal{L}_\ell n_a = 0$ (and maintaining
$\ell^a n_a=-1$).

We then introduce the tensor $S_{ab}$
\begin{equation}
  S_{ab} = \mathcal{D}_an_b\,.
\end{equation}
Without loss of generality, we can take $\mathcal{D}_{[a}n_{b]} = 0$
so that $S_{ab}$ is symmetric.  It is easy to verify that
$\omega_a = S_{ab}\ell^b$.  The remaining information in $S_{ab}$ is
thus obtained by projecting to the cross-section:
\begin{equation}\label{eq:S-cross-section}
  \widetilde{S}_{ab} = \widetilde{q}_a{}^c\widetilde{q}_b{}^d S_{cd}\,.
\end{equation}
The trace and tracefree parts of $\widetilde{S}_{ab}$ yield
respectively the expansion and shear of $n_a$.  The complete
characterization of $\Delta$ requires a specification of
$\widetilde{S}_{ab}$ everywhere on $\Delta$.  It can be shown
that $\widetilde{S}_{ab}$ satisfies the following
constraint equation \cite{Ashtekar_2002}
\begin{equation}
  \label{eq:ihconstraint}
  \mathcal{L}_\ell\widetilde{S}_{ab} = -\kappa_{(\ell)}\widetilde{S}_{ab} + \mathcal{\widetilde{\mathcal{D}}}_{(a}\widetilde{\omega}_{b)} + \widetilde{\omega}_a\widetilde{\omega}_b - \frac{1}{2}\widetilde{\mathcal{R}}_{ab}\,.
\end{equation}
Here $\widetilde{\mathcal{R}}_{ab}$ is the Ricci tensor on the cross-section
calculated from the 2-metric $\widetilde{q}_{ab}$.  Thus, by
specifying $\widetilde{S}_{ab}$ on some initial cross-section, a
solution of this constraint equation then yields $\widetilde{S}_{ab}$
everywhere on $\Delta$.  

These geometric quantities and identities can be employed to choose a
suitable equivalence class $[\ell^a]$ on a NEH and thereby find an
admissible IH.  As shown in \cite{Ashtekar_2002}, a suitable condition
is to require that the expansion of $n^a$ is time-independent.  Under
this condition, one can choose an equivalence class $[\ell^a]$ if the
following elliptic operator is invertible:
\begin{equation}
  L := \widetilde{\mathcal{D}}^2 + 2\widetilde{\omega}^a\widetilde{\mathcal{D}}_a + \widetilde{\mathcal{D}}^a \widetilde{\omega}_a + \widetilde{\omega}^a\widetilde{\omega}_a - \frac{1}{2}\widetilde{\mathcal{R}}\,.
\end{equation}
It is interesting to note that the invertibility of a very similar
operator appears in the stability analysis of marginally trapped
surfaces \cite{Andersson:2005gq,Andersson:2007fh}.  When the
cross-section is taken to be a marginally trapped surface lying on a
Cauchy surface, then the stability of the MOTS under deformations is
shown to be equivalent to the invertibility of $L$.

Apart from the constraint equation Eq.~\eqref{eq:ihconstraint}, it
turns out that there are additional constraints on $\Psi_2$ appearing
due to the algebraic nature of the Weyl tensor.  If one imposes
constraints on the other Weyl tensor components, it turns out that the
Bianchi identities restrict $\Psi_2$ as well. This is important for us
because, as we have mentioned earlier, the geometric multipoles of an
IH are determined by $\Psi_2$.  Therefore, such constraints
potentially limit the type of tidal perturbations that are allowed on
an IH.  As an example, it was shown in
\cite{Dobkowski-Rylko:2018usb,LEWANDOWSKI_2002,Lewandowski:2002ua}
that if the Weyl tensor is time dependent at $\Delta$ and is of Petrov
type D, i.e. if we can find a frame in which $\Psi_2$ is the only
non-vanishing Weyl tensor component on the horizon, then it cannot be
specified freely but must satisfy
\begin{equation}
  \label{eq:psi2constraint}
  \eth\eth\Psi_2^{-1/3} = 0\,.
\end{equation}
Here $\eth$ is the spin-weighted angular derivative operator~\cite{Goldberg:1966uu}, which we will formally introduce in Sec.~\ref{sec:npsummary} (for the action of the $\eth$ operator on the spin-weighted spherical harmonics see Appendix~\ref{sec:spin-weighted-SH}).  We will see that this condition implies that
if we require non-trivial tidal perturbations, then $\Psi_3$ and/or
$\Psi_4$ cannot vanish at the horizon. These components of the Weyl
tensor indicate the presence of gravitational radiation at the horizon
which is transverse to $\Delta$, i.e. not infalling into the black
hole. This result shows that such radiation must be present for a
tidally disturbed black hole.  The Robinson-Trautman solutions
\cite{Robinson:1962zz,Chrusciel:1992rv} furnish good examples of
spacetimes with such transverse radiation in the vicinity of an IH;
these solutions will be discussed in Sec.~\ref{sec:robinson-trautman}.

\section{Constructing the near horizon spacetime}
\label{sec:nearhorizon}

\subsection{The Newman-Penrose formalism}
\label{sec:npsummary}

With the intrinsic geometry of an isolated horizon understood, here we
shall summarize the construction of the near horizon geometry.  It
will be convenient to work with the Newman-Penrose
formalism.  For this, as mentioned earlier, we complete the null normals $(\ell^a,n^a)$ to a null-tetrad $(\ell^a,n^a,m^a,\bar{m}^a)$
satisfying
\begin{equation}
  \ell\cdot n = -1\,, m\cdot\bar{m} = 1\,,
\end{equation}
with all other inner products vanishing.  The directional covariant
derivatives along these basis vectors are denoted as
\begin{equation}
  D:= \ell^a\nabla_a\,,\quad, \Delta := n^a\nabla_a\,,\quad \delta := m^a\nabla_a\,.
\end{equation}
The connection is explicitly represented as a set of 12 complex
functions known as the spin coefficients. These are typically
represented in terms of the directional derivatives of the basis
vectors:
\begin{subequations}
  \label{eq:spincoeffs}
  \begin{align}
    D\ell& = (\epsilon + \bar{\epsilon})\ell - \bar{\kappa}m -
    \kappa\bar{m} \,,\\
    Dn& = -(\epsilon + \bar{\epsilon})n + \pi m + \bar{\pi}m \,,\\
    Dm& = \bar{\pi}\ell - \kappa n + (\epsilon - \bar{\epsilon})m \,,\\
    \Delta \ell& = (\gamma + \bar{\gamma})\ell - \bar{\tau}m -
    \tau\bar{m}\,,\label{eq:Deltal}\\
    \Delta n & = -(\gamma + \bar{\gamma})n + \nu m + \bar{\nu}\bar{m}\,,\label{eq:Deltan}\\
    \Delta m & = \bar{\nu}\ell - \tau n + (\gamma-\bar{\gamma})m\,,\label{eq:Deltam}\\
    \delta\ell & = (\bar{\alpha} + \beta)\ell -\bar{\rho}m -
    \sigma\bar{m}\,,\\
    \delta n & = -(\bar{\alpha} + \beta)n + \mu m +
    \bar{\lambda}\bar{m}\,,\\
    \delta m & = \bar{\lambda}\ell - \sigma n + (\beta-\bar{\alpha})m\,,\label{eq:deltam}\\
    \bar{\delta}m & = \bar{\mu}\ell - \rho n +
    (\alpha-\bar{\beta})m\,. \label{eq:deltabarm}
  \end{align}
\end{subequations}
A technical benefit of tetrad formalisms is that the covariant
derivatives (here the spin coefficients) can be calculated using only
exterior derivatives.  This is useful in practical calculations
because, given a metric, the calculation of the spin coefficients
require a fewer number of derivatives and no Christoffel symbols are
required.  It can be shown that the exterior derivatives of the basis
1-forms are:
\begin{widetext}
  \begin{subequations}
    \label{eqs:spincoeffs-exterior}
    \begin{align}
      d\ell &= (\epsilon+\bar{\epsilon})\ell\wedge n + [\bar{\tau} - (\alpha+\bar{\beta})]\ell\wedge m + [\tau - (\bar{\alpha}+\beta)]\ell\wedge\bar{m} + \bar{\kappa}n\wedge m + \kappa n\wedge\bar{m} + (\bar{\rho} - \rho)m\wedge\bar{m}\,,\\
      dn &= (\gamma+\bar{\gamma})\ell\wedge n - \nu\ell\wedge m - \bar{\nu}\ell\wedge\bar{m} - [\pi - (\alpha+\bar{\beta})]n\wedge m - [\bar{\pi}- (\bar{\alpha}+\beta)]n\wedge\bar{m} + (\bar{\mu} - \mu)m\wedge\bar{m}\,,\\
      dm &= (\tau + \bar{\pi})\ell\wedge n - [\bar{\mu}+ (\gamma-\bar{\gamma})]\ell\wedge m + \bar{\lambda}\ell\wedge\bar{m} + [\rho - (\epsilon-\bar{\epsilon})]n\wedge m + \sigma n\wedge\bar{m} + (\bar{\alpha} - \beta)m\wedge\bar{m}\,.
    \end{align}
  \end{subequations}
\end{widetext}

Some important spin coefficients for us are: the real parts of
$\rho$ and $\mu$ are the expansion of $\ell$ and $n$ respectively; the imaginary parts yield the twist;
$\sigma$ and $\lambda$ are the shears of $\ell$ and $n$ respectively;
the vanishing of $\kappa$ and $\nu$ implies that $\ell$ and $n$ are
respectively geodesic; $\epsilon + \bar{\epsilon}$ and $\gamma +
\bar{\gamma}$ are respectively the accelerations of $\ell$ and $n$,
$\alpha-\bar{\beta}$ yields the connection in the $m$-$\bar{m}$ plane
and thus the curvature of the manifold spanned by $m$-$\bar{m}$.

Since the null tetrad is typically not a coordinate basis, the above definitions
of the spin coefficients lead to non-trivial commutation relations:
\begin{widetext}
  \begin{subequations}
    \label{eqs:commutation}
    \begin{align}
      (\Delta D - D\Delta)f &= (\epsilon+\bar{\epsilon})\Delta f + (\gamma
                              + \bar\gamma)Df - (\bar\tau + \pi)\delta f -(\tau + \bar{\pi})\bar{\delta}f\,,\\
      (\delta D-D\delta)f &= (\bar{\alpha}+\beta-\bar\pi)Df + \kappa\Delta
                            f - (\bar{\rho}+\epsilon-\bar{\epsilon})\delta f
                            -\sigma\bar{\delta}f\,,\\
      (\delta\Delta-\Delta\delta)f &= -\bar\nu Df +
                                     (\tau - \bar{\alpha}-\beta)\Delta f + (\mu-\gamma+\bar\gamma)\delta f + \bar{\lambda}\bar{\delta}f\,,\\
      (\bar{\delta}\delta-\delta\bar{\delta})f &= (\bar{\mu}-\mu)Df +
                                                 (\bar{\rho}-\rho)\Delta f + (\alpha-\bar{\beta})\delta f -
                                                 (\bar{\alpha}-\beta)\bar{\delta}f\,.
    \end{align}
  \end{subequations}
\end{widetext}
The Weyl tensor $C_{abcd}$ breaks down into 5 complex scalars
\begin{subequations}
  \begin{align}
  \Psi_0 &= C_{abcd}\ell^am^b\ell^cm^d\,,\quad  \Psi_1 = C_{abcd}\ell^am^b\ell^cn^d\,,\\
  \Psi_2 &=  C_{abcd}\ell^am^b\bar{m}^cn^d\,,\quad  \Psi_3 = C_{abcd}\ell^an^b\bar{m}^cn^d\,,\\
  \Psi_4 &= C_{abcd}\bar{m}^an^b\bar{m}^cn^d\,.
  \end{align}
\end{subequations}
Similar decompositions apply for the Ricci tensor or Maxwell fields,
but since we deal with vacuum spacetimes in this paper, we do not need these here.

The relation between the spin coefficients and the curvature
components lead to the so-called Newman-Penrose field equations which
are a set of 16 complex first order differential equations.  The
Bianchi identities, $\nabla_{[a}R_{bc]de} = 0$, are written explicitly
as 8 complex equations involving both the Weyl and Ricci tensor
components, and 3 real equations involving only Ricci tensor
components.  See
\cite{Penrose:1985jw,Chandrasekhar:1985kt,Stewart:1991} for the full
set of field equations and Bianchi identities (but beware that they
use slightly different conventions such as the sign for the metric
signature and normalization of the null tetrad, leading to possible
minus sign changes).

It will also be useful to use the notion of spin weights and the
$\eth$ operator for derivatives in the $m$-$\bar{m}$ plane (which will
be angular derivatives in our case).  A tensor $X$ projected on the
$m$-$\bar{m}$ plane is said to have spin weight $s$ if under a spin
rotation $m\rightarrow e^{i\psi}m$, it transforms as $X\rightarrow
e^{i s\psi}X$.  Thus, $m^a$ itself has spin weight $+1$ while
$\bar{m}^a$ has weight $-1$.  For  instance, the scalar $X= m^{a_1}\cdots
m^{a_p}\bar{m}^{b_1}\cdots \bar{m}^{b_q}X_{{a_1}\cdots {b_q}}$ 
has spin weight $s=p-q$ and the Weyl tensor component
$\Psi_k$ has spin weight $2-k$.  

The $\eth$ and $\bar{\eth}$ operators are defined as
\begin{eqnarray}
  \label{eq:19}
  \eth X = m^{a_1}\cdots m^{a_p}\bar{m}^{b_1}\cdots
  \bar{m}^{b_q}\delta X_{{a_1}\cdots {b_q}}\,,\\
  \bar{\eth} X = m^{a_1}\cdots m^{a_p}\bar{m}^{b_1}\cdots
  \bar{m}^{b_q}\bar{\delta} X_{{a_1}\cdots {b_q}}\,.
\end{eqnarray}
From Eqs.~(\ref{eq:deltam}) and (\ref{eq:deltabarm}), after projecting
on to the $m$-$\bar{m}$ plane, we get
\begin{equation}
  \delta m^a = (\beta-\bar{\alpha})m^a\,,\qquad \bar{\delta}{m}^a =
  (\alpha-\bar{\beta})m^a\,. 
\end{equation}
A short calculation shows that
\begin{equation}
  \label{eq:55}
  \eth X = \delta X + s(\bar{\alpha}-\beta)X\,,\qquad \bar{\eth}X =
  \bar{\delta}X - s(\alpha-\bar{\beta})X\,.
\end{equation}
It is clear that $\eth$ and $\bar{\eth}$ act as spin raising and
lowering operators.  See \cite{Goldberg:1966uu} for further properties of the $\eth$ operator and its connection to representations of the rotation group.

The transformations of the null tetrad which preserve their inner product 
are
\begin{description}
\item[Boosts and Spin Rotations]
\begin{equation}
  \label{eq:32}
  l\rightarrow Al\,,\quad n\rightarrow A^{-1}n\,,\quad m\rightarrow e^{i\theta}m\,,
\end{equation}

\item[Null rotations] 
  \begin{equation}
    \label{eq:53}
    \ell\rightarrow \ell\,,\quad m\rightarrow m + a\ell\,,\quad n
    \rightarrow n + \bar{a}m + a\bar{m} + |a|^2\ell\,,
  \end{equation}
and the null rotations around $n$ (obtained by interchanging $\ell$
  and $n$ in Eq.~(\ref{eq:53})).
\end{description}
We refer to \cite{Penrose:1985jw,Chandrasekhar:1985kt,Stewart:1991}
for a more complete discussion.

\subsection{A general construction of the near horizon spacetime}
\label{sec:General-near-horizon}
The construction of the near horizon geometry follows, in principle,
the same philosophy as the standard 3+1 decomposition: we prescribe
initial data on certain hypersurfaces and use the Einstein equations
to obtain the spacetime metric in a neighborhood.  The difference is
that instead of specifying data on a spatial hypersurface, we use a
characteristic initial value problem and prescribe data on a pair of
transverse null hypersurfaces
\cite{Friedrich:1981at,Friedrich:1983vi}.  We refer here also to the
seminal work by Bondi and collaborators on constructing the spacetime
near null infinity \cite{Bondi:1962px} following a similar procedure.  

In the characteristic formalism, the field equations (i.e. the
Einstein equations and the Bianchi identities) are written as
first-order quasi-linear equations of the form
\begin{equation}
  \sum_{J=1}^N A^a_{IJ}(x,\psi)\partial_a\psi_J + F_I(x,\psi) = 0\,.
\end{equation}
Here $x^a$ are coordinates on a manifold, and we have $N$ dependent
variables $\psi_I$.  In the usual Cauchy problem, we specify $\psi_I$
at some initial time, and solve these equations to obtain $\psi_I$ for
later times.  Alternatively, within the characteristic formulation, we
have a pair of null surfaces $\mathcal{N}_0$ and $\mathcal{N}_1$ whose
intersection is a co-dimension-2 spacelike surface $S$.  It turns out
to be possible to specify appropriate data on the null surfaces and on
$S$ such that the above system of equations is well-posed and has a
unique solution, at least locally near $S$.  We briefly summarize this
construction in our present case.

\begin{figure}
  \begin{center}
    \includegraphics[angle=-0,width=\columnwidth]{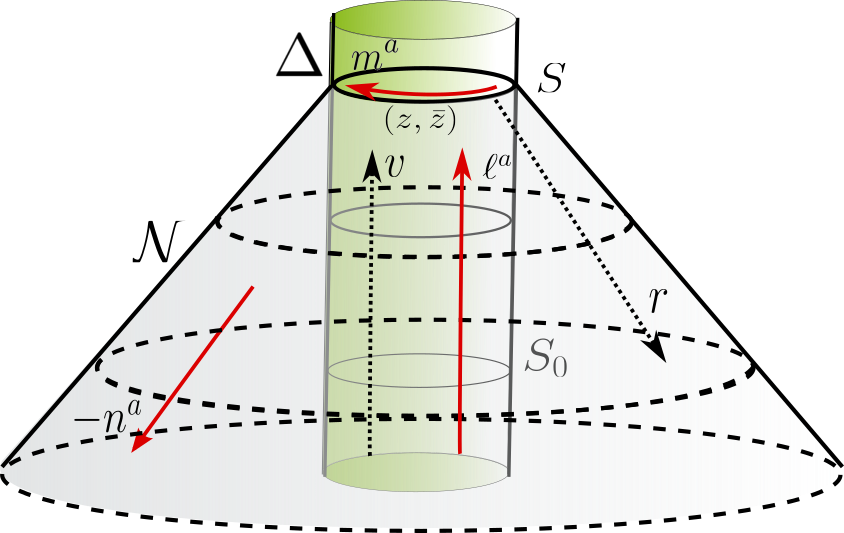}
    \caption{The characteristic initial value problem for constructing
      the near horizon geometry.  Here $\Delta$ is the horizon whose
      geometry is shown in Fig.~\ref{fig:projections}.  The null
      surface $\mathcal{N}$ is generated by past-directed null
      geodesics starting from a cross-section $S$ with coordinates
      $(z,\bar{z})$. The affine parameter along the geodesics is $r$,
      and the null vector is $n^a\partial_a = -\partial_r$.  The
      spacetime metric is constructed starting with suitable data on
      $\Delta$, $\mathcal{N}$ and $S$.
    }
    \label{fig:nearhorizon}
  \end{center}
\end{figure}

One of the null surfaces will be the isolated horizon $\Delta$, while
the other null surface $\mathcal{N}$ is generated by past-directed
null geodesics emanating from a cross-section of $\Delta$ as shown in
Fig.~\ref{fig:nearhorizon}.  This construction was first proposed in
\cite{Lewandowski:1999zs,Ashtekar:2000sz} and further elaborated upon
in \cite{Krishnan:2012bt,Lewandowski:2018khe,Scholtz_2017}.  We start
with the past-directed null vector $-n^a$ at the horizon obtained from
a particular cross-section $S_0$.  Integrating the geodesic equation
(till the conjugate point) gives us the null geodesics generated by
$-n^a$, and thus yields a null surface $\mathcal{N}$ generated from
$S_0$.  The spacetime metric is calculated in a characteristic
formulation by prescribing initial data on the null surfaces
$\mathcal{N}$ and $\Delta$.  The data on $\mathcal{N}$ is the
Weyl tensor component $\Psi_4$ while the data on $\Delta$ consist
of the geometric information required for an IH,
i.e. $(q_{ab},\mathcal{D},[\ell^a])$.  If we have coordinates $(v,\theta,\phi)$
on $\Delta$ such that $S_0$ is a surface of constant $v$ and
$(\theta,\phi)$ are coordinates on $S_0$, and $r$ is the affine
parameter along $-n^a$, then this construction yields a coordinate
system $(v,r,\theta,\phi)$ in a neighborhood of $\Delta$.  For
technical convenience, instead of real angular coordinates
$(\theta,\phi)$, we shall work on the stereographic plane with complex
coordinates $(z,\bar{z})$.

The above construction implies that we can choose
\begin{equation}
  \label{eq:2}
  n_a = -\partial_a v \qquad \textrm{and} \qquad n^a\nabla_a
  :=\Delta = -\frac{\partial}{\partial r}\,.
\end{equation}
To satisfy the inner product relations $\ell^a n_a = -1$ and
$m^an_a=0$, the other basis vectors must be of the form:
\begin{subequations}\label{eq:tetrad-general}
  \begin{align}
  \label{eq:3}
  \ell^a\nabla_a &:= D = \frac{\partial}{\partial v} + U\frac{\partial}{\partial r}
                   + X\frac{\partial}{\partial z} + \bar{X}\frac{\partial}{\partial \bar{z}}\,,\\
    m^a\nabla_a&:= \delta =
  \Omega\frac{\partial}{\partial r} + \xi_1\frac{\partial}{\partial z} + \xi_2\frac{\partial}{\partial \bar{z}}\,.
  \end{align}
\end{subequations}
The frame function $U$ is real while $X,\Omega,\xi_i$ are complex.  We
wish to now specialize to the case when $\ell^a$ is a null normal of
$\Delta$ so that the null tetrad is adapted to the horizon.  Since
$\partial_v$ is tangent to the null generators of $\Delta$, this
clearly requires that $U,X^i$ must vanish on the horizon.  Similarly,
we want $m^a$ to be tangent to the spheres $S_v$ at the horizon, so
$\Omega$ should also vanish on $\Delta$.

Since $n^a$ is an affinely parameterized geodesic, and $\ell$ and $m$
are parallel propagated along $n^a$, we have
$\Delta n = \Delta \ell = \Delta m = 0$.  From Eqs.~(\ref{eq:Deltal}),
(\ref{eq:Deltan}) and (\ref{eq:Deltam}), this leads to
\begin{equation}
  \label{eq:gamma-tau-nu}
  \gamma = \tau = \nu = 0\,.
\end{equation}
We first impose these conditions to the commutation relations in
Eqs.~(\ref{eqs:commutation}).  Then, setting $f=v$ in those equations
leads to
\begin{equation}
  \label{eq:pi-mu}
  \pi = \alpha + \bar{\beta}\,,\quad \mu = \bar{\mu}\,.
\end{equation}
These must hold throughout the region where the coordinate system is
valid.

The rest of the discussion can be separated into three parts: i)
equations which involve the time derivatives along $\ell^a$ and
include a description of the horizon geometry, ii) the radial
derivatives along $n^a$ which propagate geometric information away
from $\Delta$, and iii) equations which exclusively involve angular
derivatives and yield the ``shape'' of the 2-sphere cross-sections.
At the horizon, since the expansion, shear, and twist of $\ell^a$
vanish, we have:
\begin{equation}
  \label{eq:rho}
  \rho \triangleq 0,\quad\kappa\triangleq 0\,,\quad \sigma\triangleq 0\,.
\end{equation}
(Equations which hold only on $\Delta$ are indicated by
`$\triangleq$' instead of the usual `$=$'). These three conditions at the horizon further imply 
\begin{equation}\label{eq:psi0-and-psi1-horizon}
   \Psi_0\triangleq0\,,\quad \Psi_1\triangleq0\,.
\end{equation} These two equations can be interpreted as the absence of ingoing transverse and longitudinal radiation at the horizon. Further, we can require $m^a$ to be
 Lie-dragged along $\ell^a$ so that $\mathcal{L}_\ell m^a = 0$. This
leads to
\begin{equation}\label{eq:epsilon-H}
  \epsilon - \bar{\epsilon} \triangleq 0\,.
\end{equation}
In terms of the Newman-Penrose spin coefficients, the connection 1-form $\omega_a$ is written as 
\begin{equation}
  \label{eq:40}
  \omega_a = -n_b\nabla_a\ell^b ={-} (\epsilon+\bar\epsilon) n_a + \pi m_a +
  \bar{\pi}\bar{m}_a\,. 
\end{equation}
Thus, $\Delta$ will be a WIH if we choose
\begin{equation}
  \label{eq:38}
  \tilde{\kappa}_{(\ell)} \triangleq  \epsilon + \bar{\epsilon} \triangleq
  \textrm{constant}\,,\quad D\pi \triangleq  0\,.
\end{equation}
The first of the above is just the zeroth law of black hole mechanics stating that the surface gravity is constant. Notice that Eqs.~\eqref{eq:rho},~\eqref{eq:psi0-and-psi1-horizon} and~\eqref{eq:38} further imply that 
\begin{equation}
    D\Psi_2\triangleq 0\,,\quad D a\triangleq 0\,,
\end{equation} where $a=\alpha-\bar{\beta}$ is the connection compatible with $\widetilde q_{ab}$. This last equation specifies that the geometry of $\tilde \Delta$ is constant in time. 

With the above conditions on the spin coefficients at hand, we now
impose them in the commutator relations, the field equations, and the
Bianchi identities.  The functions $U,X^i,\Omega,\xi^i$ are determined
by the commutation relations (\ref{eqs:commutation}) by substituting,
in turn, $r$ and $x^i$ for $f$, and imposing Eqs.~(\ref{eq:gamma-tau-nu}) and
(\ref{eq:pi-mu}) on the spin coefficients.  First, the radial derivatives for the coefficients of the tetrad are
\begin{subequations}
  \label{eqs:frame}
  \begin{align}
    \Delta U &= -(\epsilon+\bar{\epsilon}) - \pi\Omega -
               \bar{\pi}\bar{\Omega}\,,\\ 
    \Delta X^i &= -\pi\xi^i - \bar{\pi}\bar{\xi}^i\,,\\
    \Delta\Omega &= -\bar{\pi} - \mu\Omega -\bar{\lambda}\bar{\Omega}
                   \,,\\ 
    \Delta\xi^i &= -\mu\xi^i -\bar{\lambda}\bar{\xi}^i\,, 
  \end{align}
\end{subequations}
while their propagation equations along $v$ are
\begin{subequations}
  \label{eqs:alongv}
  \begin{align}
    D\Omega - \delta U&= \kappa + \rho\Omega + \sigma\bar{\Omega}\,,\\ 
    D\xi^i - \delta X^i &= (\bar{\rho}+\epsilon-\bar{\epsilon})\xi^i + \sigma\bar{\xi}^i\,.
  \end{align}
\end{subequations}
Let us now turn to the field equations.  After imposing
Eqs.~(\ref{eq:gamma-tau-nu}) and (\ref{eq:pi-mu}) on the spin
coefficients and ignoring matter terms, the field equations involving
radial derivatives are:
\begin{subequations}
  \label{eqs:radial}
  \begin{align}\label{eqs:radial:1}
    \Delta\lambda &= -2\lambda\mu -\Psi_4\,,\\\label{eqs:radial:2}
    \Delta\mu &= -\mu^2 - |\lambda|^2\,,\\\label{eqs:radial:3}
    \Delta\rho &= -\mu\rho -\sigma\lambda -\Psi_2\,,\\\label{eqs:radial:4}
    \Delta\sigma &= -\mu\sigma -\bar{\lambda}\rho\,.\\\label{eqs:radial:5}
    \Delta\kappa &= -\bar{\pi}\rho - \pi\sigma -\Psi_1\,,\\\label{eqs:radial:6}
    \Delta\epsilon &= -\bar{\pi}\alpha - \pi\beta - \Psi_2\,,\\\label{eqs:radial:7}
    \Delta\pi &= -\pi\mu-\bar{\pi}\lambda -\Psi_3\,,\\\label{eqs:radial:8}
    \Delta\beta &= -\mu\beta -\alpha\bar{\lambda}\,,\\\label{eqs:radial:9}
    \Delta\alpha &= -\beta\lambda -\mu\alpha-\Psi_3\,.
  \end{align}  
\end{subequations}
The time evolution equations become:
\begin{subequations}
  \label{eqs:timeevolution}
  \begin{align} 
    D\rho -\bar{\delta}\kappa &= \rho^2 + |\sigma|^2 +
                                (\epsilon+\bar{\epsilon})\rho -2\alpha\kappa\,,\label{eq:drho}\\ 
    D\sigma-\delta\kappa &= (\rho+\bar{\rho} +\epsilon+\bar{\epsilon})\sigma
                           -2\beta\kappa + \Psi_0\,,\label{eq:dsigma}\\ \label{eqs:timeevolution:1}
    D\alpha-\bar{\delta}\epsilon &= (\rho+\bar{\epsilon}-2\epsilon)\alpha +
                                   \beta\bar{\sigma} - \bar{\beta}\epsilon -\kappa\lambda +
                                   (\epsilon+\rho)\pi\,,\\ \label{eqs:timeevolution:2}
    D\beta -\delta\epsilon &= (\alpha+\pi)\sigma +
                             (\bar{\rho}-\bar{\epsilon})\beta -\mu\kappa
                             -(\bar{\alpha}-\bar{\pi})\epsilon + \Psi_1\,,\\ \label{eq:timeevolution:lambda}
    D\lambda -\bar{\delta}\pi &= (\rho-2\epsilon)\lambda +
                                \bar{\sigma}\mu + 2\alpha\pi\,,\\ \label{eqs:timeevolution:mu}
    D\mu - \delta\pi &= (\bar\rho - \epsilon-\bar{\epsilon})\mu + \sigma\lambda +
                       2\beta\pi + \Psi_2\,.
  \end{align}
\end{subequations}
The angular field equations are:
\begin{subequations}
  \label{eqs:angular}
  \begin{align}
    \delta\rho - \bar\delta\sigma &= \bar{\pi}\rho -
                                    (3\alpha-\bar{\beta})\sigma -\Psi_1\,,\\
    \delta\alpha - \bar{\delta}\beta &= \mu\rho -\lambda\sigma +
                                       |\alpha|^2 + |\beta|^2 -2\alpha\beta -\Psi_2\,,\\
    \delta\lambda -\bar{\delta}\mu &= \pi\mu +
                                     (\bar{\alpha}-3\beta)\lambda -\Psi_3\,.
  \end{align}
\end{subequations}
Finally, we have the Bianchi identities which, in the NP formalism,
are written as a set of nine complex and two real equations; in the
absence of matter, only 8 complex equations survive. The radial
Bianchi identities are reduced to:
\begin{subequations}
  \label{eqs:radial-bianchi}
  \begin{align} \label{eqs:radial-bianchi:0}
    \Delta\Psi_0 - \delta\Psi_1 &= -\mu\Psi_0 -2\beta\Psi_1 +
                                  3\sigma\Psi_2\,,\\\label{eqs:radial-bianchi:1}
    \Delta\Psi_1 -\delta\Psi_2 &= -2\mu\Psi_1 + 2\sigma\Psi_3\,,\\\label{eqs:radial-bianchi:2}
    \Delta\Psi_2 -\delta\Psi_3 &= -3\mu\Psi_2 + 2\beta\Psi_3 +
                                 \sigma\Psi_4\,,\\ \label{eqs:radial-bianchi:3}
    \Delta\Psi_3 - \delta\Psi_4 &= -4\mu\Psi_3 + 4\beta\Psi_4\,.
  \end{align}
\end{subequations}
Note that there is no equation for the radial derivative of $\Psi_4$.
Among all the fields that we are solving for, this is in fact the only
one for which this happens.  This means that $\Psi_4$ (in this case,
its radial derivatives) is the free data that must be specified on the
null cone $\mathcal{N}_0$ originating from $S_0$. Notice that if the spacetime is algebraically special,  $\Psi_4$ might satisfy further constraints, which need to be accounted for in the previous statement. For instance, for type $D$ spacetimes $\Psi_4$ (and its radial derivatives) are related to $\Psi_2$ and $\Psi_3$ through $2\Psi_2\Psi_4=3\Psi_3^2$~\cite{Dobkowski-Rylko:2018usb}.  

Finally, we have the components of the Bianchi equations for evolution
of the Weyl tensor components:
\begin{subequations}\label{eqs:evolution-bianchi}
\begin{align}\label{eqs:evolution-bianchi:1}
  D\Psi_1 -\bar\delta\Psi_0 &= (\pi-4\alpha)\Psi_0 +
  2(2\rho+\epsilon)\Psi_1 -3\kappa\Psi_2\,,\\\label{eqs:evolution-bianchi:2}
  D\Psi_2-\bar\delta\Psi_1 &= -\lambda\Psi_0 +2(\pi-\alpha)\Psi_1 +
  3\rho\Psi_2 -2\kappa\Psi_3\,,\\  \label{eqs:evolution-bianchi:3}
  D\Psi_3 -\bar\delta\Psi_2 &= -2\lambda\Psi_1 + 3\pi\Psi_2
  +2(\rho-\epsilon)\Psi_3 -\kappa\Psi_4 \,,\\\label{eqs:evolution-bianchi:4}
  D\Psi_4-\bar\delta\Psi_3 &= -3\lambda\Psi_2 +2(\alpha+2\pi)\Psi_3 +
  (\rho-4\epsilon)\Psi_4 \,.
\end{align}
\end{subequations}

Before proceeding to apply the above equations for a tidally distorted
black hole, it will be instructive to look at two illustrative
examples.

\subsection{Example 1: Constructing the Schwarzschild spacetime}
\label{subsec:schwarzschild}

The reader will be familiar with the Schwarzschild metric of mass $M$
in ingoing Eddington-Finkelstein coordinates
$(v,\mathfrak{r},z,\bar{z})$.  Here we distinguish between the radial
coordinate $r$ defined previously, which vanishes at the horizon, and
the coordinate $\mathfrak{r}$, which is the usual Schwarzschild radial
coordinate (at the horizon, $\mathfrak{r} = 2M$):
\begin{equation}
\label{eq:ds2-Schwarschild}
  ds^2 = -f(\mathfrak{r})dv^2 + 2dv\,d\mathfrak{r} + \frac{2\mathfrak{r}^2}{P_0^2}dz\,d\bar{z}\,.
\end{equation}
Here, as usual
\begin{equation}
  f=1-\frac{2M}{\mathfrak{r}}\,.
\end{equation}
Instead of the usual spherical coordinates, let us use complex
coordinates $z=e^{i\phi} \cot{\frac{\theta}{2}}$. The expressions for
the stereographic projection yield
\begin{equation}\label{eq:P-sch}
  P_0 = \frac{1}{\sqrt{2}}(1+z\bar{z})\,.
\end{equation}
Starting with just the data on the horizon, i.e. a spherically
symmetric horizon, taking $\mathcal{N}$ to be a constant-$v$ surface,
and setting $\Psi_4=0$ everywhere on $\mathcal{N}$, can we reconstruct
the Schwarzschild metric?  In particular, we have the usual null
tetrad and basis 1-forms:
\begin{subequations}
  \label{eq:tetrad-schwarzschild}
  \begin{align}
    \ell^a\nabla_a &= \partial_v + \frac{f}{2}\partial_{\mathfrak{r}}\,,\quad \ell_a = -\frac{f}{2}\partial_av + \partial_a \mathfrak{r}\,,\\
    n^a\nabla_a &= -\partial_{\mathfrak{r}}\,,\quad n_a = -\partial_av\,,\\
    m^a\nabla_a &= \frac{P}{\mathfrak{r}}\partial_z\,,\quad m_a = \frac{ \mathfrak{r}}{P}\partial_a\bar{z}\,.
  \end{align}
\end{subequations}
It is straightforward to calculate the spin coefficients everywhere.
But we want to instead just start with the spin coefficients at the
horizon and recover their values everywhere following the construction
outlined in the previous section.

This is in fact straightforward and instructive and we shall see in
fact the resulting spacetime is asymptotically flat as it should be.
We begin with the Weyl tensor components.  We shall first assume that
the metric is type D at the horizon, i.e.
\begin{equation}
  \Psi_0\triangleq\Psi_1\triangleq\Psi_3\triangleq\Psi_4 \triangleq 0\,.
\end{equation}
We shall assume further that $\Psi_2$ is spherically symmetric so that
the constraint of Eq.~(\ref{eq:psi2constraint}) is
satisfied. Moreover, let us take the simplest choice of $\Psi_4=0$ on
the transverse null surface $\mathcal{N}$.  Next, choose the sphere
$S_0 = \mathcal{N}\cap \Delta$ to be spherically symmetric, in the
sense that the expansion of $n_a$, i.e. $\mu$, is
constant on $S_0$ and its shear, $\lambda \triangleq 0$.

The choice of $\Psi_2$ determines the horizon source multipole
moments, and in this case we just have a mass monopole. First we note
that if $\Psi_2$ is constant, it must be real because from
Eq.~(\ref{eq:psi2-omega}) 
\begin{equation}
  \oint_{S_0}\im{\Psi_2}\, \widetilde{\epsilon} = \oint_{S_0}d\omega = 0\,.
\end{equation}
On the other hand, if $\im{\Psi_2}$ is constant then the above
equation shows that it must vanish.  Similarly, from
Eq.~(\ref{eq:J-Psi2}) the angular momentum $J$ must also
vanish, and thus from Eq.~\eqref{eq:mass-ih}, the horizon mass is
$M = R/2$.

The real part $\rm{\Psi_2}$ is determined by the topology of $S_0$ and
the Gauss-Bonnet theorem.  Since $\mathcal{R} = -4\re{\Psi_2}$:
\begin{eqnarray}
  8\pi &=& \oint_{S_0}\mathcal{R} \widetilde{\epsilon} = -4\re{\Psi_2}A\,,\\
  \implies && \Psi_2 = -\frac{2\pi}{A} =
  -\frac{1}{2R^2} = -\frac{1}{8M^2}\,.\label{eq:psi2-boundarycondition}
\end{eqnarray}
We can now in fact determine the constant value of $\mu$ on $\Delta$.
Use the last of the evolution equations
Eqs.~(\ref{eqs:timeevolution}) on $\Delta$, use $\kappa=0$
and impose $D\mu=0$ to obtain
\begin{equation}
  \widetilde{\kappa}_{(\ell)}\mu = \Psi_2\,.
\end{equation}
Using the canonical value $\widetilde{\kappa}$
(see the discussion around Eq.~(\ref{eq:kerrkappa})), we conclude that
$\mu \triangleq -1/(2M)$.

To obtain the Schwarzschild metric in the
usual coordinates, let us take the radial coordinate such that $\mathfrak{r}=2M$
at the horizon.  We begin with the first two radial equations from
Eqs.~(\ref{eqs:radial}) for the shear and expansion on $n^a$:
\begin{subequations}
  \begin{align}
  \Delta\lambda &= -2\lambda\mu \implies \Delta (\lambda\bar{\lambda}) = -4\mu|\lambda|^2\,.\\
  \Delta\mu &=-\mu^2 -|\lambda|^2\,.
  \end{align}
\end{subequations}
Note that $\Delta$ is $-\partial/\partial \mathfrak{r}$.  At the horizon, we have
$\lambda=0$, and therefore the first equation yields
\begin{equation}
  |\lambda|^2 = |\lambda_0|^2\exp\left(\int_{2M}^{\mathfrak{r}}4\mu\,d\mathfrak{r} \right) \,.
\end{equation}
We conclude immediately that $\lambda=0$ everywhere if $\lambda_0=0$, as is the case given that $\lambda \triangleq 0$.
Substituting this into the equation for $\mu$ yields
\begin{equation}
  \frac{\d\mu}{\d\mathfrak{r}} = \mu^2 \implies \frac{1}{\mu_0} - \frac{1}{\mu} = \mathfrak{r}-2M\,.
\end{equation}
Since $\mu_0 = -1/(2M)$, we find the solution everywhere on
$\mathcal{N}$:
\begin{equation}
  \mu = -\frac{1}{\mathfrak{r}}\,,
\end{equation}
as it should be.  With the solution for $\mu,\lambda$ at hand, and
using $\sigma\triangleq 0$, the fourth radial equation from
Eqs.~(\ref{eqs:radial}) yields the solution $\sigma=0$ everywhere on
$\mathcal{N}$.

Proceeding similarly, we now consider the radial Bianchi identities
Eqs.~(\ref{eqs:radial-bianchi}) starting from the last to the first.
The above boundary conditions on the Weyl tensor are sufficient to
determine it everywhere on $\mathcal{N}$. 
With $\Psi_4=0$, the last of Eqs.~(\ref{eqs:radial-bianchi}) becomes
\begin{equation}
  \Delta\Psi_3 = -4\mu\Psi_3 \implies \frac{\d\ln\Psi_3}{\d \mathfrak{r}} = 4\mu\,.
\end{equation}
This has the solution
\begin{equation}
  \Psi_3(\mathfrak{r}) = \Psi_3(\mathfrak{r}=2M)\exp\left(\int_{2M}^{\mathfrak{r}}4\mu\,\d\mathfrak{r}\right)\,.
\end{equation}
The boundary condition $\Psi_3(\mathfrak{r}=2M)=0$ then implies that
$\Psi_3(\mathfrak{r}) = 0$ everywhere on $\mathcal{N}$.  The third radial Bianchi
identity yields
\begin{equation}
  \frac{\d\Psi_2}{\d\mathfrak{r}} = 3\mu\Psi_2\,.
\end{equation}
Using the solution $\mu=-1/\mathfrak{r}$ derived above, we get
\begin{equation}
  \frac{\d\ln\Psi_2}{\d\mathfrak{r}} + \frac{3}{\mathfrak{r}} = 0\implies \Psi_2 \mathfrak{r}^3 = \textrm{constant}\,.
\end{equation}
Using the boundary condition Eq.~(\ref{eq:psi2-boundarycondition})
yields the solution
\begin{equation}
  \Psi_2 = -\frac{M}{\mathfrak{r}^3}\,.
\end{equation}
This is, again, as expected from the full Schwarzschild solution.
Finally, the first two Bianchi identities (and the solution $\sigma=0$)
give $\Psi_0(\mathfrak{r}) = \Psi_1(\mathfrak{r}) = 0$ everywhere on $\mathcal{N}$.

We can now proceed with the remaining radial equations in
Eqs.~(\ref{eqs:radial}), which all involve the Weyl tensor components.  We
conclude straightforwardly that $\pi=0$ which in turn gives
$\kappa=0$. For $\epsilon$ and $\rho$ (with the boundary condition
$\epsilon \triangleq 1/8M$ and $\rho\triangleq 0$), we get:
\begin{eqnarray}
  \Delta\epsilon &=& \frac{M}{\mathfrak{r}^3} \implies \epsilon(\mathfrak{r}) = \frac{M}{2\mathfrak{r}^2}\,,\\
  \Delta\rho &=& \frac{\rho}{\mathfrak{r}}  + \frac{M}{\mathfrak{r}^3} \implies \rho(\mathfrak{r}) = -\frac{1}{2\mathfrak{r}}\left(1-\frac{2M}{\mathfrak{r}}\right)\,.
\end{eqnarray}
We are finally left with $\alpha$ and $\beta$. These are related
to the ``shape'' of the cross-section $S_0$ and the intrinsic
2-dimensional Ricci scalar $\widetilde{R}$. These will depend on the
angular coordinates $(z,\bar{z})$.  From Eq.~\eqref{eq:pi-mu} and
$\pi=0$ we get $\alpha+\bar{\beta} = 0$.  The combination
$\alpha-\bar{\beta}$ is determined just by angular derivatives and can be obtained from the 2-metric on $S_0$. Let us denote
$a=\alpha-\bar{\beta}$. 
From the third of
Eqs.~(\ref{eqs:spincoeffs-exterior}), we have
\begin{equation}\label{eq:connection-horizon}
  a \triangleq \frac{\partial_{\bar{z}} P}{2M }\,,
\end{equation} where the stereographic function $P$ is defined in Eq.~\eqref{eq:P-sch}.
This is the boundary conditions for the radial equations involving
$\beta$ and $\alpha$.  From the radial equations:
\begin{equation}
  \Delta a = -\mu a = \frac{a}{\mathfrak{r}} \implies a(\mathfrak{r}) = \frac{\partial_{\bar{z} }P}{\mathfrak{r}}\,. 
\end{equation}
Finally, since the Weyl tensor, along with $\mu,\lambda,\sigma$ are
all time-independent on $\Delta$, the above analysis can be repeated
on all the null surfaces starting from other spherically symmetric
sections of $\Delta$.  Thus, the expressions obtained above for the
spin coefficients and Weyl tensor components are valid everywhere
outside the horizon for all $v$. The metric itself is obtained by
integrating the radial equations for the frame functions,
i.e. Eqs.~(\ref{eqs:frame}) and then combining the tetrad to obtain
the metric.  We leave this to the reader to verify that we do indeed
obtain the Schwarzschild null tetrad given in
Eqs.~(\ref{eq:tetrad-schwarzschild}), and thus the Schwarzschild
metric in ingoing Eddington-Finkelstein coordinates.

This concludes our derivation of the Schwarzschild solution using the
characteristic initial value problem.  This might seem to be a rather
convoluted derivation of a simple and well-known metric. Nevertheless
it illustrates the general procedure and clarifies the role played by
the different quantities and equations (for this reason we have not
spared any of the details). The payoff has been a very detailed
understanding of the spacetime with explicit expressions for the
curvature, connection (and, of course, also the metric if desired).
These features will hold in more general physical situations as
well.  All aspects of the classical isolated horizon formalism are
seen to be essential: 
a) The Hamiltonian calculations gave us
appropriate values for mass and surface gravity, 
b) The geometric
constraints on the isolated horizon needed to be satisfied in
accordance with the algebraic properties of the Weyl tensor, c) The multipole moments yielded $\Psi_2$, and 
d) The radial and angular field equations accomplished the rest. All of these features will carry over when we introduce tidal distortions.

We can also remark on the asymptotic properties of the solution as $r\rightarrow\infty$ and its global stationarity. We have obtained an asymptotically flat and stationary solution but it is clear that this will not hold generally for other choices of boundary conditions. In
fact, from the black hole uniqueness theorems, we should expect
to obtain asymptotically flat stationary solutions only for Kerr data
on the horizon and with $\Psi_4=0$ on $\mathcal{N}$.  This issue has
been studied in \cite{Dobkowski-Rylko:2018ahh,Lewandowski:2014nta}.
When tidal perturbations are introduced, the Weyl tensor will not be
algebraically special. The metric will not be asymptotically flat,
corresponding to an external tidal field acting on the black hole.

\subsection{Example 2: Schwarzschild with (non-falling) radiation --
  The Robinson-Trautman spacetime}
\label{sec:robinson-trautman}

In general, the local geometry constructed from the above procedure
will contain radiation.  Let us now consider the simplest
generalization to the Schwarzschild construction above by including a
non-vanishing $\Psi_4$ in the transverse null surface $\mathcal{N}$,
but still maintaining the intrinsic geometry on $\Delta$ to be the
Schwarzschild data.  In this way, we would obtain a spacetime
corresponding to a Schwarzschild black hole with constant area, but
possibly with radiation arbitrarily close to the horizon propagating
parallel to the horizon.

We start with the first two equations in Eqs.~\eqref{eqs:radial} which
describe the radial behavior of $\mu,\lambda$, i.e. the expansion and
shear of $n^a$. Previously, with vanishing $\lambda$ and $\Psi_4$, we
could explicitly solve for $\mu$.  Following \cite{Newman:1961qr}, we
note that these two equations can be written as a Ricatti equation: 
\begin{equation}
  \frac{\partial \mathcal{P}}{\partial \mathfrak{r}} = \mathcal{P}^2 + \mathcal{Q}
\end{equation}
where
\begin{equation}
  \mathcal{P} = \begin{pmatrix}
    \mu & \lambda\\
    \bar{\lambda} & \mu
  \end{pmatrix}\,,
  \quad \mathcal{Q} = \begin{pmatrix}
    0 & \Psi_4\\
    \bar{\Psi}_4 & 0
  \end{pmatrix}\,.
\end{equation}
The Ricatti equation can be cast in terms of a linear second-order
equation by the substitution $\mathcal{P} = -(\partial_{\mathfrak{r}} Y)Y^{-1}$ where
\begin{equation}
  Y =
  \begin{pmatrix}
    y_1& y_2\\
    \bar{y}_1 & \bar{y}_2
  \end{pmatrix}\,.
\end{equation}
Then it can be shown that $Y$ satisfies the linear equation
\begin{equation}
  \frac{\partial^2 Y}{\partial \mathfrak{r}^2} = -\mathcal{Q}Y\,.
\end{equation}
Thus, with a choice of $\Psi_4$ (i.e. $\mathcal{Q}$), initial conditions at
$\mathfrak{r}=2M$ on $\mu$ as in Schwarzschild, and $\lambda\triangleq 0$, we can
solve this second order equation for $Y$, and hence obtain $\mathcal{P}$.  

\begin{figure}
  \begin{center}
    \includegraphics[trim={0 7cm 0 5cm},clip,angle=-0,width=\columnwidth]{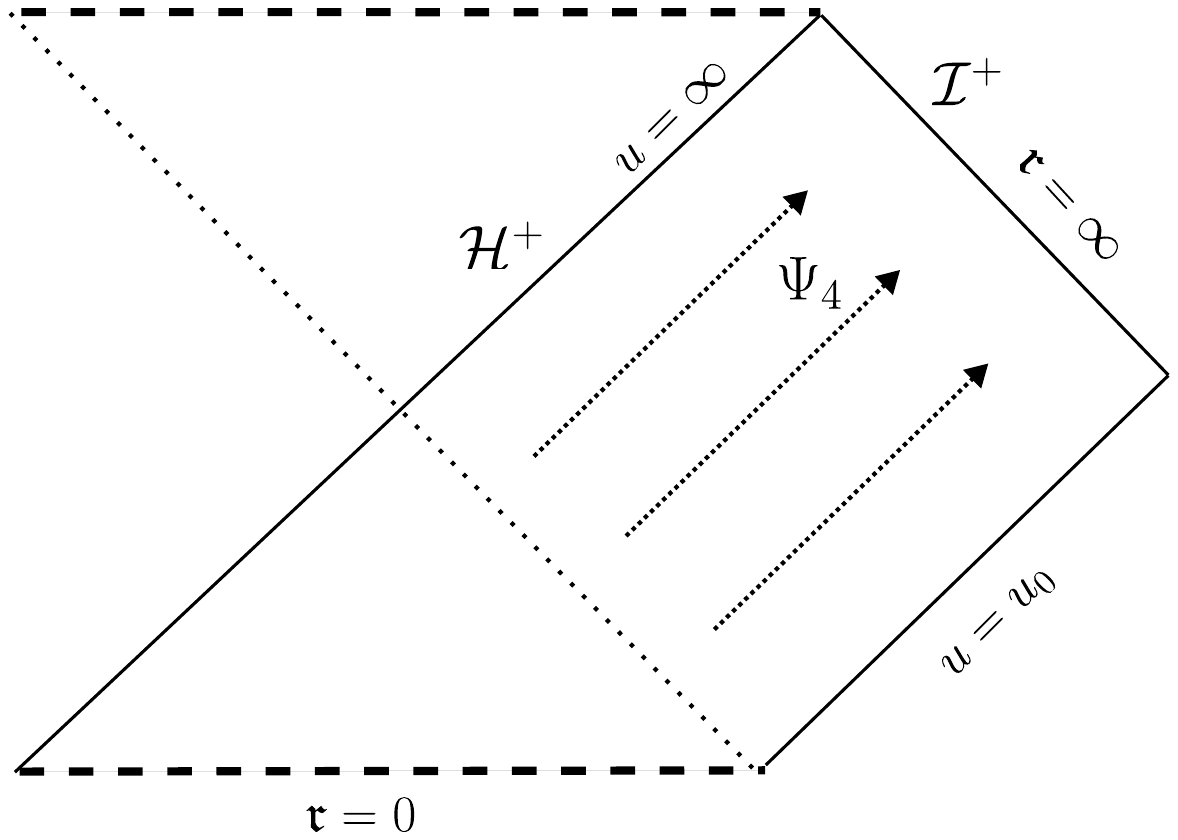}
    \caption{Penrose diagram for the Robinson-Trautman spacetime.}
    \label{fig:rt}
  \end{center}
\end{figure}

The Robinson-Trautman solutions
\cite{Robinson:1962zz,Chrusciel:1992rv} provide an illustrative
example of such an exact solution where $\Psi_2$ is unchanged and of
the Weyl scalars, only $\Psi_3$ and $\Psi_4$ are modified from its
Schwarzschild values~\footnote{In the perturbative limit, this is an
  example of the algebraically special perturbations studied in
  \cite{doi:10.1098/rspa.1984.0021}.}.  The standard form of the
Robinson-Trautman solution is written in terms of \emph{outgoing} null
coordinates $(u,\mathfrak{r},z,\bar{z})$ 
\begin{equation}
  ds^2 = -f(u,\mathfrak{r},z,\bar{z})\d u^2 - 2du\,\d \mathfrak{r} + \frac{2\mathfrak{r}^2}{P(z,\bar{z},u)}\d z\,\d\bar{z}\,.
\end{equation}
Using the vacuum Einstein equations, specifically
$R_{ab}m^a\bar{m}^b = 0$, it can be shown that
\begin{equation}
  \label{eq:rt-f}
  f = \Delta_P\ln P - 2\mathfrak{r}\frac{\partial}{\partial u}\ln P - \frac{2M}{\mathfrak{r}}\,.
\end{equation}
Here $\Delta_P := 2P^2\partial_z\partial_{\bar{z}}$ is the unit 2-sphere
Laplacian; note also that $\Delta_P\ln P$ is the Gaussian curvature of
the 2-sphere.  The parameter $M$ is a positive constant, namely the
mass.  When $P=P_0$ (see Eq.~\eqref{eq:P-sch}) is the time-independent round 2-sphere metric,
then we recover the Schwarzschild solution.  More generally, it can be
shown that $P$ satisfies the Robinson-Trautman equation:
\begin{equation}
  \Delta_P\Delta_P\ln P + 12M\frac{\partial}{\partial u}\ln P = 0\,.
\end{equation}
This follows from the expression for $f$ given in Eq.~\eqref{eq:rt-f}
combined with the Raychaudhuri equation along the future-directed
ingoing null direction $\ell^a$ given below.  Turning to the Weyl
tensor, we use the following null tetrad:
\begin{eqnarray}
  \ell^a\partial_a &=& \frac{\partial}{\partial u} - \frac{f}{2}\frac{\partial}{\partial \mathfrak{r}}\,,\\
  n^a\partial_a &=& \frac{\partial}{\partial \mathfrak{r}}\,, \qquad m^a\partial_a = \frac{P}{\mathfrak{r}}\frac{\partial}{\partial \bar{z}}\,.\end{eqnarray}
With this tetrad, the Weyl tensor components are (see
e.g. \cite{Podolsky:2009an})
\begin{subequations}
  \begin{align}
  \Psi_0 &= \Psi_1 = 0\\
  \Psi_2 &= -\frac{M}{\mathfrak{r}^3}\\
  \Psi_3 &= -\frac{P}{2\mathfrak{r}^2}\frac{\partial}{\partial z}\Delta_P\ln P \\
  \Psi_4 &=  -\frac{1}{\mathfrak{r}}\frac{\partial}{\partial z}\left(P^2\frac{\partial^2}{\partial u\partial z}\ln P \right) + \frac{1}{2\mathfrak{r}^2}\frac{\partial}{\partial z}\left(P^2\frac{\partial}{\partial z} \Delta_P\ln P\right)\,.
  \end{align}
\end{subequations}
We see that $\Psi_2$ is the same as for Schwarzschild while $\Psi_4$
is non-vanishing.  In this sense, the solution represents a
Schwarzschild black hole with non-infalling radiation as claimed
before.  In terms of a characteristic initial value formulation, the
solution can be constructed by prescribing the conformal factor of the
2-metric on a constant $u$ surface, say at the $u=u_0$ surface in
Fig.~\ref{fig:rt}.

There is however one issue which we have not addressed: Since $u$ is
an outgoing null coordinate, the horizon appears in the limit
$u\rightarrow \infty$ as shown in the Penrose diagram in
Fig.~\ref{fig:rt}.  Can the solution be extended beyond the future
horizon $\mathcal{H}^+$ at $u=\infty$?  For the Schwarzschild case,
it is clear that this can be done, and one obtains the usual extended
Schwarzschild spacetime.  As shown by Chru\'sciel
\cite{Chrusciel:1992rv}, this can indeed be done.  To go beyond
$\mathcal{H}^+$ we can attach the interior Schwarzschild spacetime and
the metric turns out to be sufficiently smooth (though not
$C^\infty$). The radiation decays exponentially when we approach
$\mathcal{H}^+$ (as we shall shortly see) and there is non-vanishing
transverse radiation arbitrarily close to the horizon in the exterior.
Since $\Psi_2$ is unchanged, this radiation does not perturb the
horizon geometry and its source multipole moments.
  
The Robinson-Trautman solution given above is an exact solution to the
Einstein equations.  It is instructive to consider the perturbative
limit wherein the amplitude of $\Psi_4$ is small~\cite{Qi:1993ey}.  Let us take a
perturbation given by
\begin{equation}
  P(z,\bar{z},u) = P_0 \, e^{W(z,\bar{z},u)}
\end{equation}
with $W$ taken to be small.  The linearization of the
Robinson-Trautman equation yields
\begin{equation}
  \Delta_0\Delta_0 W + 2\Delta_0W = -12M\frac{\partial W}{\partial u}\,.
\end{equation}
We can write a solution for $W$ as a linear superposition of spherical
harmonics (eigenfunctions of $\Delta_0$).  When we take
\begin{equation}
  W(z,\bar{z},u) :=\sum_{l,m} W_{l m}= \sum_{l=0}^\infty \sum_{m=-l}^l Y_{l}^m(z,\bar{z}) \mathcal{V}_l(u) \; ,
\end{equation}
we obtain exponentially decaying solutions
\begin{equation}
  \mathcal{V}_l(u) = A_l e^{-k_l u}\,,\quad   k_l = \frac{l(l+1)(l+2)(l-1)}{12M}\,,
\end{equation}
and $A_l$ is the amplitude of the $l$-mode.  Thus, the radiation
decays exponentially as we approach $\mathcal{H}^+$ as claimed above.
A little algebra yields the metric function $f$ as
\begin{equation}
  f = 1-\frac{2M}{\mathfrak{r}} +\sum_{l,m} (l-1)(l+2)\left[1+\frac{\mathfrak{r}}{6M} l(l+1)\right]W_{l m}\,.
\end{equation}
As for the Weyl tensor, $\Psi_2$ is unchanged.  It is straightforward
to check that $\Psi_3$ and $\Psi_4$ are modified and exponentially
decaying as $u\rightarrow \infty$ while $\Psi_0$ and $\Psi_1$ vanish
identically.

\section{Perturbations of the intrinsic horizon geometry}
\label{sec:ihperturb}
In the following, we study perturbations of the horizon detailed in the previous sections. We restrict ourselves to tidal perturbations so that the area of the perturbed cross-section is unchanged from the unperturbed one. Similarly, through Eq.~\eqref{eq:kerrkappa}, the surface gravity of the perturbed horizon coincides with the unperturbed one.  The
perturbations to the Weyl scalars $\Psi_0$ and $\Psi_1$ are taken to vanish at
the horizon, so the
perturbed horizon remains isolated to first order.  Consequently, the perturbed horizon is still characterized by
a surface of vanishing expansion (and shear). In this construction, we choose a convenient gauge motivated by how we want to slice the horizon, and using this gauge, we derive all quantities at the horizon for a general tidal perturbation. 

It is useful to keep in mind that we construct our coordinate system such that the horizon is located at $r=0$. Further, we select the perturbed null normal $\ell$ to be
tangent to the horizon, whence it remains geodetic. The affine parameter $v$ along $\ell$ will be chosen as before, so the null normal $\ell$ is only perturbed away from the horizon. Nonetheless, the perturbation modifies the geometry of the cross-section, as well as how it is embedded in the NEH.  

To construct a perturbed NEH, which forms the basis for perturbing the
near-horizon spacetime, we shall proceed in two steps.  The first is
to perturb a cross-section, which could be either a given
cross-section of the NEH, or the base-space $\widetilde{\Delta}$
arising from the projection
$\Pi: \widetilde{\Delta} \times\mathbb{R} \rightarrow
\widetilde{\Delta}$.  This perturbed cross-section will then be
embedded within the NEH and will determine time derivatives along the
horizon.  The main result of this section can be stated as follows.
Perturbations of the horizon geometry are specified by a perturbation
of $\Psi_2$: $\Psi_2 \to \Psi_2 + \widehat{\Psi}_2$.  From our
discussion of the multiples, this is equivalent to a perturbation of
the source multipole moments.  Since $\Psi_2$ has spin weight zero,
the perturbation $\widehat{\Psi}_2$ can be expanded in the usual
spherical harmonics. Keeping the mass fixed, we shall see that we only
need to consider multipoles beyond the dipole:
\begin{equation}
    \widehat{\Psi}_2 \triangleq \sum_{l\geq 2,m} \widehat{k}_{lm} Y_{lm}(z,\bar{z})
\end{equation}
where $\widehat{k}_{lm} = \widehat{e}_{lm}+i \widehat{b}_{lm}$.  Given
the coefficients $\widehat{k}_{lm}$, we shall show how the complete
geometry of the horizon can be reconstructed. As a by-product, it will
also become clear that such a perturbation of $\Psi_2$ necessary
implies that $\Psi_4$ cannot vanish so that the horizon cannot be of
Petrov type $D$ and that the spacetime must be radiative.  The
coefficients $\widehat{e}_{lm},\widehat{b}_{lm}$ are respectively
related to the electric and magnetic moments of the external field, as we will show in Sec.~\ref{sec:tidal-bh}.

\subsection{Perturbing a horizon cross-section}
\label{sec:perturbing-cross-section}

The examples shown in the previous section have left the horizon
geometry identical to the Schwarzschild case. Generically, however,
one would expect a perturbation to modify the horizon multipole
moments and therefore the near horizon geometry. In this section, we
detail the perturbation of the intrinsic horizon geometry.  We start
with the 2-metric $\widetilde{q}_{ab}$ defined on a 2-sphere $S_0$.
This could be any regular Riemannian 2-metric but for the
astrophysical applications that we have in mind, this would be a
distorted Schwarzschild 2-sphere metric.  Though not essential, it
will be useful to use complex coordinates $(z, \bar{z})$ for this
purpose so that the 2-metric has the form
\begin{equation} \label{eq:cross-section-P}
  \tilde{ds}^2 = \frac{2R^2}{P^2(z,\bar{z})}dz\,d\bar{z}\,,
\end{equation}
with $R$ being the area radius of the horizon.  The complex coordinate
$z$ can be obtained from the usual spherical coordinates
$(\theta,\phi)$ using a stereographic projection.  In this form, the
2-metric is conformally flat.

The Ricci scalar $\widetilde{\mathcal{R}}$ of this 2-metric is given in terms of $P$
as
\begin{equation}\label{eq:Ricci-scalar-Sphere}
  \widetilde{\mathcal{R}} = \frac{4P^2}{R^2}\frac{\partial^2}{\partial z\partial\bar{z}}\ln P = 2\Delta_P\ln P\,,
\end{equation}
with $\Delta_P= 2 P^2/R^2 \partial_z\partial_{\bar{z}}$ the Laplacian
of the sphere of radius $R$.

For the Kerr metric, it is relatively straightforward to work out the
transformation to arrive at the $(z,\bar{z})$ coordinates and the
conformal factor. We start with the expression for the Kerr metric with mass $M$ and
specific angular momentum $a$ in the usual in-going Eddington-Finkelstein
coordinates $(v,\mathfrak{r},\theta,\phi)$: 
\begin{widetext}
\begin{equation}
  ds^2 = -\left(1-\frac{2M \mathfrak{r}}{\rho^2}\right)\d v^2 + 2\d v\,\d \mathfrak{r}  - 2a\sin^2\theta \d \mathfrak{r}\,d\varphi 
  - \frac{4aM\mathfrak{r} \sin^2\theta}{\rho^2} \d v\,\d\varphi + \rho^2 \d\theta^2 +  \frac{\Sigma^2\sin^2\theta}{\rho^2}\,\d\varphi^2\,,
\end{equation}
where 
\begin{equation}
  \rho^2 = \mathfrak{r}^2 + a^2\cos^2\theta\,, \qquad
  \Delta = \mathfrak{r}^2-2M\mathfrak{r}+a^2\,,\qquad 
  \Sigma^2 = (\mathfrak{r}^2+a^2)\rho^2 + 2a^2M\mathfrak{r}\sin^2\theta\,.
\end{equation}
\end{widetext}
Here $\Delta$ should not be confused with the directional derivative
along $n^a$ as defined earlier, and the distinction should be clear
from the context.  The horizon is located at $\Delta = 0$, i.e. at
$\mathfrak{r}=\mathfrak{r}_+$, where $\mathfrak{r}_+ =M+\sqrt{M^2-a^2}$. 
  The volume form on a
cross-section of the horizon ($\mathfrak{r}=\mathfrak{r}_+$ and constant $v$) is
$\tilde\epsilon = (\mathfrak{r}_+^2 + a^2)\sin\theta d\theta \wedge d\varphi$ 
. Thus,
the area of the horizon is $A= 4\pi (\mathfrak{r}_+^2 + a^2)$ and the area radius
is $R = \sqrt{\mathfrak{r}_+^2+ a^2}$.

The metric within a cross-section of the horizon can be written as
\cite{Smarr:1973zz,Ashtekar:2004gp}
\begin{equation} \label{eq:line-element-surface}
    \text{d} \tilde{s}^2 = R^2\left(\frac{\text{d} \zeta^2}{f(\zeta)} + f(\zeta) \text{d} \phi^2\right)\,.
\end{equation}
where $\zeta = \cos\theta$ and
\begin{equation}\label{eq:f-Kerr}
  f(\zeta) = \frac{(1-\zeta^2)}{1 -b^2 (1- \zeta^2)}
\end{equation}
with
$b=\frac{a}{\sqrt{2M
    \mathfrak{r}_+}}=\frac{a}{\sqrt{\mathfrak{r}_+^2+a^2}}$.  The
complex coordinate $z$ is then
\begin{equation}
  z =  e^{i\phi-b^2 \zeta} \sqrt{\frac{1+\zeta}{1-\zeta}}  \,. \label{eq:z-to-zeta-and-phi}
\end{equation}
Finally, the 2-metric takes the manifestly conformally flat form as
desired:
\begin{equation}
  ds^2 = \frac{R^2 f}{ z  \bar{z}} \d z\d \bar{z}\,.
\end{equation}
Thus, the function $P(z,\bar{z})$ is
\begin{equation}\label{eq:P-z-zbar-f}
  P(z,\bar{z}) = \sqrt{\frac{2z \bar{z}}{f}}\,.
\end{equation}
The expression~\eqref{eq:z-to-zeta-and-phi} is invertible in the small
spin limit $a\ll 1$, so the metric function~\eqref{eq:f-Kerr} (combined
with $z\bar{z}$ as it appears in the metric) can be expressed in terms
of the new coordinates $\{z,\bar{z}\}$ as
\begin{equation}
    \frac{f(z,\bar{z})}{2z\bar{z}} = \frac{2}{(1+z\bar{z})^2}\left\{1 +\frac{3a^2}{4M^2} \frac{(1-z\bar{z})}{(1+z\bar{z})^2}\right\}+\mathrm{O}[a^4]\,.
\end{equation}

The construction above is more general than just for the Kerr horizon.
In fact, any axisymmetric 2-sphere can be expressed in the form of
Eq.~\eqref{eq:line-element-surface}, and
Eq.~\eqref{eq:z-to-zeta-and-phi}
yields the complex coordinate $z$.

Going now beyond axisymmetry, while the 2-metric can no longer be
expressed as Eq.~\eqref{eq:line-element-surface}, the conformal
representation still remains valid.  Thus, a perturbation of an
axisymmetric metric can be written as a perturbation of $P$ even when
the perturbation is non-axisymmetric. Thus, starting from $P_0$, we
shall perturb the 2-metric by
\begin{equation}\label{eq:perturbation-cross-section}
  P(z,\bar{z}) \rightarrow P_0(z,\bar{z})(1 + \widehat{P}(z,\bar{z}))
\end{equation}
where $\widehat{P}$ is a small perturbation.  We will only keep the terms of
linear order in $\widehat{P}$ in the remainder of this section.  In a
given concrete physical situation the perturbation $\widehat{P}$ will
depend on a small parameter.  The prototypical example is a binary
system with the small parameter being a combination of the mass of the
binary companion and the separation between the two masses.

With the above construction, we still have the freedom to perform a
complex coordinate transformation $z\rightarrow z^\prime$.  The form
of the metric is unchanged under a fractional linear transformation
corresponding to a $SU(2)$ matrix $A$: 
\begin{equation}\label{eq:Moebius-transormation}
  A = \begin{pmatrix}
    a & b\\
    c & d
  \end{pmatrix}\,, \quad
  z \rightarrow z^\prime = \frac{az + b}{cz + d} \; .
\end{equation}
The round 2-sphere metric (denoted by $P_\circ$ --- note the difference with $P_0$~\footnote{Notice that we use the subscript ${}_0$ to refer to any axisymmetric background, while the quantities with ${}_\circ$ refer specifically to the Schwarzschild background with $P_\circ$ given by Eq.~\eqref{eq:P-sch}. }) is invariant under this transformation:
\begin{equation}\label{eq:isometries-sphere-metric}
  \frac{2}{P^2_\circ(z,\bar{z})}dz\,d\bar{z} = \frac{2}{P^2_\circ(z^\prime,\bar{z}^\prime)}dz^\prime\,d\bar{z}^\prime\,.
\end{equation}
Thus, an arbitrary 2-metric $\widetilde{q}_{ab}$ is conformally
equivalent to a 3-parameter family of round 2-sphere metrics
corresponding to the allowed $SU(2)$ fractional linear
transformations.

In summary: any 2-sphere metric is written as
\begin{equation}
  \widetilde{ds}^2 = \frac{2\psi^2(z,\bar{z})}{P_\circ^2(z,\bar{z})}dz\,d\bar{z}\,,
\end{equation}
and we have a 3-parameter family of allowed complex coordinates
$(z,\bar{z})$ corresponding to the $SU(2)$ matrix $A$.

A procedure for choosing a canonical round 2-sphere metric
from this 3-parameter family is given in \cite{Ashtekar:2021kqj},
based on requiring the dipole ``area'' moment to
vanish; see also \cite{Korzynski:2007hu}.  

Under a
tidal perturbation, the geometry of the cross-section of the horizon
is modified. The geometry of $\Delta$ is determined by
$(\widetilde{q}_{ab},\mathcal{D})$ or, as discussed earlier, by the
2-metric $\widetilde{q}_{ab}$ and $\omega_a$. The tidal perturbation
will modify both of these.  As far as $\widetilde{q}_{ab}$ is
concerned, the gauge-invariant information of the tidal perturbation
is contained in variations of the scalar curvature
$\widetilde{\mathcal{R}}$.  The scalar curvature is given by
Eq.~\eqref{eq:Ricci-scalar-Sphere}
with $\Delta_P$ the Laplace-Beltrami compatible with the metric of the
cross-section $S_0$.

A perturbation of $P$ away from $P_0$ according to
$P(z,\bar{z}) = P_0(1+\widehat{P})$ leads in general to a perturbation of
the area and the scalar curvature
$\mathcal{R} = \mathcal{R}_0+\widehat{\mathcal{R}}$.  We assume
that the area is unchanged under the perturbation which can be shown,
at linear order in $\widehat{P}$, to be equivalent to
\begin{equation}
  \oint_S \widehat{P} = 0\,.
\end{equation}
Thus, for a round 2-sphere, when we expand $\widehat{P}$ in terms of
spherical harmonics, this implies that the monopole term of $\widehat{P}$ should
vanish. For the scalar curvature, we obtain using
Eq.~\eqref{eq:Ricci-scalar-Sphere} 
\begin{equation}\label{eq:ricci-perturbed}
  \widehat{\mathcal{R}} = -2(\Delta_0\widehat{P} + \mathcal{R}_0\widehat{P})\,.
\end{equation}
Thus, for a perturbation of a round 2-sphere of radius $R$ 
(i.e.
$\mathcal{R}_0 = 2/{R^2}$), the perturbation leaves the scalar curvature
unaffected if
\begin{equation}
  \Delta_0 \widehat{P} + \frac{2}{ R^2}\widehat{P} = 0\,.
\end{equation}
This happens if $\widehat{P}$ is a dipole perturbation, i.e. it is a linear
combination of the three $\ell=1$ spherical harmonics.  This leads to
a 3-parameter class of perturbations which do not affect the scalar
curvature.  Any quadrupolar or higher $\ell$ perturbations leads to a
genuine perturbation of the scalar curvature.  The above interpretations of
the monopole and dipole parts of $\widehat{P}$ continue to hold under any
M\"obius transformation with a $SU(2)$ matrix.  Then, we can define
the equivalence class of cross-sections with curvature perturbation
$\widehat{\mathcal{R}}$ as
\begin{equation}\label{eq:equivalent-class}
  \widehat{P}\sim \widehat{P}^\prime\,, \text{ if } \widehat{\mathcal{R}}=\widehat{\mathcal{R}}^\prime\,.
\end{equation}
Different choices of perturbation within this equivalence class
characterized by $\widehat{\mathcal{R}}$ give rise to different gauge
choices on the cross-section.

Apart from the curvature $\mathcal{R}$, the other ingredient which
specifies the horizon geometry is the derivative operator
$\mathcal{D}$. Hence, we need to discuss how the perturbation changes
the cross-section's connection, and for later convenience, the
directional derivatives on the sphere.  Since
$m^a\partial_a \triangleq P/c \,\partial_{z}$, when $P\to P(1+\widehat{P})$,
it is clear that $m^a\to m^a + \widehat{P}m^a$. Similarly, using again
the notation $a:= \alpha - \bar{\beta}$, we will have
$a\to a+\widehat{a}$.  It is easy to show that
\begin{equation}\label{eq:a-P-hat}
  \widehat{a} \triangleq a_0 \widehat{P} + \delta_0 \widehat{P} \,,\quad \widehat{\delta} \triangleq  \widehat{P} \delta_0\,.
\end{equation}
The perturbed cross-section, characterized by this
connection~\eqref{eq:a-P-hat}, is taken to be a cross-section of a NEH
$\Delta$. Notice that we have not added a tilde on these expressions
to avoid cumbersome notation. However, it should be clear from the
context that the derivative operator $\widehat{\delta}$ and the
cross-section's connection are computed for the 2-dimensional
spacelike manifold $\tilde{\Delta}\sim S$.

The angular field equations~\eqref{eqs:angular} relate the connection of the cross-section with the Weyl scalar $\Psi_2$. Perturbing to first order the real and imaginary part of the second equation in~
Eq.~\eqref{eqs:angular} yields expressions for the perturbation to the
Weyl scalar $\Psi_2$ as a function of the perturbed connection $\widehat{a}$ and the spin coefficient $\widehat{\pi}$
\begin{subequations}\label{eq:re-im-Psi2-prime}
\begin{align} \label{eq:re-im-Psi2-prime:1}
     -2\re \widehat{\Psi}_2 &\triangleq\delta_0 \widehat{a}  +\bar{\delta}_0 \widehat{\bar{a}} -2 \widehat{a}\bar{a}_0+({}_0 \leftrightarrow \widehat{} \,)\,,\\\label{eq:re-im-Psi2-prime:2}
    -2 i \im\widehat{\Psi}_2 &\triangleq \delta_0\widehat{\pi}-\bar{\delta}_0\widehat{\bar{\pi}}-\widehat{\pi} \bar{a}_0+\widehat{\bar{\pi}}a_0+({}_0 \leftrightarrow \widehat{} \,)\,.
\end{align} 
\end{subequations} Expressing Eq.~\eqref{eq:ricci-perturbed} in terms of the perturbation to the connection~\eqref{eq:a-P-hat}, and comparing with Eq.~\eqref{eq:re-im-Psi2-prime:1} yields
\begin{equation}
    \widehat{\mathcal{R}} = -4\re \widehat{\Psi}_2\,.
\end{equation} 
Therefore, we see that a perturbation to the real part of the Weyl
scalar $\Psi_2$ is fully determined by the perturbation of
$P$. However, not all of the data on the NEH is determined by these
perturbations.  For instance, Eq.~\eqref{eq:re-im-Psi2-prime:2}
relates a perturbation to the imaginary part of $\Psi_2$ with the
perturbations of the connection and the spin coefficient $\pi$. The
spin coefficient $\pi$ cannot be uniquely determined on the
cross-section. Rather, we need to specify the foliation of the horizon
to find the dependence of $\widehat{\pi}$ on the perturbation to the
geometry. We turn to this in the next subsection.

\subsection{Embedding a perturbed cross-section within a NEH}
\label{sec:embedding-perturbed-cross-section}

In the previous section, we detailed how the connection and curvature
of the cross-section are altered by a perturbation. However, to
construct the perturbed NEH we still need to specify
how a perturbation alters the foliation of the horizon. In this
subsection, we detail this construction, together with the gauge choices
we make. 

To embed a perturbed cross-section into the structure of the isolated horizon, we need to specify the foliation of the horizon in 2-spheres. Different choices of the one-form $\widetilde\omega_a$, which is the pullback of $\omega_a$ to $S$ (see Fig.~\ref{fig:projections}),
\begin{equation}\label{eq:w-1}
    \widetilde\omega_a\triangleq \iota_\star \omega_a \triangleq \pi \widetilde{m}_a+\bar{\pi}\bar{\widetilde m}_a\,,
\end{equation} 
can correspond  to different foliations.  Hence, to construct the
isolated horizon we need to specify how the horizon is foliated and
whether the perturbation changes the foliation. As discussed
in~\cite{Ashtekar_2002}, there exists a preferred foliation of the
unperturbed horizon. In the following, we will review this
construction and choose a foliation for the perturbed horizon that is
convenient to deal with tidal perturbations.  Recall that there are no
harmonic 1-forms on a sphere, and so any 1-form can be uniquely
decomposed in terms of its exact and co-exact parts as
\begin{equation}\label{eq:w-2}
  \widetilde \omega = -\star \d \mathcal{U} + \d \mathcal{V} \,,
\end{equation}
where $\mathcal{U}$ and $\mathcal{V}$ are smooth, real functions on
the cross-section. By taking the exterior derivative of
Eqs.~\eqref{eq:w-1} and~\eqref{eq:w-2}, we see that $\mathcal{U}$ and $\mathcal{V}$ are
potentials for the divergence and curl of $\widetilde{\omega}$:
\begin{eqnarray}
  \Delta_P \mathcal{U} &=& \star  \,d\widetilde{\omega} = 2\im\Psi_2\,,\label{eq:Delta-U}\\
  \Delta_P\mathcal{V} &=& \textrm{div} \,\widetilde{\omega}\,.\label{eq:Delta-V}
\end{eqnarray}
Recall that $\Delta_P$ is the Laplace-Beltrami operator associated with the
metric~\eqref{eq:cross-section-P}.

We perturb $\im\Psi_2 \to \im\Psi_2^0 + \im\widehat{\Psi}_2$, the
rotational potential as
$\mathcal{U}\to \mathcal{U}_0 + \widehat{\mathcal{U}}$, and also $P$
according to Eq.~\eqref{eq:perturbation-cross-section} (which
transforms $\Delta_P$). Under these transformations,
Eq.~\eqref{eq:Delta-U} leads to
\begin{equation}
    \Delta_{P_0} \widehat{\mathcal{U}}=2 \im\widehat{\Psi}_2-4 \widehat{P} \; \im\Psi_{2}^0\,.
\end{equation}  
For perturbations of a non-rotating background spacetime, i.e.
$\im \Psi_{2}^0 =0$, the perturbation to the rotational scalar
potential can be expanded in spherical harmonics as
\begin{equation}
    \mathcal{U}=-2 R^2\sum_{l,m} \frac{\widehat{b}_{lm}}{l(l+1)} Y_{lm}\,.
\end{equation}
The coefficients $\widehat{b}_{lm}$ are related to perturbations of
the spin multipole moments.  

In Eq.~\eqref{eq:w-2}, $\mathcal{V}$ represents the gauge freedom in
the choice of $\widetilde \omega$. A commonly used gauge choice is
$\d\mathcal{V}=0$~\cite{Ashtekar_2002}, which is related to choosing
the so-called good cuts. We shall use this same gauge choice for the
unperturbed background, i.e. $\mathcal{V}_0=\text{const}$. The
divergence of the perturbed $\widehat{\widetilde \omega}$ can then be
expressed as
\begin{align}\label{eq:divergence-omega}
  \d \star\widehat{ \widetilde\omega}
  &= \Delta_{P_0} \widehat{\mathcal{V}} \tilde \epsilon \nonumber \\
  & = \left\{ \delta_0\widehat{\pi}+\bar{\delta}_0\widehat{\bar{\pi}}-\widehat{\pi} \bar{a}_0-\widehat{\bar{\pi}}a_0+ ({}_0 \leftrightarrow \widehat{} \,)\right\} \tilde\epsilon_0\,,
\end{align} 
where $\tilde\epsilon_0=i \tilde{m}_0\wedge \bar{\tilde m}_0$ is the unperturbed area element. The second expression in Eq.~\eqref{eq:divergence-omega} has been obtained using Eqs.~\eqref{eq:w-1} and~\eqref{eqs:spincoeffs-exterior}. The real function  $ \widehat{\mathcal{V}}$  
characterizes the change of foliation with respect to the unperturbed slicing in  $v=\text{const}$ surfaces. In other words, if $\Delta_{P_\circ} \widehat{\mathcal{V}} \triangleq 0$, the perturbed horizon is still foliated by the good cuts of the unperturbed horizon.
However,  for this paper, it is convenient to choose instead  
\begin{equation}\label{eq:cuts}
  \Delta_{P_0} \widehat{\mathcal{V}} \triangleq \textrm{div} \,\widetilde{\omega} \triangleq -2\re[\widehat{\Psi}_2]\,,
\end{equation}
so that the slicing of the perturbed horizon changes if its geometry
is altered. This choice guarantees that the vector $n$ and its
expansion are not modified regardless of the perturbation. In other words, the ``perturbed" radial coordinate
coincides with the unperturbed one.  As we will see in Sec.~\ref{sec:tidally-perturbed-sch},
this choice facilitates the comparison of our tidally perturbed black
hole with the existing literature on tidally perturbed black holes
(see for instance,~\cite {Poisson_2010,Poisson:2021yau,Binnington:2009bb,LeTiec:2020bos,LeTiec:2020spy,Charalambous:2021mea,Pani:2015hfa,Pani:2018inf,Damour:2009vw,Damour:2009wj}). This gauge~\eqref{eq:cuts}
also simplifies the expressions for the perturbed Weyl scalars and spin
coefficients in terms of the spin-weighted spherical harmonics.

Finally, it is also worth noting the link between this gauge condition
and quasi-local notions of ``momentum'' and ``force'' on a black hole.
It is of interest, especially in the context of binary black hole
simulations, to calculate linear momentum quasi-locally
\cite{Krishnan:2007pu}.  This is interesting, for example, when
calculating the ``kick'' imparted to the remnant black hole.  From the
perspective of the quasi-local horizon, momentum is connected with the
foliation of the horizon. A clear example is a 
``boosted'' Kerr black hole in Kerr-Schild coordinates, and it is easy
to check that the foliation is then determined by the boost parameter
\cite{Huq:2000qx}.  The foliation, as we have seen, is determined by
$\textrm{div} \, \widetilde{\omega}$ and thus must be connected with the
boost, or linear momentum; the curl of $\omega$ determines angular
momentum while its divergence determines linear momentum.  Our gauge
condition links this to $\re\widehat{\Psi}_2$, which is just the
external tidal force acting on the black hole; for a binary companion
of mass $M_2$ at a distance $d$, we would have
$\re\widehat{\Psi}_2 \sim M_2/d^3$.  The external reference frame in
which we determine the momentum is specified by the properties of the
past lightcone, namely the expansion of $-n^a$.

\subsection{The geometry of a perturbed Schwarzschild horizon}
\label{subsec:schwarzschild-perturb}

The discussion so far has been for perturbations of any background
cross-section characterized by $P_0$.  However, before proceeding to
express the perturbed horizon data in terms of the perturbation, we
choose the background to be a Schwarzschild background (denoted with
the sub-index ${}_\circ$ instead of ${}_0$), which has a round
background cross-section. This simplification allows us to set the
following background quantities to zero
\begin{equation}\label{eq:Schwarzschild-vanishing-data}
  \pi_\circ = \lambda_\circ=\im\Psi_2^\circ=\Psi_3^\circ=\Psi_4^\circ=\Psi_1^\circ=\Psi_0^\circ= 0\,,
\end{equation}
which will simplify the discussion of the perturbed data.

We start writing the perturbation to the spin coefficient
$\widehat{\pi}$ in a more concise form using the $\eth$ operator.
First note that in general, since $\pi = \bar{m}^a\omega_a$, a short
calculation shows that
\begin{equation}
  \eth\pi = \bar{m}^a\delta\widetilde{\omega}_a = \frac{1}{2}\textrm{div}\widetilde{\omega} -\frac{i}{2}\star d\widetilde{\omega}\,.
\end{equation}
Combining the equations for the curl ~\eqref{eq:re-im-Psi2-prime:2}
and divergence of
$\omega$~\eqref{eq:divergence-omega} 
yields
$ \eth_\circ \widehat{\pi} \triangleq \Delta_{P_\circ}
\widehat{\mathcal{V}}/2-i \im[\widehat{\Psi}_2]$. Considering now the
perturbations of $\pi$ and $\widetilde{\omega}$, noting that these are
already first order quantities, using the gauge
condition~\eqref{eq:cuts}, the differential equation for
$\widehat{\pi}$ can be concisely written as
\begin{equation}\label{eq:edth-pi}
  \eth_\circ \widehat{\pi} \triangleq -\widehat{\Psi}_2\,,
\end{equation}
so that it is manifest that $\widehat{\pi}$ can be easily solved in
terms of $\widehat{\Psi}_2$ using the properties of the $\eth$
operator~\footnote{Notice that our convention for the $\eth$  operator is slightly different to the one presented in~\cite{Goldberg:1966uu}. The action of $\eth$  over the spherical harmonics is given by $\bar{\eth}\eth{}_s Y_{lm}=-\frac{(l-s)(l+s+1)}{2c^2} {}_s Y_{lm}$. }. The definition of the $\eth$ operator and its action on the spin-weighted spherical harmonics are summarized in Appendix~\ref{sec:spin-weighted-SH}. 

The third angular equation in Eq.~\eqref{eqs:angular} defines the perturbation of the Weyl scalar $\widehat{\Psi}_3$ at the horizon
\begin{equation}\label{eq:psi_3-perturbed}
   \widehat{\Psi}_3 \triangleq   \bar{\delta}_{\circ} \widehat{\mu}+\widehat{\bar{\delta}} \mu_{\circ}- \delta_{\circ} \widehat{\lambda}+ \widehat{\pi} \mu_{\circ}+(\bar{\alpha}_\circ-3 \beta_\circ) \widehat{\lambda}\,.
\end{equation}
The perturbed evolution equations at the horizon (see Eqs.~\eqref{eqs:evolution-bianchi} and~\eqref{eqs:timeevolution}) imply that the following quantities are such that
\begin{equation}\label{eq:D-conditions-horizon}
    D_\circ \widehat{\Psi}_2 \triangleq D_\circ\widehat{\pi}\triangleq D_\circ\widehat{\alpha}\triangleq D_\circ\widehat{\beta}\triangleq D_\circ \widehat{a} \triangleq 0\,.
\end{equation}

Combining the equations for $\widehat{\epsilon}$~\eqref{eqs:timeevolution:1} and~\eqref{eqs:timeevolution:2}
together with Eq.~\eqref{eq:D-conditions-horizon} we obtain
\begin{align}
   & \delta_\circ (\widehat{\epsilon}+\widehat{\bar{\epsilon}})\triangleq -2\bar{\epsilon}_\circ (\widehat{\bar{\pi}}+\widehat{\bar{\alpha}}-\widehat{\beta})-2\widehat{\bar{\epsilon}}(\bar{\alpha}_\circ+\beta_\circ) \\
   &\delta_\circ (\widehat{\epsilon}-\widehat{\bar{\epsilon}}) \triangleq 2\bar{\alpha}_\circ (\widehat{\epsilon}-\widehat{\bar{\epsilon}}) \; .
\end{align} 
Analogously to the general gauge conditions for an isolated horizon detailed in Sec.~\ref{sec:General-near-horizon}, we choose a gauge such that the condition $\pi \triangleq \alpha + \bar{\beta}$ holds also to first order, i.e. 
$\widehat{\pi}\triangleq\widehat{\alpha}+\widehat{\bar{\beta}}$  and using  $\alpha_\circ=-\bar{\beta}_\circ$, we see that $\delta_\circ (\widehat{\epsilon}+\widehat{\bar{\epsilon}})\triangleq 0$. The trivial solution to this equation is $(\widehat{\epsilon}-\widehat{\bar{\epsilon}})\triangleq 0$, which we shall choose. Therefore $(\widehat{\epsilon}+\widehat{\bar{\epsilon}})\triangleq\text{const}$ at the horizon. In this gauge, $\widehat{\epsilon}$ is related to the perturbation of the surface gravity at the horizon, which we will choose to vanish $\widehat{\epsilon}\triangleq\widehat{\bar{\epsilon}}\triangleq0$. This last condition 
is not an extra restriction in our construction, rather, it follows from us limiting our study to linear, tidal perturbations of isolated horizons. Choosing the area of the perturbed horizon to coincide with the area of the unperturbed horizon makes the comparison between these two horizons more transparent.    
Therefore, we consider that the radius of the perturbed horizon does not change with respect to the unperturbed one, and by Eq.~\eqref{eq:mass-ih}, its mass is perturbed quadratically with the perturbation to $J$. Similarly, using Eq.~\eqref{eq:kerrkappa}
, we see that the perturbation to the surface gravity is at least quadratic in the perturbation.  Therefore, we can set $\widehat{\kappa}_{(l)}=\widehat{\epsilon}+\widehat{\bar{\epsilon}}\triangleq0$ without loss of generality.         

Finally, we can now show that with our gauge choice
Eq.~\eqref{eq:cuts}, $\mu$ remains unaffected by the perturbation,
i.e.  $\widehat{\mu} \triangleq 0$.  The spin coefficients
$\widehat{\lambda}$ and $\widehat{\mu}$ satisfy the equations
\begin{subequations}\label{eq:evolution-mu-lambda}
  \begin{align}
    & D_{\circ} \widehat{\lambda} + \widetilde{\kappa}_{(\ell)} \widehat{\lambda}\triangleq\bar{\delta}_{\circ} \widehat{\pi}+ a_{\circ} \widehat{\pi} \triangleq \bar{\eth}_\circ \widehat{\pi}\,, \\
    & D_{\circ} \widehat{\mu} + \widetilde{\kappa}_{(\ell)} \widehat{\mu}\triangleq\delta_{\circ} \widehat{\pi}- \bar{a}_\circ \widehat{\pi}+\widehat{\Psi}_{2} \triangleq 0 \,.
  \end{align}
\end{subequations}
where we have used Eq.~\eqref{eq:edth-pi} and our choice of
cuts~\eqref{eq:cuts} in the last equation. Notice that the right-hand
side of these expressions is ``time-independent''. This means that the
spin coefficients $\widehat{\mu}$ and $\widehat{\lambda}$ have
solutions of the form $(1 -e^{- \widetilde{\kappa}_{(\ell)}v})F[z,\bar{z}]$, where the
integration constant is chosen so that
$\widehat{\mu}=\widehat{\lambda}\triangleq0$ at $v=0$.  When the
horizon is isolated, the extrinsic
curvature~\eqref{eq:S-cross-section} is ``time-independent''
$D \tilde{S}_{ab}\triangleq 0$ (or equivalently
$D\mu\triangleq D\lambda\triangleq 0 $) and
$\widehat{\mu} \triangleq0$.

Using Eqs.~\eqref{eq:evolution-mu-lambda}, we see that the evolution equation for the Weyl scalar $\widehat{\Psi}_3$ in Eq.~\eqref{eqs:evolution-bianchi:3}
\begin{equation}
    \label{eq:evolution-psi3}D\widehat{\Psi}_3+ \widetilde{\kappa}_{(\ell)}\widehat{\Psi}_3\triangleq \bar{\eth} \widehat{\Psi}_2+3\widehat{\pi}\Psi_2^\circ
\end{equation}
is equivalent to Eq.~\eqref{eq:psi_3-perturbed}.  Notice that the perturbed spin coefficients $\widehat{\mu}$, and $\widehat{\lambda}$, and the Weyl scalar $\widehat{\Psi}_3$, depend on the foliation of the horizon (and therefore on our choice of $\widehat{\mathcal{V}}$ 
in Eq.~\eqref{eq:cuts}). However, the perturbed $\Psi_4$ is independent of the foliation~\eqref{eq:cuts}, and depends uniquely on the background quantities and $\widehat{\Psi}_2$.   

The fact that $\widehat{\Psi}_4$ is foliation independent becomes manifest by taking the $D$ derivative of the time evolution equation for $\Psi_4$ in~\eqref{eqs:evolution-bianchi} and eliminating the terms $D\widehat{\lambda}$ and $D\widehat{\Psi}_3$ using Eqs.~\eqref{eq:evolution-mu-lambda} and~\eqref{eq:evolution-psi3}. Simplifying and rearranging the terms, we obtain the following differential equation for $\widehat{\Psi}_4$ 
\begin{equation}\label{eq:DPsi4-horizon}
    D^2\widehat{\Psi}_4 +3 \widetilde{\kappa}_{(\ell)}D\widehat{\Psi}_4 +2 \widetilde{\kappa}_{(\ell)}^2 \widehat{\Psi}_4 \triangleq \bar{\eth}_\circ^2 \widehat{\Psi}_2 +8\pi_\circ \bar{\delta}_\circ \widehat{\Psi}_2 +12\pi_\circ^2 \widehat{\Psi}_2\,.
\end{equation} 
For perturbations of the Schwarzschild horizon, the right-hand side of this expression simplifies to $\bar{\eth}^2_\circ \widehat{\Psi}_2$
. Further, notice that the right-hand side of this equation is time-independent by Eq.~\eqref{eq:D-conditions-horizon}, while $D\widehat{\Psi}_4\neq 0$ in general.  The form of Eq.~\eqref{eq:DPsi4-horizon} suggests a solution for $\widehat{\Psi}_4$ at the horizon of the form $\widehat{\Psi}_4 \triangleq T(v) Y(z,\bar{z})$. Using this ansatz we can separate Eq.~\eqref{eq:DPsi4-horizon} in two independent differential equations for $T(v)$ and $Y(z,\bar{z})$
\begin{equation}\label{eq:relation-psi2-psi4-H}
    D^2T +3 \widetilde{\kappa}_{(\ell)}DT +2 \widetilde{\kappa}_{(\ell)}^2 T \triangleq K\,,\quad
    K Y \triangleq \bar{\eth}^2 \widehat{\Psi}_2\,,
\end{equation} where $K$ is a separation constant.  Therefore, the angular dependence of $\widehat{\Psi}_4$   can only be freely specified at the horizon when $K=0$, which limits the perturbation to  $\Psi_2$ to be a solution of $\bar{\eth}^2 \widehat{\Psi}_2\triangleq 0$, i.e.  $\widehat{\Psi}_2$ can only be monopolar or dipolar. As already been discussed in Sec.~\ref{sec:perturbing-cross-section}, a dipolar perturbation of the real part of $\Psi_2$ is pure gauge, and we impose the monopolar perturbation of $\widehat{\Psi}_2$ to vanish since this term would be related to a black hole's mass perturbation. 
Hence, the only physically relevant case corresponds to a dipolar perturbation of the imaginary part of $\Psi_2$, which will be discussed in more detail in Sec.~\ref{sec:Sch+dipole}.  Equivalently,  Eq.~\eqref{eq:DPsi4-horizon} implies that any quadrupolar (or higher) perturbation of a type D horizon yields at least a type II horizon~\footnote{The type D horizon condition reads $3\Psi_2\Psi_4=2\Psi_3^2$~\cite{Lewandowski:2018khe}. A type D isolated horizon is characterized by the existence of at least one tetrad where $\Psi_4$ and $\Psi_3$ vanish and $\Psi_2\neq0$. Using this tetrad and perturbing linearly all the Weyl scalars, we see that the perturbed horizon is type D if $\hat{\Psi}_4\triangleq 0$
The relationship between the spin coefficients $\hat{\Psi_4}$ and $\hat{\Psi_2}$ at the horizon~\eqref{eq:DPsi4-horizon} imposes a real restriction on the form of the perturbation, namely $\bar{\eth}_0^2\hat{\Psi}_2\triangleq 0$. $\Psi_2$ has spin zero, so $\hat{\Psi}_2$ can be spanned using the spherical harmonics $\hat{\Psi}_2 \propto Y_{lm}$. The condition above implies $ \bar{\eth}_0^2Y_{lm} = [(l-1)l(l+1)(l+2)]^{1/2} {}_{-2}Y_{lm}\triangleq0$, and therefore that $ {}_{-2}Y_{lm}=0\quad\forall l\geq 2$. Consequently, $\hat{\Psi}_2$  can only have a monopolar or dipolar contribution for the perturbed horizon to be type D. These perturbations represent a change in the mass and center of mass of the black hole, which we set to zero by considering a center-of-mass coordinated system for the isolated horizon.
} with $\widehat{\Psi}_4\neq 0$ at the horizon.

In sum, since $\Psi_2$ has spin-weight zero, it can be spanned using
spherical harmonics, in particular,
\begin{equation}\label{eq:psi2-prime}
    \widehat{\Psi}_2 \triangleq \sum_{l\geq2,m} \widehat{k}_{lm} Y_{lm}(z,\bar{z})
\end{equation}
where $\widehat{k}_{lm} = \widehat{e}_{lm}+i \widehat{b}_{lm}$. When we
consider an axisymmetric perturbation, the expression above simplifies
to
$ \widehat{\Psi}_2 \triangleq \sum_{l} \widehat{k}_{l0} P_l[(z
\bar{z}-1)(z\bar{z}+1)^{-1}]$, where $P_l$ are the Legendre
polynomials.  Notice that we have reabsorbed the constant
$\sqrt{\frac{2l+1}{4\pi}}$ in the constant $\widehat{k}_{l0}$ to
simplify notation.
Eqs.~\eqref{eq:re-im-Psi2-prime}-\eqref{eq:DPsi4-horizon} make
explicit that $\widehat{\pi}$ and $\widehat{\Psi}_3$,
$\widehat{\lambda}$ and $\widehat{\Psi}_4$, can be expanded in terms
of spin -1 and -2 weighted spherical harmonics respectively using the
properties of the $\eth$
operator:
\begin{subequations}\label{eqs:initial-data-perturbed}
\begin{align}
    \widehat{\pi}&\triangleq -\sum_{l\geq 1,m} \frac{ \sqrt{2} c\widehat{k}_{lm}}{\sqrt{l(l+1)}} {}_{-1}Y_{lm}(z,\bar{z})\\
    \widehat{\lambda}&\triangleq \frac{1-e^{- \widetilde{\kappa}_{(\ell)}v}}{ \widetilde{\kappa}_{(\ell)}}\sum_{l\geq2,m} \widehat{k}_{lm}\sqrt{\frac{(l+2)(l-1)}{l(l+1)}} {}_{-2}Y_{lm}(z,\bar{z})\\
    \widehat{\Psi}_3&\triangleq-\frac{1-e^{- \widetilde{\kappa}_{(\ell)}v}}{\sqrt{2}c \widetilde{\kappa}_{(\ell)} } \sum_{l\geq1,m} \frac{\widehat{k}_{lm}}{\sqrt{l(l+1)}} [l(l+1)-3] {}_{-1}Y_{lm}(z,\bar{z})\\
    \begin{split}
        \widehat{\Psi}_4&\triangleq  \sum_{l\geq2,m} \left(1-\frac{l^2+l+1}{l(l+1)}e^{- \widetilde{\kappa}_{(\ell)}v}+\frac{l^2+l+2}{l(l+1)} e^{-2 \widetilde{\kappa}_{(\ell)}v}\right) \times\\
        &\frac{\widehat{k}_{lm}}{4 \widetilde{\kappa}_{(\ell)}^2c^2} \sqrt{(l-1)l(l+1)(l+2)} {}_{-2}Y_{lm}  \; . 
    \end{split}
 \end{align}
\end{subequations} 
Eqs.~\eqref{eqs:initial-data-perturbed} satisfy the initial data equations for a Weakly Isolated Horizon~\eqref{eq:edth-pi},~\eqref{eq:psi_3-perturbed},~\eqref{eq:D-conditions-horizon},~\eqref{eq:evolution-mu-lambda},~\eqref{eq:evolution-psi3}, and~\eqref{eq:DPsi4-horizon} under the assumption that a perturbed spin coefficient or Weyl scalar $\widehat{X}$ admits a decomposition $\widehat{X}=\sum_{lm} T_{lm}^{\widehat{X}}(v) \mathcal{Y}_{lm}^{\widehat{X}}(z,\bar{z})$.  The data for an isolated horizon can be easily obtained from these equations by replacing $e^{- \widetilde{\kappa}_{(\ell)}v}\to0$ and will be used extensively in the next sections~\footnote{The only difference between the WIH and IH is that for the IH we restrict the extrinsic curvature ( $\lambda$ and $\mu$)  to be time-independent. However, by Eq.~\eqref{eq:psi_3-perturbed}, $\Psi_3$ is also time-independent. }.

Finally, recall that the metric perturbation $\widehat{P}$  is also sourced by the perturbation to the real part of $\Psi_2$. Explicitly, 
\begin{equation}
    \widehat{P} \triangleq -2 R^2 \sum_{l\geq 2,m} \frac{\widehat{e}_{lm}}{(l+2)(l-1)} Y_{lm}\,.
\end{equation}
Therefore, specifying the perturbation constants $\widehat{e}_{lm}$ and $\widehat{b}_{lm}$ at the horizon determines fully the free data at the horizon given the gauge choices we implemented (Eq.~\eqref{eq:cuts}, $\widehat{\epsilon} =0$, $\widehat{\pi}\triangleq \widehat{\alpha} + \widehat{\bar{\beta}}$, and $\widehat{a} =\widehat{a}-\widehat{\bar{\beta}}$). The constants $\widehat{e}_{lm}$ and $\widehat{b}_{lm}$ are directly related to the electric and magnetic moments of the tidal field in the standard metric formulation in \cite{Poisson:2021yau,Binnington:2009bb,LeTiec:2020bos,LeTiec:2020spy,Charalambous:2021mea,Pani:2015hfa,Pani:2018inf}, as we shall see later.

\section{The integration of the radial equations and perturbing the near horizon geometry}
\label{sec:radial-perturb}

Having obtained a perturbed NEH in the previous section, we are now
ready to use it to perturb the near horizon geometry.  For this
purpose, we will need to integrate the radial equations~\eqref{eqs:radial}.

We propagate the tetrad basis and the coordinate system defined at the
horizon by parallel propagating all fields along the inward-pointing
future-directed null vector $n^a$~\cite{Krishnan:2012bt}. In our
construction, the directional derivative $\Delta = n^a\nabla_a$ is not
affected by the tidal perturbation: $\Delta=\Delta^\circ$.
To obtain the metric in any spacetime point, we need to integrate the coupled system of radial differential equations~\eqref{eqs:radial} and~\eqref{eqs:radial-bianchi}.

We begin with the expansion $\mu$.  Our gauge conditions on the
foliation of the horizon ensure that $\mu$ is unaffected by the
perturbation.  Moreover, in the equation for $\Delta \mu$
(i.e. Eq.~\eqref{eqs:radial:2}),  note that $|\lambda|^2$ is second order in the smallness parameter and can be ignored.  It is thus evident that the radial equation for $\mu$ is unchanged along with its boundary value at the horizon.  Thus, $\mu$ is unchanged even away from the horizon and we get: 
\begin{equation}
   \widehat{ \mu} = \frac{c^2 \widehat{\mu}_{\Delta}}{(r+c)^2}=0 \, ,
\end{equation} 
where the $\Delta$ subscript denotes the perturbed data at the horizon (given by the solution to Eq.~\eqref{eq:evolution-mu-lambda}). The constant
\begin{equation}
  c=2 \widetilde{\kappa}_{(\ell)}R^2
\end{equation}
naturally arises when building an isolated horizon from a
non-rotating, round cross-section~\footnote{ Plugging the values of
  the surface gravity and horizon radius for a Schwarzschild black
  hole we would obtain $ \tilde{\kappa}_{(\ell)}=\frac{1}{2 c}$, and $c=R$.}, and as before,  $r+c$ is the ``standard'' Schwarzschild radial coordinate (see Sec.~\ref{subsec:schwarzschild}).  
  
To discuss the radial dependence of $\widehat{\Psi}_4$, notice first that the time-derivative operator $D$~\eqref{eq:3} only evolves the fields in the temporal direction along the horizon. That is because $U\triangleq X^A\triangleq0$. Away from the horizon, $U\neq 0$ in general, and the operator $D$ contains a radial derivative as well. Therefore, the radial dependence of $\widehat{\Psi}_4$ is specified
implicitly through the evolution equation for $\widehat{\Psi}_4$~\eqref{eqs:evolution-bianchi:4} away from the horizon. 
This implies that the temporal, radial, and angular dependence of
$\widehat{\Psi}_4$ are coupled non-trivially. Hence, we need to first
solve the differential equation for $\widehat{\Psi}_4$ before solving
Eqs.~\eqref{eqs:radial-bianchi} and~\eqref{eqs:radial}. Taking the
$\bar{\delta}$ and $\Delta$ directional derivatives of
Eqs.~\eqref{eqs:radial-bianchi:3} and ~\eqref{eqs:evolution-bianchi:4}
respectively, and combining them to eliminate the
$\Delta\bar{\delta}\widehat{\Psi}_3$ terms, we obtain the Teukolsky
equation~\cite{Teukolsky:1972my,Teukolsky:1973ha} for
$\widehat{\Psi}_4$
  \begin{equation}
     \mathcal{O}_T^\circ \widehat{\Psi}_4=0
\end{equation} with
\begin{equation}\label{eq:Teukolsky-operator}
\begin{split}
 \mathcal{O}_T^\circ&= [(\Delta D-\bar{\delta}\delta) +(4\epsilon-\rho)\Delta+5\mu D +2\bar{a}\bar{\delta}+2\bar{\delta}\bar{a}
      \\
      &  -a\delta+\Delta (4\epsilon-\rho) +2\bar{a}a+5\mu (4\epsilon -\rho) -3\Psi_2]_\circ\,.
\end{split}
\end{equation}
Using the previous ansatz for $\widehat{\Psi}_4$, i.e.
\begin{equation}
  \widehat{\Psi}_4 = T(v) X(r)Y(z,\bar{z}) \,,
\end{equation}
with the radial function $X(r=0)=1$, and the Schwarzschild values for the spin coefficients, Weyl scalars and unperturbed tetrad components appearing in the operator $\mathcal{O}_T^\circ$,
\begin{subequations} \label{eq:Schwarzschild-nonvanishing-data}
    \begin{align} 
        \mu_\circ&= -\frac{1}{c+r}\,,\quad  a_\circ  = \frac{z}{\sqrt{2}(r+c)} \,,
        \quad\epsilon_\circ = \frac{c}{4(c+r)^2}\,,\\
        \rho_\circ &= -\frac{ r}{2(c+r)^2 }\,, \quad\Psi_2^\circ =-\frac{c}{2(c+r)^3}\,, 
    \end{align} and 
\end{subequations} 
\begin{subequations}\label{eq:tetrad-unperturbed}
    \begin{align}
        l_\circ^a &= \partial_v +\frac{ r}{2(c+r)} \partial_r\,,\\
        n_\circ^a &= -\partial_r\,,\\
        m_\circ^a & = \frac{P_0}{ (c+r)} \partial_z\,,
    \end{align}
\end{subequations}
we can separate the Teukolsky equation in the following three
differential equations
\begin{subequations}\label{eq:separation-teukolsky}
    \begin{align}\label{eq:separation-teukolsky:1}
     &   \frac{\partial_v T}{T} = -\chi\,,\\ \label{eq:separation-teukolsky:2}
     &  [r(r+c) \partial_r^2 X +3(c+2r) \partial_r X -X (k-4)] \notag \\
     &       =2\chi (c+r)  [5X +(r+c) \partial_r X]\,,\\ \label{eq:separation-teukolsky:3}
     &- \frac{k}{2c^2} Y
     = (r+c)^2\bar{\eth}_\circ\eth_\circ Y\,.
    \end{align}
\end{subequations} 
We denote by  $\chi$ and $k$ the separation constants and the directional derivatives are those of the unperturbed basis vectors (given in Eq.~\eqref{eq:tetrad-unperturbed}). Notice that the last equation~\eqref{eq:separation-teukolsky:3} is independent of $r$ given that the $\eth$ operator ``has a factor of $1/(r+c)$'' (use the definition of this operator $\eth_\circ \eta=\delta_\circ \eta+s\bar{a}_\circ \eta$ together with Eqs.~\eqref{eq:Schwarzschild-nonvanishing-data} and~\eqref{eq:tetrad-unperturbed}). 

Eqs.~\eqref{eq:separation-teukolsky} need to satisfy the boundary conditions at the horizon given by Eqs.~\eqref{eq:relation-psi2-psi4-H}. Combining the temporal equations, we arrive at 
\begin{equation}
    \label{eq:equation-teukolsky-psi4-horizon}
    T(v) (\chi- \widetilde{\kappa}_{(\ell)})(\chi-2 \widetilde{\kappa}_{(\ell)}) \triangleq K\,,
\end{equation} which has two different solutions
\begin{align}
    \text{sol 1: \qquad } \chi&= \widetilde{\kappa}_{(\ell)} \text{ or } 2 \widetilde{\kappa}_{(\ell)}\,,\quad K=0\,,\\
    \text{sol 2: \qquad } \chi&=0\,,\quad \quad\quad\quad \quad \text{  } K=2 \widetilde{\kappa}_{(\ell)}^2 T(v)\,.
\end{align}
In the first solution, the angular behavior of $\widehat{\Psi}_4$ and
$\widehat{\Psi}_2$ are independent.  The perturbation
$\widehat{\Psi}_4$ is obtained by solving
Eqs.~\eqref{eq:separation-teukolsky} with $\chi= \widetilde{\kappa}_{(\ell)}$ or
$2 \widetilde{\kappa}_{(\ell)}$. In particular, the angular part of the perturbation
$\Psi_4$, $Y$ can be spanned using spin $s=-2$ spherical harmonics, so
Eq.~\eqref{eq:separation-teukolsky:3} is solved by choosing the
constant $k=(l+1)(l-2)$.  The radial equation
in~\eqref{eq:separation-teukolsky} can be solved in terms of confluent
Heun functions. Then, the general form of $\widehat{\Psi}_4$ is
\begin{equation}\label{eq:psi4-perturbed-Heun}
    \widehat{\Psi}_4 = \sum_{l\geq2,m}\sum_{n=1,2} \widehat{y}_{lm} b_n H_l^n(r)e^{-n \widetilde{\kappa}_{(\ell)}v} {}_{-2}Y_{lm}
\end{equation}
where $H_l^n = H[l(l-1)+5n -6,5n,3-n,3,n,-r/c]$ is the confluent Heun
function and $\widehat{y}_{lm}, b_n \in \mathbb{C}$ are constants.
Notice that $\widehat{\Psi}_2$ is ``time-independent'' in this case
and satisfies $\bar{\eth}^2\widehat{\Psi}_2\triangleq0$, so this
decoupling only occurs for the monopolar and dipolar modes of
Eq.~\eqref{eq:psi2-prime}.

Here we focus instead on the second solution to
Eq.~\eqref{eq:equation-teukolsky-psi4-horizon}, which corresponds to
an isolated horizon perturbed by a generic tidal perturbation that is
not monopolar or dipolar. The second solution represents a
``time-independent'' perturbation to $\Psi_4$ with
$T(v)= K/(2 \widetilde{\kappa}_{(\ell)}^2)$. The angular part of $\widehat{\Psi}_4$
is given by $Y(z,\bar{z})= \bar{\eth}^2\widehat{\Psi}_2/2 \widetilde{\kappa}_{(\ell)}^2$, and the radial differential equation in Eq~\eqref{eq:separation-teukolsky} can be solved in terms of the associated Legendre polynomials:
\begin{equation}
    X(r) = \frac{k_1}{r(r+c)} P^2_{\gamma} \left(1+\frac{2r}{c}\right) +  \frac{k_2}{r(r+c)} Q^2_{\gamma} \left(1+\frac{2r}{c}\right) \,,
\end{equation} where $P_\gamma^m$ and $Q^m_\gamma$ are the associated Legendre functions of the first and the second kind, and $\gamma=\frac{1}{2}(-1+\sqrt{9+4k})$.   $k_1$ and $k_2$ are integration constants. We set $k_1 =-\frac{2c^2}{k(k+2)} $ and $k_2=0$ so that $X(r)$ is regular and normalized at the horizon $X(r=0)=1$. Using again the properties of the $\eth$ operator, it is straightforward to show that the angular function $Y(z,\bar{z})=\bar{\eth}^2\widehat{\Psi}_2/2 \widetilde{\kappa}_{(\ell)}^2$ is a solution of the Teukolsky equation~\eqref{eq:separation-teukolsky} for $l\geq 2$ and $k=(l-1)(l+2)$. This choice for $k$ yields
\begin{align}
  \widehat{\Psi}_4 &= \sum_{l\geq2,m}  \frac{\widehat{k}_{lm}\sqrt{l(l+1)(l-1)(l+2)}}{4 \widetilde{\kappa}_{(\ell)}^2(r+c)^2} \times \nonumber \\
  &\qquad \qquad {}_2 F_1[-l,l+1,3,-r/c] {}_{-2}Y_{lm}
\end{align}
Notice that for $l=2$, $X(r) =1$, so $\widehat{\Psi}_4 = \widehat{\Psi}_4(z,\bar{z})$. For $l>2$
 , $X(r)\sim r^{l-2}$ diverges as $r\to \infty$ 
 . Given $\widehat{\Psi}_4$, we can proceed to integrate the radial differential equations~\eqref{eqs:radial} and~\eqref{eqs:radial-bianchi}
 to first order in the perturbation. We use the data at the horizon discussed in the previous section, see~\eqref{eqs:initial-data-perturbed} for an isolated horizon (with $e^{-\kappa_{(\ell)}v}\to 0$), as the boundary condition at $r=0$. In the following, we present the results for the Weyl scalars, spin coefficients, tetrad components, and the metric of a tidally perturbed isolated horizon.

 \begin{widetext}
   Introducing the notation shortcut
   \begin{equation}\label{eq:shortcut-function}
     F_{n}^l(r) := \left((l-1) {}_2F_1 [1-l,l+2,n,-\frac{r}{c}] +3 {}_2F_1 [2-l,l+2,n,-\frac{r}{c}]\right)\,,
   \end{equation}
   the perturbed Weyl scalars are
   \begin{subequations}\label{eq:tidal-perturbed-Weyl-scalars}
     \begin{align}
       \widehat{\Psi}_0 & = -\frac{1}{r+c} \int_0^r \d r^\prime (r^\prime+c) (\eth_\circ \widehat{\Psi}_1 (r^\prime) +3\widehat{\sigma}(r^\prime) \Psi_2^\circ(r^\prime)) \\ 
       \widehat{\Psi}_1 &= -\frac{r}{2\sqrt{2}  \widetilde{\kappa}_{(\ell)} (r+c)^2}\sum_{l,m}\widehat{k}_{lm}  \frac{\sqrt{l(l+1)}   }{(l+2)} {}_{1}Y_{lm}F_2^l(r)  -\frac{3c}{2(r+c)^2}\int_0^r \d r^\prime \frac{\widehat{\Omega}(r^\prime)}{(r^\prime+c)^2}
       \\
       \widehat{\Psi}_2 & = \sum_{l,m} \frac{\widehat{k}_{lm} Y_{lm}}{2 \widetilde{\kappa}_{(\ell)} (l+2)(r+c)} F_1^l(r)\\
       \widehat{\Psi}_3 & = \sum_{l,m}\frac{\widehat{k}_{lm} {}_{-1} Y_{lm}}{c  \widetilde{\kappa}_{(\ell)} \sqrt{2} \sqrt{l(l+1)}} \left({}_2F_1[l+3,2-l,1,-\frac{r}{c}] -(l+2)(l-1) {}_2F_1[l+3,2-l,2,-\frac{r}{c}]\right)\\
       \widehat{\Psi}_4& = \sum_{l\geq2,m}  \frac{\widehat{k}_{lm}\sqrt{l(l+1)(l-1)(l+2)}}{4 \widetilde{\kappa}_{(\ell)}^2(r+c)^2}  {}_2 F_1[-l,l+1,3,-r/c] {}_{-2}Y_{lm}\; .
     \end{align}
   \end{subequations}
   The perturbed spin coefficients are
   \begin{subequations}\label{eq:tidal-perturbed-spin-coefs}
     \begin{align}
       \widehat{\lambda}& = \sum_{l,m} \frac{c^2 \widehat{k}_{lm} }{ \widetilde{\kappa}_{(\ell)} (r+c)^2} \sqrt{\frac{(l-1)(l+2)}{l(l+1)}} {}_2F_1[-l-1,l,2,-\frac{r}{c}] {}_{-2}Y_{lm} \\
       \widehat{\rho} &= \sum_{l,m} \frac{\widehat{k}_{lm} r}{2 \widetilde{\kappa}_{(\ell)} (l+2) (r+c)} Y_{lm}F_2^l(r)\\
       \begin{split}
         \widehat{\pi} &= \sum_{lm} \frac{\widehat{k}_{lm} {}_{-1}Y_{lm}}{\sqrt{2}  \widetilde{\kappa}_{(\ell)} (l+2)(l(l+1))^{3/2}(r+c)} \left\{-[c(1+l^2+l^3)+(l-1)(l+1)^2 r] {}_2F_1[1-l,2+l,1,-\frac{r}{c}]\right.\\
         & \left.+[1-2l(1+l)](c+r) {}_2F_1[2-l,2+l,1,-\frac{r}{c}]\right\}\\
         \widehat{\sigma} & = -\frac{r^2}{2 (r+c)}\sum_{lm} \widehat{\bar{k}}_{lm} \sqrt{\frac{(l+2)(l-1)}{l(l+1)}} {}_{2}Y_{lm} {}_2F_1[2-l,3+l,3,-\frac{r}{c}] \\   \widehat{\epsilon}+\widehat{\bar{\epsilon}} & =2r \sum_{l,m} \widehat{e}_{lm} Y_{lm} {}_2F_1 [2-l,1+l,2,-\frac{r}{c}] \\
         \widehat{\epsilon}-\widehat{\bar{\epsilon}} & =2ir \sum_{l,m} \widehat{b}_{lm} Y_{lm} {}_2F_1 [2-l,1+l,2,-\frac{r}{c}]+\int_0^r\d r^\prime (a_0(r^\prime) \hat{\bar{\pi}}(r^\prime)-\bar{a}_0(r^\prime)\hat{\pi}(r^\prime)) \\
         \widehat{\kappa} & = \int_0^r \d r^\prime (\rho_\circ(r^\prime
         ) \widehat{\pi}(r^\prime) + \widehat{\Psi}_1(r^\prime))\,.
       \end{split}
     \end{align} 
   \end{subequations}
   Finally, the perturbed tetrad functions~\eqref{eq:tetrad-general} are
   \begin{subequations}\label{eq:tidal-perturbed-tetrad}
    \begin{align}
      \widehat{\Omega} &=\sum_{l,m} (-1)^m\frac{\widehat{\bar{k}}_{lm} \quad{}_{1}{Y}_{lm}}{\sqrt{2} \widetilde{\kappa}_{(\ell)}\sqrt{l(l+1)} (l+2)}r F_2^l(r)\\
      \widehat{\xi}^z & = \frac{\widehat{P} P_\circ}{(r+c)}\\
      \widehat{\xi}^{\bar{z}}&=\frac{P_0}{ (r+c)} \sum_{l,m}(-1)^m \frac{\widehat{\bar{k}}_{lm} \quad{}_{2}Y_{lm}}{\sqrt{l(l+1)(l+2)(l-1)}} \left(2c^2 [-1+{}_2F_1[1-l,2+l,1,-\frac{r}{c}]]+(l+2)(l-1) r^2 {}_2F_1[2-l,3+l,3,-\frac{r}{c}]\right)\\
      X^z &= P_0\int_0^r \d r^\prime\frac{\widehat{\pi}(r^\prime)}{(r^\prime+c)}\\ 
      X^{\bar{z}} &= P_0\int_0^r \d r^\prime\frac{\widehat{\bar{\pi}}(r^\prime)}{(r^\prime+c)} \\
      \widehat{U} & = 2\int_0^r \d r^\prime \int_0^{r^\prime} \d r^{\prime\prime} \re \widehat{\Psi}_2(r^{\prime\prime})= r^2\sum_{lm} \widehat{e}_{lm} Y_{lm} {}_2F_1[2-l,l+3,3,-\frac{r}{c}] \; .
    \end{align}
  \end{subequations}
\end{widetext}

The metric can be reconstructed using $g_{ab}=-2l_{(a}n_{b)}+2m_{(a}\bar{m}_{b)}$, together with the tetrad functions. To first order in the perturbation, the metric elements are then
\begin{subequations}\label{eqs:metric-reconstruction}
  \begin{align}
    g_{vv}&= -\frac{r}{r+c} -2\widehat{U}\\
    g_{rv} &=1\\
    g_{vz}&=-\frac{r+c}{P_0^2} ( (r+c) X^{\bar{z}}+P_0 \widehat{\Omega})\\
    g_{zz} & = -\frac{2(r+c)^3}{P_0^3} \widehat{\xi}^{\bar{z}}\\
    g_{z\bar{z}}& = \frac{(r+c)^2}{P_0^2} -\frac{(r+c)^3(\widehat{\bar{\xi}}^{\bar{z}}+\widehat{\xi}^{z})}{P_0^3}\,.
  \end{align}
\end{subequations}

Notice from our expressions for the Weyl scalars~\eqref{eq:tidal-perturbed-Weyl-scalars}, that the spherical harmonics we defined at the horizon can still be used at $r\gg 0$ (as long as our coordinate system remains valid). In particular, this implies that we can make the same spherical harmonic decomposition of the Weyl scalars at the horizon and away from the horizon. We have then a natural way to decompose the electric and magnetic parts of the tidal field in spherical harmonics.

\subsection{Asymptotic behavior for large $r$}
\label{subsec:asymptotics}

It is interesting to analyze the asymptotic behavior (the limit $r\to \infty$) of the quantities presented above. In particular, as we will discuss in the next section, the asymptotic behavior of $\widehat{\Psi}_2$ and $\widehat{U}$ has special relevance since it can be related to the field Love numbers. Let us start with $\widehat{\Psi}_2$. Notice that $\widehat{\Psi}_2$ can be rewritten in the compact form
\begin{equation}\label{eq:psi2-tidal}
    \widehat{\Psi}_2= \sum_{l,m} \widehat{k}_{lm} Y_{lm} {}_2F_1[2-l,l+3,1,-\frac{r}{c}]\,.
\end{equation} The proof that this expression is equivalent to the one presented in Eq.~\eqref{eq:tidal-perturbed-Weyl-scalars} can be found in Appendix~\ref{sec:Asymptotics}. However, it is interesting to highlight that this simplification occurs only because we can factor out a term $(r+c)$ from the function $F_1^l(r)$. Interestingly, despite the similar structure between $F_1^l(r)$ and $F_2^l(r)$, this simplification does not occur for $F_2^l(r)$, nor any other combination of hypergeometric functions in Eqs.~\eqref{eq:tidal-perturbed-spin-coefs}-\eqref{eq:tidal-perturbed-tetrad}. We used Eq.~\eqref{eq:psi2-tidal} to integrate the expressions for $\widehat{\epsilon}+\widehat{\bar{\epsilon}}$ and $\widehat{U}$.

Recall that when one of the two first entries of the hypergeometric function is negative or zero, the hypergeometric function has finitely many terms and it is defined for any argument
  \begin{equation}\label{eq:hypergeometric-fnct-finite}
    {}_2F_1[-m, b,c,z] = \sum_{n=0}^m (-1)^n \left(\begin{matrix}
        m\\
        n
      \end{matrix}\right)\frac{(b)_n}{(c)_n} z^n\,,
  \end{equation} with $m\in \mathrm{Z}^+$ and $(b)_n=(b+n-1)!/(b-1)!$ the Pochhammer symbol.  We can use this last property~\eqref{eq:hypergeometric-fnct-finite} to expand  $\widehat{\Psi}_2$ in a finite series in $r$
\begin{equation}\label{eq:psi2hat-asymptotic}
  \widehat{\Psi}_2= \sum_{l,m} \widehat{k}_{lm} Y_{lm} \sum_{n=0}^{l-2} \left(\begin{matrix}
      l-2\\
      n
    \end{matrix}\right) \frac{(l+3)_n}{(1)_n} \left(\frac{r}{c}\right)^n\,.
\end{equation}
From this expression we can see that the dominant asymptotic behavior
is $r^{l-2}$, while the least dominant term is constant. In other
words,
\begin{equation}
     \lim_{r\to\infty}\widehat{\Psi}_2 \sim r^{l-2},\quad r^{l-3}\,,\,.\,.\,.\quad r^0\,.
\end{equation} Similarly, expanding the hypergeometric function in $\widehat{U}$, we see that
\begin{equation}
    \lim_{r\to\infty} \widehat{U}\sim r^l\,,\quad r^{l-1}\,,\,.\,.\,.\quad r^2\,
\end{equation} the leading order terms goes like $r^l$ and the subdominant one as $r^2$.

For completeness, we also provide the asymptotic behaviors of the Weyl scalars, spin coefficients, and tetrad components presented in Eqs.~\eqref{eq:tidal-perturbed-spin-coefs}-\eqref{eq:tidal-perturbed-tetrad}. Using Eqs.~\eqref{eq:tidal-perturbed-Weyl-scalars}, it is straightforward to check that all of the Weyl scalars have the same asymptotic behavior, given by 
\begin{equation}
    \widehat{\Psi}_0\,,\quad \widehat{\Psi}_1\,,\quad\widehat{\Psi}_2\,,\quad\widehat{\Psi}_3\,,\quad\widehat{\Psi}_4\sim r^{l-2}\,.
\end{equation}

Finally, the spin coefficients and tetrad functions in Eqs.~\eqref{eq:tidal-perturbed-spin-coefs} and~\eqref{eq:tidal-perturbed-tetrad} have the asymptotic behavior
  \begin{equation}
      \widehat{\lambda}\,, \quad\widehat{\pi}\,,\quad\widehat{\sigma}\,,\quad\widehat{\epsilon}\,,\quad \kappa\sim r^{l-1}\,;\quad \widehat{\rho} \sim r^{l-2}\,,
  \end{equation} and
  \begin{equation}
      \widehat{\Omega}\,,\widehat{\xi}^{\bar{z}}\,, X^z\,, X^{\bar{z}} \sim r^{l-1}\,,\quad \widehat{U}\sim r^l\,,\quad \widehat{\xi}^z\sim r^{-1}\,.
  \end{equation}

In Sec.~\ref{sec:tidally-perturbed-sch}, we specialize the solution to a quadrupolar perturbation of $\Psi_2$. We will further show that the isolated horizon coincides with the known solution of a tidally perturbed Schwarzschild black hole in the literature derived using the metric formulation~\cite{Poisson:2021yau,Binnington:2009bb,LeTiec:2020bos,LeTiec:2020spy,Charalambous:2021mea,Pani:2015hfa,Pani:2018inf}.

\section{Tidally perturbed black hole spacetime}
\label{sec:tidal-bh}

With the general expressions obtained in the previous sections at
hand, we are now ready to construct the metric of a tidally perturbed
black hole. In this section, we first consider a non-spinning tidally perturbed
black hole followed by the slowly spinning case.

\subsection{Tidally perturbed non-spinning black hole}
\label{sec:tidally-perturbed-sch}

We specialize the general equations \eqref{eq:tidal-perturbed-Weyl-scalars}-\eqref{eqs:metric-reconstruction} in the previous section to the case of a quadrupolar $l=2$ perturbation to the Weyl scalar $\widehat{\Psi}_2$. Since we consider perturbations of the Schwarzschild spacetime we can set $m=0$ without loss of generality.  Integrating the expressions for $\widehat{\Psi}_0$ and $\widehat{\Psi}_1$ and simplifying the above expressions, we obtain for the Weyl scalars
\begin{subequations}
  \begin{align}\label{eqs:psi-non-spinning}
   \Psi_0&= \sqrt{\frac{3}{2}} \frac{r^2}{(r+c)^2} \widehat{k}_{20}\, {}_2Y_{20}\\
   \begin{split}
       \Psi_1 &= -\frac{\sqrt{3}r \, {}_1Y_{20}}{4(r+c)^3} [(2r^2+7rc+4c^2)\widehat{e}_{20} +\\
        &i (2r^2+5rc+4c^2)\widehat{b}_{20}]
   \end{split}\\
   \Psi_2 & = -\frac{c}{2 (c+r)^3} + \widehat{k}_{20} Y_{20}\\
   \Psi_3 & = -\sqrt{3} \widehat{k}_{20}\, {}_{-1}Y_{20}\\
   \Psi_4 & = 2\sqrt{6} \widehat{k}_{20}\, {}_{-2}Y_{20}\,,
  \end{align}
\end{subequations} where the spin-weighted spherical harmonics are given explicitly in Appendix~\ref{sec:spin-weighted-SH}, and the constant $c=2 \widetilde{\kappa}_{(\ell)} R^2=2M$. Similarly, the same procedure for the spin coefficients yields
\begin{subequations}
    \begin{align}
        \rho =& -\frac{r}{2(r+c)^2} + \widehat{k}_{20}\frac{r(r+2c)}{2(r+c)} Y_{20}\\
        \pi  =&- \widehat{k}_{20} \frac{3r^2+6cr+2c^2}{2\sqrt{3}} \, {}_{-1} Y_{20}\\
        \kappa  =&-\frac{r^2 \, {}_1Y_{20}}{4\sqrt{3} (r+c)^2} [(7c+9r)\widehat{e}_{20}+i(5c+3r)\widehat{k}_{20}] \\
   \sigma  =&-\frac{r^2 \widehat{\bar{k}}_{20}}{\sqrt{6}(r+c)}\, {}_2Y_{20}\\
   \begin{split}
        \epsilon  =&\frac{c}{4(r+c)^2} +r\left(\widehat{e}_{20}-i\frac{r}{2(r+c)} \widehat{b}_{20}\right)Y_{20}\\
        &-i\frac{\widehat{b}_{20}r (3r+2c)}{4(r+c)} \sqrt{\frac{5}{\pi}} \frac{3z\bar{z}-1}{(1+z\bar{z})2}
   \end{split}\\
   \mu  =& -\frac{1}{r+c}\\
   \lambda  = &2 \widehat{k}_{20}\sqrt{\frac{2}{3}} (r+c) \,{}_{-2}Y_{20}
    \end{align}
 Finally, using the tetrad functions~\eqref{eq:tidal-perturbed-tetrad} for the case $l=2$, together with Eq.~\eqref{eqs:metric-reconstruction}, we can reconstruct the metric of a non-rotating isolated horizon with a quadrupolar perturbation
\end{subequations}
   \begin{subequations}\label{eq:metric-quadrupolar}
    \begin{align}
        g_{vv} &= - \left(\frac{r}{r+c}+2 r^2 \widehat{e}_{20} Y_{20}\right)\\
   g_{vr} & = 1\\
   g_{vz} & = -\frac{2}{3}  \frac{r(r+c)^2}{P_\circ} \widehat{\bar{k}}_{20} \, {}_1Y_{20}\\
   g_{zz} & =-2\sqrt{\frac{2}{3}} \widehat{\bar{k}}_{20} \frac{r(r+c)^2 (r+2c)}{P_\circ^2}\, {}_2Y_{20}\\
   g_{z\bar{z}} &=  \frac{ 
   (r+c)^2}{P_\circ^2}-\sqrt{\frac{5}{\pi}}\frac{2
     c^2 (r+c)^2 \left( 1-2 
    z \bar{z}\right)}{\left(1+z
   \bar{z}\right)^4}\,.
    \end{align}
\end{subequations}
The complex coordinates $\{z,\bar{z}\}$ are related to the usual spherical coordinates of the background Schwarzschild spacetime $\{\theta,\phi\}$ though 
\begin{equation}\label{eq:z-to-zeta-phi-Poisson}
  z=\sqrt{\frac{1+\zeta}{1-\zeta}} e^{i\phi} \left(1+\frac{R^2}{4}\sqrt{\frac{5}{\pi}}\left(\widehat{e}_{20} (1-\zeta)-i \widehat{b}_{20}\zeta\right) \right)\,,
\end{equation}
where $\zeta=\cos\theta$ and $R$ is the radius of the unperturbed horizon. Notice that in the absence of the perturbation $\widehat{e}_{20}=\widehat{b}_{20}=0$, the coordinate transformation coincides with Eq.~\eqref{eq:z-to-zeta-and-phi}. Further, this transformation is not unique as discussed in Sec.~\ref{sec:perturbing-cross-section}. Transforming the metric~\eqref{eq:metric-quadrupolar} to the coordinates $\{v,\mathfrak{r},\theta,\phi\}$ (with $\mathfrak{r}=r+c$) yields the metric of a tidally perturbed Schwarzschild black hole presented in Eqs.~(1.5)-(1.7) of Ref.~\cite{Binnington:2009bb} upon the identification
\begin{equation}\label{eq:tidal-moments-Poisson}
  \widehat{e}_{20} = \frac{1}{2} \mathscr{E}^{(2)}_0\,,\quad \widehat{b}_{20} = \frac{1}{2} \mathscr{B}^{(2)}_0\,. 
\end{equation}
Here $\mathscr{E}_0^{(2)}$ and $\mathscr{B}_0^{(2)}$ are the components of the electric and magnetic tidal moments spanned in a basis of spherical harmonics, i.e., 
\begin{equation}
  \mathscr{E}_L x^L= r^l\sum_m \mathscr{E}_m^{(l)} Y_{lm} = r^l \mathscr{E}^{(l)}\,,
\end{equation}
where $ \mathscr{E}_L$ is an $l\times l$ symmetric trace-free tensor with multi-index $L=a_1\,, ... a_l$ defined in a quasi-cartesian system with $x^a=\{\cos\phi\sin\theta,\sin\phi\sin\theta,\cos\theta\}$, and  $\mathscr{E}^{(l)}$ is the electric tidal scalar potential. Recall that for us the constant $\widehat{e}_{20}$ and $\widehat{b}_{20}$ have a transparent geometric meaning: they encode the magnitude of the real and imaginary parts of the quadrupolar perturbation to the Weyl scalar $\widehat{\Psi}_2$, which we have connected in Sec.~\ref{sec:perturbing-cross-section} to the deformation of the cross-section, and how it is embedded in the isolated horizon structure (see Sec.~\ref{sec:embedding-perturbed-cross-section}). 

Notice that the coordinate transformation~\eqref{eq:z-to-zeta-phi-Poisson} and the identification~\eqref{eq:tidal-moments-Poisson} only take the above forms when the perturbation is purely quadrupolar. For a general perturbation, our constants $\widehat{e}_{lm}$ and $\widehat{b}_{lm}$ are related to the components of the tidal electric and magnetic fields defined in~\cite{Poisson:2005pi, Binnington:2009bb} through
\begin{align}
    \widehat{e}_{lm} &=\frac{(l-2)!(l+2)!}{2(2l)!} c^{l-2} \mathscr{E}_m^{(l)}\,,\\
    \widehat{b}_{lm} &=\frac{(l+1)(l-2)!(l+2)!}{3!(2l)!} c^{l-2} \mathscr{B}_m^{(l)}\,.
\end{align} The coordinate transformation~\eqref{eq:z-to-zeta-phi-Poisson} can be easily worked out for each multipole using the above identification. Notice that similarly to~\cite{Binnington:2009bb}, we could also define our ``electric'' tidal potential $\mathrm{E}_l$ through 
\begin{equation}
     \mathrm{E}_l= \sum_m \widehat{e}_{lm} Y_{lm} = \frac{(l-2)!(l+2)!}{2(2l)!} c^{l-2}\mathscr{E}^{(l)}\,,
\end{equation} which is proportional to $\mathscr{E}^{(l)}$. However, contrary to~\cite{Binnington:2009bb}, we can also define a ``magnetic'' scalar tidal potential $\mathrm{B}_l$ through
\begin{equation}
    \mathrm{B}_l =\sum_m \widehat{b}_{lm} Y_{lm}\,,
\end{equation} so that our perturbation to the Weyl scalar $\Psi_2$ reads
\begin{equation}
    \widehat{\Psi}_2 = \sum_l (\mathrm{E}_{lm}+ i\mathrm{B}_{lm}) {}_2F_1[2-l,l+3,1,-\frac{r}{c}]\,.
\end{equation}

\subsection{Slowly rotating tidally perturbed black hole}
\label{sec:Sch+dipole}

Next, we consider perturbations of the Schwarzschild horizon such that $\Psi_4= 0$.  As discussed above, this choice of initial data at the horizon only allows for a monopolar or a dipolar perturbation of the Weyl scalar $\Psi_2$, ($\widehat{k}_{lm}\triangleq 0$ for $l\geq 2$ 
in Eq.~\eqref{eq:psi2-prime}).  Recall that the real and imaginary parts of $\Psi_2$ are related to the mass and angular momentum multipole moments respectively. Therefore, since we consider isolated horizons, the only physically relevant perturbed horizon left corresponds to
\begin{equation}\label{eq:condition-perturbation-spin}
  \widehat{b}_{00}=\widehat{e}_{00}=0\,,\quad \widehat{e}_{1m}=0\,,\quad \widehat{k}_{lm}=0 \text{ for }l\geq 2\,.
\end{equation}
Since we describe spacetimes without infalling flux of matter or
radiation into the black hole, its mass is not modified. This
implies $\widehat{e}_{00} =0$ and an angular momentum monopole is not
physical, so $\widehat{b}_{00}=0$. Further, the mass dipole moment is
related to the rest frame of the black hole. Setting
$\widehat{e}_{1m}\neq0$ `kicks' the hole out of the rest frame, but it
does not modify its geometry.  In other words, the only physically
relevant perturbation is to the angular momentum dipole
$ \widehat{b}_{1m}\neq0$, which yields
\begin{equation}
    \widehat{\Psi}_2\triangleq i \sum_m\widehat{b}_{1m} Y_{1m}\,.
\end{equation}
Given the symmetries of the Schwarzschild spacetime, without loss of generality, we can choose a gauge in which the perturbation is only given in terms of the $m=0$ spherical harmonic
\begin{equation}
    \widehat{\Psi}_2 \triangleq i \widehat{b}_{10}\sqrt{\frac{3}{4\pi}} \frac{z\bar{z}-1}{z\bar{z}+1}\,.
\end{equation}
This perturbation, together with the isolated horizon assumption, gives rise to the slowly rotating limit of the Kerr isolated horizon, as we will show. Notice that this initial data would not be suitable to describe the Kerr horizon with arbitrary spin, since the rotation of the hole deforms the cross-section's geometry of the Kerr black hole,  which we have not accounted for in  $\re[\widehat{\Psi}_2] $. However, at linear order in these perturbations, this is a consistent solution as the horizon's geometry is only deformed at the second order.

 As we discussed in Sec.~\ref{sec:ihperturb}, we take the Schwarzschild isolated horizon data as our unperturbed spacetime, so the background Weyl scalars and spin coefficients in Eq.~\eqref{eq:Schwarzschild-vanishing-data} vanish. 
The nontrivial spin coefficients and Weyl scalars for the basis Schwarzschild spacetime are given in Eq.~\eqref{eq:Schwarzschild-nonvanishing-data}, and the unperturbed tetrad in Eq.~\eqref{eq:tetrad-unperturbed}.

The perturbations to the initial data are given by Eqs.~\eqref{eq:psi2-prime} and~\eqref{eqs:initial-data-perturbed}, together with the condition in the expansion constants discussed above~\eqref{eq:condition-perturbation-spin}. Using Eqs.~\eqref{eqs:initial-data-perturbed}, together with Eq.~\eqref{eq:psi_3-perturbed}, we see that $\widehat{\lambda}\triangleq0$, $\widehat{P}\triangleq0$, $\widehat{\Psi}_3$ needs to be time-independent, and the most general form for $\widehat{\Psi}_4$ is given by Eq.~\eqref{eq:psi4-perturbed-Heun} with $b_1=0$, i.e.,
\begin{equation}
    \widehat{\Psi}_4 \triangleq \sum_{l\geq 2,m} \widehat{y}_{lm} H_l^2(r) e^{-2 \widetilde{\kappa}_{(\ell)}v} {}_{-2}Y_{lm}\,.
\end{equation} This initial data would yield a perturbative version of the Robinson Trautman spacetime discussed in Sec.~\ref{sec:robinson-trautman} with a slow rotating horizon in the gauge discussed in Sec.~\ref{sec:ihperturb}. However, here we will restrict ourselves to the simplest nontrivial case, i.e., since $\widehat{\Psi}_4$ is independent of the perturbation to $\Psi_2$, we can simply set it to zero $\widehat{\Psi}_4=0$. 

Recall that $\re[\widehat{\Psi}_2]\triangleq 0 $, so the horizon is foliated by good cuts $\textrm{div}\widetilde{\omega}\triangleq 0$, and the connection on the two-sphere is not modified  $\widehat{a} \triangleq0$.  
Similarly, $\widehat{\mu}\triangleq \widehat{\lambda} \triangleq 0$.  Explicitly, the non-trivial perturbed spin coefficients and Weyl scalars at the horizon are
    \begin{equation}
        \widehat{\pi} \triangleq -i c\widehat{b}_{10} {}_{-1} Y_{10}\,,\quad\widehat{\Psi}_3 \triangleq i \widehat{b}_{10} {}_{-1}Y_{10}\,.
    \end{equation} 

    Integrating Eqs.~\eqref{eqs:radial}, we obtained $\widehat{\tau} =\widehat{\sigma}=\widehat{\gamma}=\widehat{\nu}=\widehat{\lambda} = 0$ and $\widehat{\Psi}_0=\widehat{\Psi}_4=0$. The tetrad is modified with the functions $\widehat{U}=\widehat{\xi}^A=0$ and 
    \begin{subequations}\label{eq:a-perturbed-tetrad}
        \begin{align}
            \widehat{X}^A &=i\frac{c\, r\, \widehat{b}_{10} (4r^4 +9c r +6 c^2)}{4\sqrt{3\pi}(c+r)^3 }\left\{ z, - \bar{z}\right\}\\
            \widehat{\Omega} & = -\frac{i c^2 r(r+2c) \widehat{b}_{10}}{2 (r+c)^2 } {}_1Y_{10}\,.
        \end{align}
    \end{subequations}
    The remaining spin coefficients and Weyl scalars are 
    \begin{subequations}\label{eq:Weyl-scalars-small-a}
         \begin{align}
        \widehat{\rho} =& \frac{i c^2 r (r+2c)  \widehat{b}_{10}}{2(c+r)^3  }Y_{10}\\
        \widehat{\alpha} =& -\frac{i c^4 \widehat{b}_{10}}{2(c+r)^3 } {}_{-1} Y_{10}\\
        \widehat{\pi} =& -\frac{i c^2 (r^2+2c r +2 c^2)  \widehat{b}_{10}}{2(c+r)^3 } {}_{-1}Y_{10}\\
        \widehat{\beta} =& -\frac{i c^2  \widehat{b}_{10}}{2 (c+r) } {}_1Y_{10}\\
        \widehat{\kappa} =& -\frac{i c^2 r^2  \widehat{b}_{10}}{4(c+r)^3 } {}_1Y_{10}\\
        \begin{split}
             \widehat{\epsilon} =&-\frac{i c r \widehat{b}_{10}}{6(r+c)^3}\left(r(2r+3c) Y_{10} \right.\\
             &\left.+\sqrt{\frac{3}{4\pi}} \frac{4r^2+9cr+6c^2}{1+z\bar{z}}\right)
        \end{split}\\
        \widehat{\Psi}_1  =& -\frac{ic^3 r(3r+4c)  \widehat{b}_{10}}{4 (c+r)^5 } {}_{1}Y_{10}\\
        \widehat{\Psi}_3  =& \frac{i c^4  \widehat{b}_{10}}{(c+r)^4 } {}_{-1}Y_{10}\\
        \widehat{\Psi}_2  =&-\frac{c}{2(c+r)^3} +\frac{ic^4  \widehat{b}_{10}}{(c+r)^4 }Y_{10}
    \end{align}
    \end{subequations}
    which yields the metric     \begin{equation}\label{eq:ds2-Kerr-small-a}
    \d s^2 = \d s^2_{\text{Sch}} + \widehat{b}_{10}\frac{4 i cr (r^2+3c r+3c^2)  }{\sqrt{3\pi}(c+r) (1+z\bar{z})^2} \d v (\bar{z} \d z-z \d \bar{z})  \; .
\end{equation} where $\d s_{\text{Sch}}^2$ is the Schwarzschild line element~\eqref{eq:ds2-Schwarschild} in $\{v,r,z,\bar{z}\}$ coordinates. 

We could show that this line element corresponds to the
slow rotating limit of the Kerr metric by direct comparison of the
line element with the small $a$ limit of Eqs.~(41),~(52), and~(55) 
in~\cite{Flandera:2016qwg}. However,  the slicing we choose~\eqref{eq:cuts} is different from the one used in Ref.~\cite{Flandera:2016qwg}, which can be computed using their Eqs.~(57) and (62a), and our Eq.~\eqref{eq:divergence-omega}. Further, the  angular coordinates used in Ref.~\cite{Flandera:2016qwg} are different from both our complex $\{z,\bar{z}\}$ and real $\{\zeta,\phi\}$ coordinates, even in the small $a$ limit. Therefore, to directly compare Eq.~\eqref{eq:ds2-Kerr-small-a} we should simultaneously modify the slicing of our horizon and our coordinates, which is quite cumbersome. 
 Instead, we show that the line
element~\eqref{eq:ds2-Kerr-small-a} corresponds to the slow rotating
limit of the Kerr black hole by analyzing its mass monopole and
angular momentum dipole, and by showing that this spacetime is indeed
type D. Using the expressions~\eqref{eq:J-ih}
and~\eqref{eq:mass-ih}, we see that the horizon has mass and spin given by
    \begin{equation}\label{eq:mass-spin-ih}
        M = \frac{R}{2}+\mathrm{O}(\widehat{b}_{10}^2)\,,\quad J = -\frac{\widehat{b}_{10} R^2}{2\sqrt{3\pi}}+\mathrm{O}(\widehat{b}_{10}^2)\,,
    \end{equation} where $\widehat{b}_{10}\sim a$ is small and the mass of the black hole is only modified to the second order in the perturbation.  All higher mass and angular momentum multiples, which do not vanish for the Kerr solution, are at least second order in the spin, so  we take them to vanish in the slowly rotating limit. 
Further, we can show that this horizon is type D by using the invariant 
\begin{equation}
  \mathcal{I} := |(\bar{\delta}+\alpha-\bar{\beta})\bar{\delta} (\Psi_2)^{-1/3}| 
\end{equation}
defined in~\cite{Dobkowski-Rylko:2018usb}. The invariant $\mathcal{I}$
measures the deviation of a generic isolated horizon from a horizon of the Kerr
family. In other words, when $\mathcal{I}=0$, the horizon is type D,
and therefore belongs to the Kerr family.
Using Eqs.~\eqref{eq:Weyl-scalars-small-a}, it can be easily shown that
\begin{equation}
    \mathcal{I} \triangleq \mathrm{O}[\widehat{b}_{10}]^2\,.
\end{equation}
Therefore, this horizon belongs to the Kerr family and has mass and spin given by Eq.~\eqref{eq:mass-spin-ih}. We can show that this is indeed the slow rotating limit of a Kerr black hole by comparing the mass and spin multipole moments with those of the Kerr black hole evaluated at the horizon.  In~\cite{Gupta:2018znn}, we see that for Kerr $I_l=0$ ($L_l=0$) for $l$ odd (even) and $L_l\,,  I_l\propto a^l$
. Hence, our solution coincides with the slow rotating limit of the Kerr horizon up to linear order in the spin.

Finally, notice that although our gauge choices for the slowly rotating limit of the Kerr black hole do not allow a straightforward comparison with~\cite{Flandera:2016qwg}, they are consistent with the gauge choices we used to describe the tidally perturbed Schwarzschild black hole. Consequently,  the tidally perturbed slowly rotating limit of the Kerr black hole follows by combining the perturbations to the spin coefficients, Weyl scalars, and metric components in Eqs.~\eqref{eq:tidal-perturbed-Weyl-scalars}-\eqref{eq:tidal-perturbed-tetrad} with those of the slowly rotating Kerr horizon Eqs.~\eqref{eq:a-perturbed-tetrad}-\eqref{eq:Weyl-scalars-small-a}. Further, although cumbersome, the 
 solution for the tidally perturbed Kerr black hole with arbitrary (subextremal) spin could in principle be obtained in a straightforward manner using our formalism. We would need to combine the equations for the tidally perturbed Schwarzschild horizon with the spin coefficients, Weyl scalars, and tetrad components resulting from higher $a$ terms in the slowly rotating spin expansion. These contributions can be obtained by expanding the multipole moments of the Kerr black hole at the horizon, computed in~\cite{Gupta:2018znn}, in a Taylor series around $a=0$. The perturbations to the real and imaginary parts of $\Psi_2$ will be of the form~\eqref{eq:psi2-prime}, so the perturbations to the initial data would follow trivially from Sec.~\eqref{sec:ihperturb}.   By resuming the infinite terms in the series, we would obtain the tidally perturbed Kerr solution. This application of our formalism will be discussed in detail elsewhere.

\section{Field vs surficial Love numbers}
\label{sec:lovenumbers}

\subsection{The vanishing of the tidal Love numbers}
\label{subsec:tidal-love-vanishing}

The notion of tidal Love numbers of stars and compact objects plays an
important role in astrophysics and it is particularly important in
gravitational wave astronomy.  The basic idea is straightforward:
under the influence of an external field, say due to a binary
companion, the shape of a star and its gravitational field can both be
distorted, and at leading order, the distortion is linearly
proportional to the strength of the external field. The value of this
proportionality constant, call it $\Lambda$, can be measured. Detailed
studies in the context of self-gravitating objects in Newtonian theory
date back to 1933 \cite{1933MNRAS..93..462C,10.1093/mnras/115.1.101}.
More recently, the measurement of the tidal Love numbers allows us to
constrain the equation of state of neutron stars from the observations
of binary neutron star mergers (see
e.g. \cite{LIGOScientific:2018hze,De:2018uhw,Capano:2019eae}); see
\cite{Lattimer:2015nhk} for a review.  Love number measurements might
also allow us to distinguish black holes from neutron stars based
purely on gravitational wave observations (see
e.g. \cite{Brown:2021seh}).  This relies on the claim that black holes
have vanishing Love numbers \cite{Binnington:2009bb,Damour:2009vw}.

Following \cite{Flanagan:2007ix,Hinderer:2007mb} (see also
\cite{Damour:1991yw,Damour:2009vw}), let $\Phi_{ext}$ be the external
gravitational potential acting on a star.  This leads to an external
quadrupolar tidal field
\begin{equation}
  \mathcal{E}_{ij} = \frac{\partial^2 \Phi_{ext}}{\partial x_i\partial x_j}\,,
\end{equation}
where the $x_i$ are the so-called asymptotically mass-centered
coordinates \cite{Thorne:1980ru,Thorne:1984mz}. In these coordinates, the
time-time component of the metric is given by
\begin{eqnarray}
  \label{eq:gtt-expansion}
  -\frac{(1+g_{00})}{2} &=& -\frac{M}{r} - \frac{3Q_{ij}}{2r^3}\left(n_in_j-\frac{1}{3}\delta_{ij} \right) + \mathcal{O}(r^{-4}) \nonumber \\&+& \frac{1}{2}\mathcal{E}_{ij}x^ix^j + \mathcal{O}(r^3)\,,
\end{eqnarray}
where $n^i := x^i/r$, $r=\sqrt{\sum_i (x_i)^2}$.  In this coordinate
system, the mass dipole moment vanishes by construction.  At leading
order, it can be shown that the quadrupole moment $Q_{ij}$ is related
linearly to the external tidal field:
\begin{equation}
  \label{eq:lambda-def}
  Q_{ij} = -\lambda\mathcal{E}_{ij}\,.
\end{equation}
The Love number $k_2$ is the dimensionless constant constructed from
$\lambda$ and the star's radius $R$ as
\begin{equation}
  k_2 = \frac{3}{2}\lambda R^5\,.
\end{equation}
The Love number turns out to depend on the equation of state, and it
vanishes for a black hole.  The Love number is typically
incorporated heuristically in the gravitational wave signal as a
contribution to the $5^{th}$ Post-Newtonian energy and flux
\cite{Hinderer:2009ca}; in the PN expansion, it appears as a
contribution to the $x^5$ term with $x=(M\omega)^{2/3}$ being the
dimensionless Post-Newtonian parameter, $\omega$ and $M$ are
respectively the angular velocity and total mass of the binary.
Alternatively, tidal effects can also be included as part of the
Effective-One-Body formalism \cite{Damour:2009wj}, or as part of
modeling based on numerical relativity simulations
\cite{Dietrich:2018uni}.  Implicit in the expansion of
Eq.~\eqref{eq:gtt-expansion} is the existence of a buffer region
typically used in the process of carrying out matched asymptotic
expansions \cite{Thorne:1980ru,Thorne:1984mz}.  Inside this buffer
region, the gravitational field is dominated by the black hole, while
outside this region the external universe dominates.

It is evident that the above discussion is based on the properties of
the gravitational field of the compact object.  The Love number thus
defined may be referred to as the ``field'' or ``gravitational'' Love
number.  Whenever we refer to ``Love number'' without any other
qualifications, we shall always refer to these field Love numbers.
However, corresponding to the distortion of the gravitational field,
the surface of the star is distorted by tidal effects as well.  Just
like the above Love number, one can introduce the ``surficial Love
numbers'' by employing source multipole moments.  For a neutron star,
these would be based on the distribution of matter fields, or for a
black hole as surface integrals like we have used in this paper.
Within general relativity, the surficial Love numbers generally differ
from the field Love numbers; see also
\cite{Landry:2014jka,Damour:2009va,Gurlebeck:2015xpa}.

Our calculations here can be used to evaluate both the field and
surficial Love numbers for black holes as we now discuss.  The main
ingredient will be the asymptotic behavior of the Weyl tensor at large
$r$.  The discussion presented below is not formulated with a
sufficient level of rigor in terms of the asymptotic conditions; it
should rather be viewed in the same spirit as \cite{Thorne:1984mz}, i.e. as
being useful for astrophysical applications. The issue is that the
tidally perturbed spacetime is not asymptotically flat, and the
curvature components do not vanish asymptotically. Neither is the
spacetime asymptotically de Sitter or some other universal class of
known solutions.  This is natural because our solutions are local and
the typical application we have in mind is, say, a binary black hole
system.  The ``asymptotic'' region with large $r$ thus includes the
region between the two black holes.  At present we therefore do not
have, say, well developed notions of asymptotic symmetries, conserved
quantities or fluxes such as we do at null and spatial infinity for asymptotically flat spacetimes.

We have already described the perturbation of the geometry of
$\Delta$, which can be encoded as perturbations of the geometric
multipole moments $(I_\ell,L_\ell)$ defined in
Eq.~\eqref{eq:geometric-multipoles}.  These can be taken to be source
multipole moments for our purposes (after rescaling them suitably to
get the right dimensions and normalizations).  The important point
here is that these moments are obtained by a spherical harmonic
decomposition of the Weyl tensor component $\Psi_2$.  In fact, it is
$\Psi_2$ that also appears in the definition of the field multipole
moments and Love numbers.  This is seen from
Eq.~\eqref{eqs:metric-reconstruction} which expresses $g_{vv}$ in
terms of $\widehat{U}$.  After accounting for the fact that our radial
coordinate starts with $r=0$ at the horizon and is thus shifted with
respect the area coordinate in the Schwarzschild solution, it is
evident that the potential $\widehat{U}$ is the analog of the quantity
$-(1+g_{00})/2$ appearing above (in our case $-\widehat{U}\sim (1+g_{vv})/2$ with $g_{vv}$ given in Eq.~\eqref{eqs:metric-reconstruction}).  Since $\widehat{U}$ is a potential
for $\widehat{\Psi}_2$, it becomes clear that we can also discuss the
Love numbers and field multipole moments in terms of $\Psi_2$.
Moreover, since our construction of the near horizon metric is based
on a Bondi-like coordinate system, it explicitly connects the horizon
with the asymptotic region and it provides thereby an unambiguous link
between the source and field multipole moments and Love numbers.
Concretely, our construction connects spherical harmonics at the
horizon and in the asymptotic region, and thus also provides the link between
the field and surficial Love numbers.  The value of the perturbation
$\widehat{\Psi}_2$ at the horizon gives us the perturbations of the
source (i.e. the surficial) multipole moments in terms of
$\widehat{k}_{20}$, which is related to the external perturbation.  In
the absence of these perturbations, the asymptotic form of the Weyl
tensor can be written schematically as
\begin{equation}
  \Psi_2 \sim \frac{\textrm{Mass monopole}}{r^3} + \frac{\textrm{Spin dipole}}{r^4} + \ldots\,.
\end{equation}
The additional terms will be higher powers of $1/r$ and also, for a
Kerr black hole, higher powers of the spin.  Therefore, since we
restrict ourselves to slowly spinning black holes, we shall only
consider the first two terms for our purposes. When we perturb
$\Psi_2\rightarrow \Psi_2 + \widehat{\Psi}_2$, asymptotically the
perturbations develop additional terms.  In the case of a non-spinning
tidally perturbed black hole, we get
\begin{equation}
  \widehat{\Psi}_2 \sim \frac{\textrm{Mass quadrupole}}{r^5} + \textrm{External quadrupole pert}\,.
\end{equation}
The constant ($r$ independent) term represents the external
quadrupolar perturbation, and the mass quadrupole term is the response
of the black hole to this perturbation.  These two are linearly
related as in Eq.~\eqref{eq:lambda-def} via the Love number. More
generally, we will have
\begin{equation}
  \widehat{\Psi}_2 \sim \sum_{\ell\geq 2} \left( \frac{A_\ell}{r^{\ell+3}} + B_\ell r^{\ell-2} \right)\,.
\end{equation}
(Here we are suppressing the angular spherical harmonics to avoid
clutter).  As before, the non-asymptotically flat terms (i.e. the
$B_\ell$) represent the external fields, while the $A_\ell$ represent
the response of the black holes.  The linear relation between these
yield the Love numbers; the real parts of the $A_\ell$ are the
(perturbations of) mass multipole moments while the imaginary parts are
the (perturbations of) spin (or ``magnetic'') moments.  From the
result of Eq.~\eqref{eqs:psi-non-spinning}, we see that there are no
additional powers of $1/r$ beyond $M/r^3$, which means that the field
tidal Love numbers vanish.  The same is true for the slowly spinning
case shown in Eq.~\eqref{eq:Weyl-scalars-small-a}. Thus we again
conclude, as elsewhere in the literature, that the tidal Love numbers
vanish for slowly spinning Kerr black holes.

\subsection{Systematic uncertainties in the field multipole moments}
\label{subsec:multipole-uncertainties}

We now turn to potential limitations of the above discussion, related
to systematic uncertainties connected with the measurements of mass,
spin (and multipole moments) based on the asymptotic behavior of the
field.  This discussion follows closely the work of Hartle \& Thorne
\cite{Thorne:1984mz}, which employs matched asymptotic expansions to
find the equation of motion of a black hole moving in an external
field (see also \cite{DEath:1975jps,damour1983gravitational}).  

Consider a black hole of mass $M$ moving in a background spacetime
with a radius of curvature $\mathscr{R}$, taken to be much larger than
$M$.  We can then construct two different expansions for the spacetime
metric in the vicinity of the black hole.  The first is the expansion
where $\mathscr{R}$ is taken to be a large parameter:
\begin{equation}
  g_{ab} = g_{ab}^{[0]} + \mathscr{R}^{-1}g_{ab}^{[1]} + \mathscr{R}^{-2}g_{ab}^{[2]}+\ldots\,.
\end{equation}
Here $g_{ab}^{[0]}$ is just the Kerr or Schwarzschild metric and the
successive terms are perturbations due to the effect of the external
universe.  The second expansion is to start with the external universe
metric at the location of the black hole, with $M$ now taken to be a
small parameter:
\begin{equation}
  g_{ab} = g_{ab}^{(0)} +  M g_{ab}^{(1)} + M^2g_{ab}^{(2)}+\ldots\,.
\end{equation}
Within this formalism, it is assumed that the mass of a black hole can
be defined (and measured) precisely only for an isolated black hole in
an asymptotically flat spacetime.  This corresponds to using the Kerr
metric $g_{ab}^{[0]}$, consider it as an expansion in powers of $M/r$,
and define the mass based on the asymptotic behavior of this metric
for large $r$, by using surface integrals/multipole decompositions.
We thus assume the existence of a \emph{buffer} zone surrounding the
black hole, with a radius much larger than $M$ but simultaneously much
smaller than $\mathscr{R}$. In this buffer region, we then attempt to
measure the physical parameters of the black hole again via surface
integrals/multipole expansions.  In this procedure, we can move terms
from one part of expansion to another.  Thus, as argued in
\cite{Thorne:1984mz}, when measuring the mass of the black hole, terms
of the form of $M^3/(r\mathscr{R}^2)$ need to be considered as
well. This leads to a systematic uncertainty in mass measurements
\begin{equation}
  \frac{\Delta M}{M} \sim \frac{M^2}{\mathscr{R}^2}\,.
\end{equation}
For our purposes here, the same argument can be applied to the
quadrupole moment.  This appears at order $1/r^3$ in
$g_{ab}^{[0]}$, which can be mimicked by terms of the form
$M^2/(r^3\mathscr{R}^2)$. Thus, the uncertainty in
measurements of the quadrupole moment is also of the form
\begin{equation}
  \frac{\Delta Q}{Q} \sim \frac{M^2}{\mathscr{R}^2}\,.
\end{equation}
For a binary system, the external universe is just the gravitational
field of the companion.  If $M_2$ is the mass of the companion and if
the separation is $d$, then $\mathscr{R}^2 \sim d^3/M_2$ so that
\begin{equation}
  \frac{\Delta Q}{Q} \sim \frac{M^2M_2}{d^3}\,.
\end{equation}
This is in practice a rather small uncertainty and not relevant for
current observations. If we have $M_2 = qM$ and, in the worst case, if
the black holes are close to the merger so that
$d\sim 2M + 2M_2 = 2M(1+q)$, then
\begin{equation}
  \frac{\Delta Q}{Q} \sim \frac{q}{8(1+q)^3}\,.
\end{equation}
This has a maximum value of $1/54$ for $q=1/2$.  This corresponds to,
at worst, $< 2\%$ uncertainty, much smaller than uncertainties for any
of the binary merger events observed thus far.  Realistic estimates
will be smaller than this, since they will apply to larger values of
$d$.  This might however be relevant for loud events observed by the
next generation of ground- and space-based gravitational wave
detectors.

\subsection{The Surficial Tidal Love numbers}
\label{subsec:surficial-love}

We can contrast the above discussion with how the source (i.e. the
``surficial'') multipole moments respond to the external perturbation.
The surficial Love numbers are indeed modified by the external
perturbation, and the corresponding Love numbers of both electric and
magnetic type, can again be read-off from $\Psi_2$, but now from its
value at the horizon.  Let us discuss how this can be done.  The
starting point for our analysis was to choose a perturbation
$\widehat{\Psi}_2$ at the horizon given in Eq.~\eqref{eq:psi2-prime}.
The coefficients
$\widehat{k}_{lm}=\widehat{e}_{lm} + i\widehat{b}_{lm}$ appearing here
are perturbations of the corresponding source multipole moments, with
$\widehat{e}_{lm}$ being the ``electric'' component and
$\widehat{b}_{lm}$ being the ``magnetic'' component.  Starting with
this horizon perturbation, we have derived the solution for
$\widehat{\Psi}_2$ away from the horizon including its asymptotic
behavior; see Eq.~\eqref{eq:psi2hat-asymptotic}.  Considering, the
dominant term for each $Y_{lm}$, we get the following asymptotic
behavior:
\begin{equation}
  \widehat{\Psi}_2^\infty \sim \sum_{l,m}\widehat{k}_{lm}^\infty \left(\frac{r}{c}\right)^{l-2}Y_{lm}\,.
\end{equation}
The value of $\widehat{k}_{lm}^\infty$ is obtained by taking the $n=l-2$
term in the sum over $n$ in Eq.~\eqref{eq:psi2hat-asymptotic}:
\begin{equation}
  \widehat{k}_{lm}^\infty = \widehat{k}_{lm}\frac{(l+3)_{l-2}}{(1)_{l-2}}\,.
\end{equation}
Turning our mathematical procedure around, we interpret the horizon
deformation as having been caused by this asymptotic external tidal
field. Thus, one possible definition of the surficial Love number --- which is
natural from this perspective --- is just the ratios of these
coefficients.  Writing
$\widehat{k}_{lm}^\infty = \widehat{e}_{lm}^\infty +
i\widehat{b}_{lm}^\infty$, the electric and magnetic surficial Love
numbers would be respectively
\begin{equation}
  \frac{\widehat{e}_{lm}}{\widehat{e}_{lm}^\infty} \quad \textrm{and} \quad  \frac{\widehat{b}_{lm}}{\widehat{b}_{lm}^\infty}\,.
\end{equation}
Our solution for $\widehat{\Psi}_2$ then shows that both of these
ratios are independent of $m$ and are equal to
\begin{equation}
  h^\prime_{l} =  \frac{(1)_{l-2}}{(l+3)_{l-2}} = \frac{(l-2)!(l+2)!}{(2l)!}\,.
\end{equation}
Numerical values of this ratio for some values of $l$ are:
\begin{equation}
  h^\prime_2 = 1\,,\quad h^\prime_3 = \frac{1}{6}\,,\quad h^\prime_4 = \frac{1}{28}\,.
\end{equation}
It will also be worthwhile to compare this with other calculations of
black hole surficial Love numbers, referred to as the ``shape'' Love
numbers $h_l$ by Damour \& Lecian, and Poisson \& Landry
\cite{Damour:2009va,Landry:2014jka}.  Their approach is based on a
study of the static, axisymmetric Weyl solution for two black
holes. The horizon distortion there is defined in terms of the moments
of the Gaussian curvature of the horizon, while the external tidal
field is taken to be the external gravitational potential (due to the
other black hole) at the unperturbed location of the horizon.  The
shape Love number is the ratio between these, and it leads to the
following result:
\begin{equation}
  h_l = \frac{l+1}{l-1}\frac{(l!)^2}{2(2l)!}\,.
\end{equation}
While these are similar to the $h_l^\prime$ (e.g. both decay rapidly
with increasing $l$), they are not identical:
\begin{equation}
  \frac{h^\prime_l}{h_l} = \frac{2(l+2)}{l}\,.
\end{equation}
It is not surprising that the two results differ, and in fact the
difference can be understood just as a different scaling of the
potential and scalar curvature components with $l$.  In this work,
$\widehat{\Psi}_2$ serves a double role: at the horizon, it provides
the distortion of the horizon geometry, while asymptotically it yields
the external tidal potential.  We have accordingly used it to define
surficial Love numbers.

On the other hand, Refs.~\cite{Damour:2009va,Landry:2014jka} use the
external potential at the location of the world-line (in the absence
of the black hole) instead of the asymptotic form of
$\widehat{\Psi}_2$.  Note that in our case, we can obtain the external
potential also by taking the limit of a vanishingly small black hole,
i.e. $c\rightarrow 0$, for fixed $r$~\footnote{ By using this convention for the external field we would naturally recover the usual definition of shape Love number $h_l$.}.  Once again, this will select
the $n=l-2$ term in Eq.~\eqref{eq:psi2hat-asymptotic} as the dominant
one.  Moreover, Refs.~\cite{Damour:2009va,Landry:2014jka} also use different
conventions.  They take the external potential to be of the form
\begin{equation}
  \sum_l \frac{r^l}{l(l-1)}f_l\,,
\end{equation}
where $f_l$ is an angular function (ignoring constant
$l$-independent terms).  On the other hand, the perturbation of the
scalar curvature at the horizon is taken to be of the form
\begin{equation}
  \sum_l\frac{4(l+2)}{l}h_lR^{l-2}f_l\,.
\end{equation}
We thus find different $l$-dependent factors in these expansions
compared to our results, and this also affects the definition of
the Love numbers.  It is easy to verify that this accounts for the
differences in the Love number definitions and that we recover the results
given in \cite{Damour:2009va,Landry:2014jka} for $h_l$ if we were to use these
redefinitions.

As we have mentioned earlier, the field multipole moments are believed
to be important for the gravitational wave signal and have thus
justifiably attracted greater attention.  We have also argued above
that the systematic uncertainties in these should be negligible for
current gravitational wave observations.  Nonetheless, loud events in
the next-generation detectors, where precision tests of general
relativity will be especially interesting, these uncertainties might
need to be taken into account.  Nevertheless, can we make a case for
the relevance of the surficial Love numbers? Here we note that it is
in fact now common in numerical relativity to calculate black hole
masses and spins from surface measurements at horizons following
Eqs.~\eqref{eq:J-ih} and \eqref{eq:mass-ih} (see
e.g. \cite{Dreyer:2002mx}).  These turn out to be reliable even in
dynamical situations close to the merger.  If one were to study tidal
deformabilities in numerical simulations, it would be difficult in
practice to work in the mass-centered coordinate system, which is
assumed in all of the current analytical work.  On the other hand,
surface deformations of horizons and the surficial multipole moments
are much easier to compute in these simulations \cite{Prasad:2021dfr}.
Since the multipole moments determine the near horizon metric as we
have seen here, it is plausible that they could appear in the
gravitational waveform as well, though it is not yet clear how.  For
the purposes of both high-precision gravitational wave astronomy and
numerical studies, it would therefore be of interest to explore if the
surficial Love numbers can be measured directly from gravitational
wave signals.

\section{Conclusion}
\label{sec:conclusion}

In this work, we have considered tidal deformations of slowly spinning
black holes and we have calculated the field and surficial Love
numbers.  Similar results have appeared in the literature previously
but what is new here is the application of the notions of isolated
horizons (and the Newman-Penrose formalism) to the problem.  This
yields the near horizon geometry in greater detail than before and
provides us with the metric, spin coefficients, and curvature
components.  Moreover, this approach clarifies in various places the
role of the horizon geometry and the various assumptions commonly
employed in these calculations.  For example, requiring time
independence at the horizon and the radiation content already imposes
strong restrictions on the allowed tidal perturbations of the Weyl
tensor component $\Psi_2$.  Apart from calculations of Love numbers
and the relation between field and surficial deformations that we have
focused on, these results can help in further applications of the near
horizon geometry.  These include the effect on the light ring,
particle orbits, black hole shadows and construction of initial data.
The extension to a general Kerr black hole with arbitrary spins is, in
principle, a straightforward extension of this work and will be
presented elsewhere.  Moreover, it should also be possible to include
additional fields within alternate theories of gravity that admit
black hole solutions.

This work can be extended in several useful directions and we mention
a few here.  The first is to extend the perturbative framework to
include small amounts of infalling radiation, i.e. non-vanishing
$\Psi_0$ at the horizon.  In the present work, our boundary conditions
impose that the horizon is precisely non-expanding and this should be
relaxed.  A non-expanding horizon can indeed be perturbed by including
infalling radiation to linear order \cite{Ashtekar:2021kqj}, and fluxes
and charges can be computed. This would allow, for instance, to explore the connection between the horizon geometry and the tidal heating~\cite{Datta_2020}.  The characteristic formulation can be
extended to encompass this situation. This analysis would open a way
to an analytical study of \emph{correlations} between the outgoing
radiation far away from the black hole (i.e. $\Psi_4$ on
$\mathcal{N}$) and the ingoing flux at the horizon (i.e. $\Psi_0$ at
$\Delta$), thus providing a link between gravitational wave
observations and horizon dynamics.  This has previously been
investigated as well in numerical studies (see
e.g. \cite{Jaramillo:2011rf,Jaramillo:2012rr,Rezzolla:2010df,Gupta:2018znn,Prasad:2020xgr}).

The second avenue for future applications is in extracting gauge
invariant information from numerically computed binary black hole
spacetimes. In a numerical simulation, it is generally useful to keep
track of black hole mass, angular momentum and higher multipole
moments.  Now in a binary black hole merger, we will have a regime
late in the inspiral (but before the merger) when the two black holes
are sufficiently distorted due to the tidal effects of its
companion. Moreover, numerical results show that, somewhat
surprisingly, the two horizons are in fact close to isolated in this
regime with insignificant area increase
\cite{PhysRevLett.123.171102}. Thus, the results of this work should
be applicable and it should be possible to model the near horizon
spacetime in a characteristic formalism like we have done here.  A
successful completion of this program should lead to more insights in
the binary black hole problem.

We finally speculate on a potential application in gravitational wave
astronomy.  An important goal of gravitational wave astronomy is to be
able to distinguish between black hole and neutron stars on the basis
of the gravitational Love number measurements (see e.g. \cite{Brown:2021seh});
vanishing gravitational Love numbers are taken to be signatures of black holes while
for neutron stars, gravitational Love number measurements are employed to
infer the equation of state for neutron star matter.  As discussed
in Sec.~\ref{sec:lovenumbers}, there might be a role for the surficial Love numbers in the
late inspiral where we may not have access to the gravitational Love
numbers.  In this regime, the gravitational wave signal might carry an imprint of the surficial Love numbers.  This is likely not important
for the currently operating gravitational wave detectors, but might be
relevant for high-precision measurements with the next generation
observatories.

\section*{Acknowledgments}

We are grateful to Abhay Ashtekar, Tanja Hinderer, Scott Hughes, and Eric Poisson for
useful discussions. We also thank Scott Hughes and Eric Poisson for their valuable comments on the manuscript. 

\bibliographystyle{apsrev4-1}
\bibliography{main}{}

\appendix

\section{Notation, conventions and some basic formulae}
\label{sec:notation}

To aid the reader in following the main text, in this appendix we
collect some of the basic formulae and notation used throughout this paper.
There are cases where the same symbol is used for different objects,
and need to be understood in context.  For example, $\Delta$ is the
null surface representing a NEH/WIH and it is also the directional
derivative along $n^a$: $\Delta = n^a\nabla_a$. 
\begin{itemize}
\item All manifolds and fields are assumed to be smooth unless stated
  otherwise. The spacetime metric is $g_{ab}$ with signature
  $(-+++)$. The spacetime derivative operator compatible with $g_{ab}$
  is $\nabla_a$, and the Riemann tensor is defined via
  $2\nabla_{[a}\nabla_{b]}X_c = {R_{abc}}^dX_d$, the Ricci tensor is
  $R_{ab} = {R_{acb}}^c$ and Ricci scalar is $R= g^{ab}R_{ab}$. We use
  the usual notation for symmetrization and anti-symmetrization of indices
  $X_{(ab)} = \frac{1}{2}(X_{ab} + X_{ba})$, and
  $X_{[ab]} = \frac{1}{2}(X_{ab} - X_{ba})$.  
\item Isolated horizon:
  $\Delta \sim \widetilde{\Delta}\times\mathbb{R}$ is the null surface
  representing the horizon while $\widetilde{\Delta}$ is the ``base
  space'' of spherical topology obtained by taking the quotient by the
  null generators.
 \item Quantities with a $\widetilde{}$ represent fields on either
  $\widetilde{\Delta}$ or a cross-section of the horizon. Thus,
  $(q_{ab},\epsilon_{ab},\omega_a)$ are respectively the metric,
  volume 2-form and connection 1-form on $\Delta$,while
  $(\widetilde{q}_{ab},\widetilde{\epsilon}_{ab},\widetilde{\omega}_a)$
  are respectively the corresponding quantities on
  $\widetilde{\Delta}$ or projected onto a cross-section $S$ of
  $\Delta$.
\item The Newman-Penrose null tetrad is $(\ell,n,m,\bar{m})$, with the
  corresponding directional derivatives
  $(D,\Delta,\delta,\bar{\delta})$. While these are defined in a
  neighborhood of the horizon following our construction of the near
  horizon geometry, when referring to the horizon, the vector $m^a$
  lives on $\widetilde{\Delta}$ or it is tangent to a cross-section of
  $\Delta$. Similarly, $\ell^a$ is a null-normal to $\Delta$ while
  $n_a$ is the 1-form orthogonal to cross-sections of $\Delta$.

\item \emph{Quantities on unit 2-spheres}: We shall frequently deal with a
  2-manifold $S$ of spherical topology, and we often work with complex
  coordinates $(z,\bar{z})$ on $S$.  The metric on a unit 2-sphere is 
  \begin{equation}
    \label{eq:appendix-2metric-P}
    ds^2 = \frac{2}{P^2(z,\bar{z})}dz\,d\bar{z}\,.
  \end{equation}
  For a ``round'' 2-sphere, we denote $P$ by $P_0$ and
  \begin{equation}
    P_0 = \frac{1}{\sqrt{2}}(1+z\bar{z})\,.
  \end{equation}
  For a sphere of area-radius $R$, we need to modify $P\to P/R$ in all
  the expressions appearing in the rest of this appendix.  The complex
  null 1-form $m_a$ and the vector $m^a$ (satisfying
  $m\cdot\bar{m} = 1$) are respectively
  \begin{equation}
    m = \frac{1}{P}dz\,,\quad m^a\partial_a = P\frac{\partial}{\partial\bar{z}}\,.
  \end{equation}
  Its exterior derivative is useful:
  \begin{equation}
    dm = \frac{\partial P}{\partial\bar{z}}m\wedge\bar{m}\,.
  \end{equation}
  The volume 2-form is $\widetilde{\epsilon} = im\wedge \bar{m}$, and
  the Hodge dual of a 1-form is
  $\star X_a = {\widetilde{\epsilon}_a}{}^bX_b$. Note that
  $\star m_a = im_a$ and $\star\star X = -X$.  As an example, for
  $\widetilde{\omega}_a = \pi m_a + \bar{\pi}\bar{m}_a$, its dual is
  \begin{equation}
    \star\widetilde{\omega}_a = i\pi m_a - i\bar{\pi}\bar{m}_a\,.
  \end{equation}
  
\item Covariant derivatives on $S$ are encapsulated by a single
  complex number which, in the Newman-Penrose notation is
  $\alpha-\bar{\beta}$ which has been called $a$ in the main
  text. This is easily calculated using
  $\delta m = (\beta-\bar{\alpha})m$ which implies
  \begin{eqnarray}
    \beta-\bar{\alpha} &=& \bar{m}^a\delta m = \bar{m}^am^b\nabla_bm_a\\
                       &=& \bar{m}^am^b(\nabla_bm_a - \nabla_am_b) = -\frac{\partial P}{\partial\bar{z}}\,.
  \end{eqnarray}
  
\item \emph{The $\eth$ operator}: We have defined the $\eth$ operator
  in Eqs.~\eqref{eq:19}-\eqref{eq:55}. Here we give some basic expressions for its action on
  the spin coefficient $\pi$, which is the quantity of most interest
  for us in this regard.  First using
  $\pi = \bar{m}^a\widetilde{\omega}_a$, we see that $\pi$ has spin
  weight $-1$ so that $\eth\pi$
  will have spin weight $0$.  From the definition it follows that
  \begin{equation}
    \eth\pi = \bar{m}^a\delta\widetilde\omega_a = \delta\pi-\frac{\partial P}{\partial\bar{z}}\pi = P^2\frac{\partial}{\partial\bar{z}}\left(\frac{\pi}{P}\right)\,.
  \end{equation}

\item\emph{The Laplace-Beltrami operator}: For the metric given in
  Eq.~\eqref{eq:appendix-2metric-P}, the Laplace-Beltrami operator is
  denoted $\Delta_P$.  Acting on a scalar $f$ it can be directly
  calculated as:
  \begin{equation}
    \Delta_P f := \frac{1}{\sqrt{\widetilde{q}}}\frac{\partial}{\partial x^a}\left(\sqrt{\widetilde{q}}\,\widetilde{q}^{ab}\frac{\partial f}{\partial x^b} \right) = 2P^2\frac{\partial^2 f}{\partial z\partial\bar{z}}\,.
  \end{equation}
  From the definitions above, this can also be written as
  \begin{equation}
    \Delta_P f = -\star d\star df\,.
  \end{equation}
  Similarly, the definition for the divergence of
  $\widetilde{\omega}_a$ and some alternate expressions used for it in
  the main text are
  \begin{eqnarray}
    \textrm{div}\; \widetilde{\omega} &=& \frac{1}{\sqrt{\widetilde{q}}}\frac{\partial}{\partial x^a}\left(\sqrt{\widetilde{q}}\,\widetilde{q}^{ab}\widetilde{\omega}_b\right)\\
                                   &=& -\star d\star\widetilde{\omega} = \eth\pi + \bar{\eth}\bar{\pi}\,.
  \end{eqnarray}
  
\end{itemize}

\section{Spin-weighted spherical harmonics}
\label{sec:spin-weighted-SH}
The spin-weighted spherical harmonics are defined in~\cite{Goldberg:1966uu} as
\begin{equation}
\begin{split}
     {}_sY_{lm} = \frac{a_{lm}}{\sqrt{(l-s)!(l+s)!}} (1+z\bar{z})^{-l} \times\\
    \sum_p \left(\begin{matrix}
        l-s\\
        p
    \end{matrix}\right)\left(\begin{matrix}
        l+s\\
        p+s-m
    \end{matrix}\right) z^p(-\bar{z})^{p+s-m}
\end{split}
\end{equation}for the complex coordinates $\{z,\bar{z}\}$, where
\begin{equation}
    l\geq0\,,\quad -l\leq m \leq l\,,\quad |s|\leq l\,.
\end{equation}
In Tab.~\ref{tab:spin-weighted-SH}, we present explicitly the three lowest harmonics $l=1,2,3$ with $m=0$ for the spins $s=0,1$ and 2, which appear in our expressions for the tidally perturbed Schwarzschild isolated horizon~\eqref{eq:tidal-perturbed-Weyl-scalars}-\eqref{eq:tidal-perturbed-tetrad}. 
\begin{table}[]
    \centering
    \begin{tabular}{c|c|c|c}
    \hline
     ${}_s Y_{lm}$ & $l=1$ & $l=2$  & $l=3$    \\
\hline
$s=0$ &$ \sqrt{\frac{3}{4\pi}} \frac{z\bar{z}-1}{z\bar{z}+1}$  &  $\sqrt{\frac{5}{4\pi}} \frac{1-4 z\bar{z}+z^2\bar{z}^2}{(z\bar{z}+1)^2}$ &$\sqrt{\frac{7}{4\pi}} \frac{-1+9 z\bar{z}-9z^2\bar{z}^2+z^3\bar{z}^3}{(z\bar{z}+1)^3}$ \\
$s=1$ & $\sqrt{\frac{3}{4\pi}} \frac{\bar{z}}{1+z\bar{z}}$ &$\sqrt{\frac{15}{2\pi}} \frac{\bar{z} (z\bar{z}-1)}{(z\bar{z}+1)^2}$  & $\sqrt{\frac{21}{\pi}} \frac{\bar{z}(1-3 z\bar{z}+z^2\bar{z}^2)}{(1+z\bar{z})^3}$ \\
$s=2$ & -  & $\sqrt{\frac{15}{2\pi}} \frac{\bar{z}^2}{(1+z\bar{z})^2}$  & $-\sqrt{\frac{105}{8\pi}} \bar{z}^2\frac{1-2z\bar{z}}{(1+z\bar{z})^3}$\\
\hline
    \end{tabular}
    \caption{Spin-weighted spherical harmonics in the complex coordinates $\{z,\bar{z}\}$ for $m=0$,  $l=1,2,3$, and $s=0,1,2$. }
    \label{tab:spin-weighted-SH}
\end{table} The spin-weighted spherical harmonics with negative spin can be easily obtained from Tab.~\ref{tab:spin-weighted-SH} using
\begin{equation}
     {}_s \bar{Y}_{lm} = (-1)^{m+s} {}_{-s} Y_{lm}\,.
\end{equation} 

The operator $\eth$ is defined in the main text in Eqs.~\eqref{eq:19}-\eqref{eq:55} for an arbitrary derivative operator $\delta$. In practice, we used the angular derivative operator of the unperturbed spacetime (Schwarzschild) to compute the expressions in Tab.~\ref{tab:spin-weighted-SH}, i.e., 
\begin{equation}
    \delta = \frac{P_\circ}{(r+c)} \partial_z\,,
\end{equation} where 
\begin{equation}
    P_\circ = \frac{1}{\sqrt{2}} (1+z\bar{z})\,.
\end{equation} At the horizon $r=0$, this operator is simply $\delta=\frac{1+z\bar{z}}{\sqrt{2}c}\partial_z$. It is useful to notice that we can express the Laplacian in terms of the $\eth$ and $\bar{\eth}$ operators for a spin-0 quantity
\begin{equation}
    \Delta_P \eta = 2\bar{\eth}\eth \eta\,.
\end{equation}
Finally, we recap some of the most important properties of the spin-$s$ spherical harmonics summarized in~\cite{Goldberg:1966uu}
\begin{align}
    \eth \, {}_s Y_{lm} & = \frac{1}{\sqrt{2} (r+c)}\sqrt{(l-s)(l+s+1)} {}_{s+1} Y_{lm}\,,\\
    \bar{\eth}\, {}_s Y_{lm} & =-\frac{1}{\sqrt{2} (r+c)} \sqrt{(l+s)(l-s+1)} {}_{s-1} Y_{lm}\,,\\
    \bar{\eth}\eth \, {}_sY_{lm}&=-\frac{(l-s)(l+s+1)}{2(r+c)^2} {}_sY_{lm}\,.
\end{align}

\section{The asymptotic behavior of $\widehat{\Psi}_2$ and $\widehat{U}$}
\label{sec:Asymptotics}

In the following, we analyze in detail the asymptotic behavior of $\widehat{\Psi}_2$ and $\widehat{U}$, which have been summarized in Sec.~\eqref{subsec:asymptotics}. For convenience, let's take the $r$-dependent piece of $\widehat{\Psi}_2$
\begin{equation}
    I_l=\frac{F_1^l(r)}{r+c}
\end{equation} defined in Eq.~\eqref{eq:shortcut-function}. Since the first argument in both hypergeometric functions is negative for $l\geq 2$, the hypergeometric series has a finite number of terms and converges for arbitrary argument, i.e., we can use 
\begin{equation}\label{eq:hypergeometric-fnct-finite-app}
    {}_2F_1[-m, b,c,z] = \sum_{n=0}^m (-1)^n \left(\begin{matrix}
        m\\
        n
    \end{matrix}\right)\frac{(b)_n}{(c)_n} z^n\,,
\end{equation} where $m\geq0$ and $(b)_n= \Gamma[b+n]/\Gamma[b]$ is the Pochhammer symbol (see for instance chapter 9.1 in~\cite{gradshteyn_table_2000}) to rewrite $I_l$ as
\begin{equation}
\begin{split}
  I_l=\frac{1}{r+c} \left((l-1) \sum_{n=0}^{l-1} \left(\begin{matrix}
        l-1\\
        n
    \end{matrix}\right) \frac{(l+2)_n}{(1)_n} \left(\frac{r}{c}\right)^n\right.\\
    \left.+ 3 \sum_{n=0}^{l-2} \left(\begin{matrix}
        l-2\\
        n
    \end{matrix}\right) \frac{(l+2)_n}{(1)_n} \left(\frac{r}{c}\right)^n  \right)  
\end{split}
\end{equation} Extracting the last term from the first sum, and simplifying the coefficients we can write the above expression as
\begin{equation}
    I_l = \frac{1}{r+c} \left(A_l \left(\frac{r}{c}\right)^{l-1}+\sum_{n=0}^{l-2} B_{n,l} \left(\frac{r}{c}\right)^n\right)\,,
\end{equation} where 
\begin{equation}\label{eq:Al-and-Bl}
    A_l=\frac{(2l)!}{(l-2)!(l+1)!}\,,\quad B_{l,n} = \frac{(l+n+1)! (l^2+l-3n-2)}{(l-n-1)! n! n! l(l^2-1)}\,.
\end{equation} 
From this expression, we can conclude that in the large $r$ limit $I_l$ has the following $r$ powers
\begin{equation}
   \lim_{r\to\infty} I_l \sim r^{l-2},\quad r^{l-3}\,,\,.\,.\,.\quad r^0\,,\quad\frac{1}{r}\,.
\end{equation} The subdominant asymptotic term $1/r$ would yield a logarithmic term in the asymptotic behavior of $\widehat{U}$ (recall that $\widehat{U} =\int\d r\int\d r \re \widehat{\Psi}_2$), implying the presence of nonzero field Love numbers for an isolated horizon. However, as we will see now, we can factor out a term $(r+c)$ from the numerator of $I_l$, i.e., the numerator of $I_l$ can be written as
\begin{equation}\label{eq:poly-factorized}
     (r+c) \sum_{n=0}^{l-2} C_{n,l} \left(\frac{r}{c}\right)^n\,.
\end{equation} 
The simplest way to show this statement is by using the following trick: the numerator of $I_l$
\begin{equation}
   N_l= A_l \left(\frac{r}{c}\right)^{l-1}+\sum_{n=0}^{l-2} B_{n,l} \left(\frac{r}{c}\right)^n
\end{equation} is just a polynomial in $r$ of degree $l-1$. As such, it is easy to test whether $r=-c$ is a root of this polynomial. If it is, then we can rewrite this expression as~\eqref{eq:poly-factorized}. Of course, our coordinate $r\geq 0$, so it cannot take the value $-c$; this is merely a trick to show that we can factorize this term. This allows us to analyze the correct asymptotic behavior of $\widehat{\Psi}_2$ and $\widehat{U}$. Then, 
\begin{equation}\label{eq:Nl-at-minus-c}
    N_l(r=-c) = A_l\left(-1\right)^{l-1}+\sum_{n=0}^{l-2} B_{n,l} \left(-1\right)^n\,.
\end{equation} Using Eq.~\eqref{eq:Al-and-Bl} we can sum the $B_{n,l}$, 
\begin{equation}
    B_l=\sum_{n=0}^{l-2} B_{n,l} (-1)^n = (-1)^l (l-1) \frac{\Gamma[1+2l]}{l(l+1)\Gamma[l]^2}\,.
\end{equation} Combining the above equation with \eqref{eq:Al-and-Bl} and~\eqref{eq:Nl-at-minus-c}, we conclude 
\begin{equation}
    N_l(r=-c) = 0\,.
\end{equation}Consequently, $I_l$ can be written as
\begin{equation}\label{eq:Il-factorized}
    I_l = \sum_{n=0}^{l-2}C_{l,n} \left(\frac{r}{c}\right)^n
\end{equation} with
\begin{equation}\label{eq:cl0-cllm2}
    C_{l,0} = \frac{B_{l,0}}{c}\,,\quad C_{l,l-2} = \frac{A_l}{c}
\end{equation} and 
\begin{equation}\label{eq:cln}
    C_{l,n}= \sum_{k=0}^n (-1)^{k+n} \frac{B_{l,k}}{c}\,, \quad 1\leq n\leq l-2 \; .
\end{equation}
From Eq.~\eqref{eq:Il-factorized}, it is straightforward to see that in the limit $r\to\infty$, the dominant term goes like $r^{l-2}$, while the least dominant term is constant, i.e.,
\begin{equation}
     \lim_{r\to\infty}I_l \sim r^{l-2},\quad r^{l-3}\,,\,.\,.\,.\quad r^0\,.
\end{equation}

Finally, combining Eqs.~\eqref{eq:cl0-cllm2} and~\eqref{eq:cln} with Eq.~\eqref{eq:Il-factorized}, we find
\begin{equation}
\begin{split}
     I_l &= \sum_{n=0}^{l-2}\left(\frac{r}{c}\right)^n\sum_{k=0}^{n}(-1)^{k+n} \frac{B_{l,k}}{c}\\
     &= \sum_{n=0}^{l-2} \frac{(l+n+2)!}{cl(l^2-1)(l-n-2)! (n!)^2}\left(\frac{r}{c}\right)^n
\end{split}  
\end{equation}  
Using Eq.~\eqref{eq:hypergeometric-fnct-finite-app}, we can rewrite $I_l$ in the compact form
\begin{equation}
    I_l=\frac{(l+2)}{c} {}_2F_1[2-l,l+3,1,-\frac{r}{c}]\,.
\end{equation} 
Integrating this expression twice with respect to $r$ (recall that $\widehat{U} =\int\d r \int \d r \, \re \widehat{\Psi}_2$), we obtain the radial dependence of $\widehat{U}$
\begin{equation}
   J_l= r^2\frac{(l+2)}{2c} {}_2F_1[2-l,3+l,3,-\frac{r}{c}]\,.
\end{equation} Using again Eq.~\eqref{eq:hypergeometric-fnct-finite-app}, we can easily analyze the asymptotic behavior of $\widehat{U}$. In series form $J_l$ reads as 
\begin{equation}
   J_l= \frac{(l+2)r^2}{2c}\sum_{n=0}^{l-2} \left(\begin{matrix}
       l-2\\
       n
   \end{matrix}\right) \frac{(l+3)_n}{(3)_n} \left(\frac{r}{c}\right)^n \,.
\end{equation}
In the limit of large $r$, $J_l$ has the dominant term $r^l$ and the least  term $r^2$, i.e., 
\begin{equation}
    \lim_{r\to\infty} \widehat{U}\sim r^l\,,\quad r^{l-1}\,,\,.\,.\,.\quad r^2\, .
\end{equation}

\end{document}